\documentclass{aa}

\newcommand{\HI}{H\,{\scriptsize I}}
\newcommand{\CII}{[C\,{\scriptsize II}]}
\newcommand{\NII}{[N\,{\scriptsize II}]}

\newcommand{\OI}{[O\,{\scriptsize I}]}  
  
\newcommand{\CI}{[C\,{\scriptsize I}]}  
\newcommand{\kms}{km s$^{-1}$}  
  
\newcommand{\av}{A$_{\rm V}$}
\usepackage{graphicx}  
\usepackage[varg]{txfonts}
\usepackage{caption}

\begin{document} 

\title{First detection of the [CII] 158 $\mu$m line in the intermediate-velocity cloud Draco}

\author{Nicola Schneider     \inst{1}
\and Volker Ossenkopf-Okada  \inst{1} 
\and Eduard Keilmann         \inst{1}
\and Markus R\"ollig         \inst{2,1} 
\and Slawa Kabanovic         \inst{1} 
\and Lars Bonne              \inst{3}    
\and Timea Csengeri          \inst{4}    
\and Bernd Klein             \inst{5,6}    
\and Robert Simon            \inst{1}
\and Fernando Comer\'on      \inst{7}
}
\institute{I. Physikalisches Institut, Universit\"at zu K\"oln, Z\"ulpicher Str. 77, 50937 K\"oln, Germany  
\email{nschneid@ph1.uni-koeln.de}
\and Physikalischer Verein, Gesellschaft für Bildung und Wissenschaft, Robert-Mayer-Str.2, 60325 Frankfurt, Germany 
\and SOFIA Science Center, NASA Ames Research Center, Moffett Field, CA 94 045, USA 
\and Laboratoire d’Astrophysique de Bordeaux, Universit\'e de Bordeaux, CNRS, B18N, 33615 Pessac, France  
\and Max-Planck Institut f\"ur Radioastronomie, Auf dem H\"ugel 69, 53121 Bonn, Germany 
\and University of Applied Sciences Bonn-Rhein-Sieg, Grantham-Allee 20, 53757 Sankt Augustin, Germany 
\and European Southern Observatory, Karl-Schwarzschild-Str. 2, 85748 Garching, Germany 
}

\date{draft of \today}

\titlerunning{[CII] detection in Draco}  
\authorrunning{N. Schneider}  
  

\abstract {High-latitude intermediate-velocity clouds (IVCs) are part
  of the Milky Way's \HI\ halo and originate from either a galactic
  fountain process or extragalactic gas infall. They are partly
  molecular and can most of the time be identified in CO. Some of
  these regions also exhibit high-velocity cloud (HVC) gas, which is
  mostly atomic, and gas at local velocities (LVCs), which is partly
  atomic and partly molecular.  We conducted a study on the IVCs Draco
  and Spider, both were exposed to a very weak UV field, using the
  spectroscopic receiver upGREAT on the Stratospheric Observatory for
  Infrared Astronomy (SOFIA).  The 158 $\mu$m fine-structure line of
  ionized carbon (\CII) was observed, and the results are as follows:
  In Draco, the \CII\ line was detected at intermediate velocities
  (but not at local or high velocities) in four out of five positions.
  No \CII\ emission was found at any velocity in the two observed
  positions in Spider. To understand the excitation conditions of the
  gas in Draco, we analyzed complementary CO and \HI\ data as well as
  dust column density and temperature maps from {\sl Herschel}.  The
  observed \CII\ intensities suggest the presence of shocks in Draco
  that heat the gas and subsequently emit in the \CII\ cooling
  line. These shocks are likely caused by the fast cloud's motion
  toward the Galactic plane that is accompanied by collisions between
  \HI\ clouds.  The nondetection of \CII\ in the Spider IVC and LVC as
  well as in other low-density clouds at local velocities that we
  present in this paper (Polaris and Musca) supports the idea that
  highly dynamic processes are necessary for \CII\ excitation in
  UV-faint low-density regions.}
      

   \maketitle
%

\bigskip

\section{Introduction}

The formation of molecular clouds is commonly defined as the
transition of atomic to molecular hydrogen in the interstellar medium
(ISM). In steady-state and chemical equilibrium models
\citep{Tielens1985,Ewine1988,Sternberg1989,Krumholz2008}, the
formation of H$_2$ depends mostly on the local radiation field
(dissociation of H$_2$ by photons in the Lyman-Werner bands versus
H$_2$ formation on dust grains) and H$_2$ shielding efficiencies. In a
more dynamic scenario, H$_2$ formation is also governed by turbulent
mixing motions in the ISM \citep{Glover2007,Bialy2017} that cause
large- and small-scale density fluctuations.  In these dynamical
models of molecular cloud formation, H$_2$ formation happens in
shock-compressed layers in converging \HI\ flows in the warm neutral
medium
\citep{Walder1998,Klessen2000,Heitsch2006,Vaz2006,Dobbs2008,Clark2012}.
This dynamic, turbulent scenario reduces the H$_2$ formation times
from a few 10 Myr to a few Megayears \citep{Glover2007,Valdivia2016}.
The flows are driven by the complex interplay between gravity and
stellar feedback effects. Additionally, they are influenced by the
thermodynamic response of the multiphase ISM.  However, the
observation of these large-scale flows presents a challenge due to the
selective nature of CO molecule formation.  Notably, CO only becomes
apparent at the shocked stagnation points within the broader turbulent
flow, making the \HI\ and H$_2$ gas "CO-dark".  In a recent study by
\citet{Schneider2023}, it was demonstrated that the 158 $\mu$m line of
ionized carbon (see below) can effectively characterize this component
and unveil high-velocity \HI\ flows within the Cygnus X region.

The formation of H$_2$ and CO critically depends on the local
interstellar radiation field\footnote{We express the far ultraviolet
(FUV) field in units of Habing G$_{\rm o}$ \citep{Habing1968} or
Draine $\chi$ \citep{Draine1978}, with $\chi$ = 1.71 G$_{\rm o}$.} and
self-shielding efficiencies. In plane-parallel photodissociation
region (PDR) models for low-column density,\footnote{We used the
conversion N(H) = 1.87 \, A$_{\rm V}$ 10$^{21}$ cm$^{-2}$ mag$^{-1}$ and
N(H$_2$) = 0.94 \, A$_{\rm V}$ 10$^{21}$ cm$^{-2}$ mag$^{-1}$
\citep{Bohlin1978} with the total hydrogen column density N(H) and the
molecular hydrogen column density N(H$_2$) and the visual extinction
A$_{\rm V}$.} the transition typically takes place at values of
A$_{\rm eff,V} \approx 0.3$ for H$_2$
\citep{Roellig2007,Glover2010,Sternberg2014,Bisbas2019,Schneider2022}.
We note that the model A$_{\rm eff,V}$ is the local visual extinction
at each point in the cloud and not directly comparable to the A$_{\rm
  V}$ that is an average along the line of sight
\citep{Seifried2022}. This is why it is difficult to observationally
trace the \HI-to-H$_2$ transition, though it has been the subject of a
number of observational studies using various tracers
\citep{Imara2016}.

The thermal state of the gas is mostly regulated by photoelectric
heating and cooling through dust and collisionally excited emission
from far-infrared (FIR) molecular and atomic fine-structure lines.
Nevertheless, cooling rates exhibit only a weak dependence on
temperature for T$<$10$^4$ K.  Consequently, the cooling process
driven by atomic fine-structure lines (primarily the \CII\ 158 $\mu$m
line) induces a thermal instability, giving rise to a multiphase ISM
\citep{Field1969,Wolfire1995}.  This multiphase ISM includes a
volume-filling warm neutral gas (referred to as the \HI\ WNM)
characterized by temperatures around T$\sim$8000 K and densities of
$n\sim1$ cm$^{-3}$ in pressure equilibrium with the cold neutral
medium (CNM) exhibiting temperatures of approximately T$\sim30$--100 K
and densities of $n\sim50$--100 cm$^{-3}$ within the atomic phase. The
ISM also encompasses H$_2$ gas with temperatures typically below 30 K
and densities $n$ exceeding a few hundred cm$^{-3}$.  Observing gas in
these thermally unstable conditions poses a considerable challenge.

One tracer for the gas conditions in very different physical
environments is the \CII\ 158 $\mu$m line. It serves as a cooling
line for gas over a large range of temperatures, typically T$\sim$100
K, and densities, typically above a few 10$^3$ cm$^{-3}$, in PDRs
\citep{Hollenbach1991, Ossenkopf2013} and at much lower temperatures
(around 20 K) in CO-dark but H$_2$-rich regions
\citep{Wolfire2010,Schneider2023} and can also arise from the warm
ionized medium and from diffuse atomic gas at lower densities
\citep{Pineda2013,Beuther2014,Kabanovic2022}.  The transition from
ionized carbon (C$^+$) to CO occurs deeper in the cloud, typically at
A$_{\rm eff,V}\sim$1
\citep{Lee1996,Visser2009,Glover2010,Seifried2020}.  Here, the
photodissociation of CO by FUV photons dominates over the production
reaction \citep{Wolfire2010,Clark2012,Glover2015}.  The \CII\ 158
$\mu$m line is easy to excite thermally by collisions with electrons
and atomic and molecular hydrogen. The critical density, defined
by the collisional de-excitation rate being equal to the effective
spontaneous decay rate, depends on the temperature.  It is
9~cm$^{-3}$, 3$\times$10$^3$~cm$^{-3}$, and
6.1$\times$10$^3$~cm$^{-3}$ for collisions with e$^-$, H, and H$_2$,
respectively, for gas temperatures $\lesssim$100 K
\citep{Goldsmith2012}. In diffuse gas, the excitation temperature of
\CII, which can go down to less than 20 K \citep{Kabanovic2022}, is
notably lower than the kinetic temperature because the densities are
too low to produce a collisional excitation rate comparable to the
spontaneous decay rate \citep{Goldsmith2012}.  In addition, the
\CII\ line can also serve as a cooling line in low to moderate
velocity C-type shocks with a low incident UV field
\citep{Lesaffre2013}.  The authors developed models of low-UV
irradiated low and moderate (up to 40 km s$^{-1}$) C- and J-type
shocks and compared the results with observations. They concluded that
\CII\ is a good tracer for the dissipation of kinetic and magnetic
energy in weakly shielded gas where it is the dominant carbon species.

While there is an increasing number of studies focusing on
\CII\ emission in bright PDRs, \CII\ studies of diffuse, translucent,
and low-column density clouds,\footnote{Clouds with \av$<$1, \av=1-5,
and \av$>$5 are typically called diffuse, translucent, and low-column
density or dark clouds \citep{Barriault2010a} respectively.} are
relatively scarce.  In a work conducted by \citet{Goldsmith2018}, a
limited number of diffuse interstellar clouds were observed using
\HI\ absorption features measured against a background quasar. This
approach allowed for the sampling of the entire line of sight through
the Galaxy. The researchers concluded that photoelectric heating
stands out as the predominant heating mechanism for these clouds, with
the \CII\ line being the principal cooling line.

In this study, we present novel observational findings pertaining to
\CII\ and CO emissions within the Draco and Spider high-latitude
diffuse clouds. We utilized archival \CII\ and CO data concerning the
quiescent Polaris cloud and the low-density Musca molecular cloud. All
of these clouds are subject to a low-incident UV field, which is,
depending on the method used to infer the UV radiation field, between
$\sim$1.5 to $\sim$10 G$_{\rm o}$.  The \CII\ emission line has been
exclusively detected in the intermediate-velocity cloud Draco,
with no detection observed at local velocities in the Spider, Polaris,
and Musca clouds. A comprehensive comparison of the physical
characteristics across these distinct clouds was conducted using
models of PDR emission, shocks, and non-LTE (Local Thermodynamic
Equilibrium) conditions, leading to an argument in favor of shock
excitation as the most plausible rationale for the presence of
\CII\ within the Draco cloud. This deduction is based on the cloud's
high-velocity descent toward the Galactic plane and its interaction
between \HI\ clouds.  Consequently, this highlights the significance
of dynamics in influencing the chemical evolution within interstellar
clouds.

We start with a description of the sources (Sect.~\ref{sec:sources})
and the observations (Sect.~\ref{sec:obs}). The observations are
presented in Sect.~\ref{sec:results} and analyzed in
Sect.~\ref{sec:analysis}. Section~\ref{discuss} provides a discussion
of the results, and Sect.~\ref{sec:summary} presents a summary of the
paper.

\begin{figure*}
\centering
\includegraphics[width=7cm,angle=0]{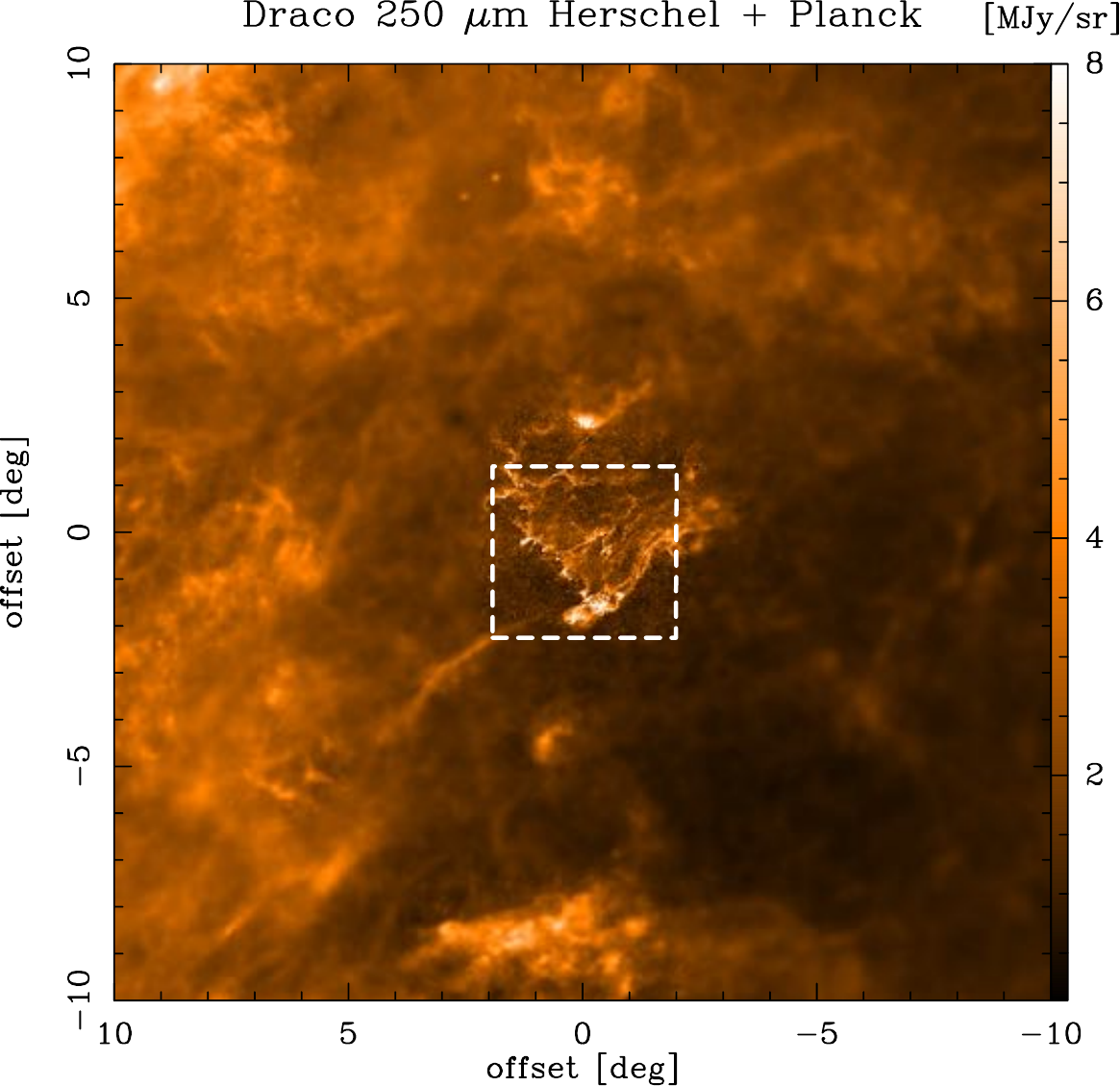}
\hspace{0.5cm}
\includegraphics[width=7cm,angle=0]{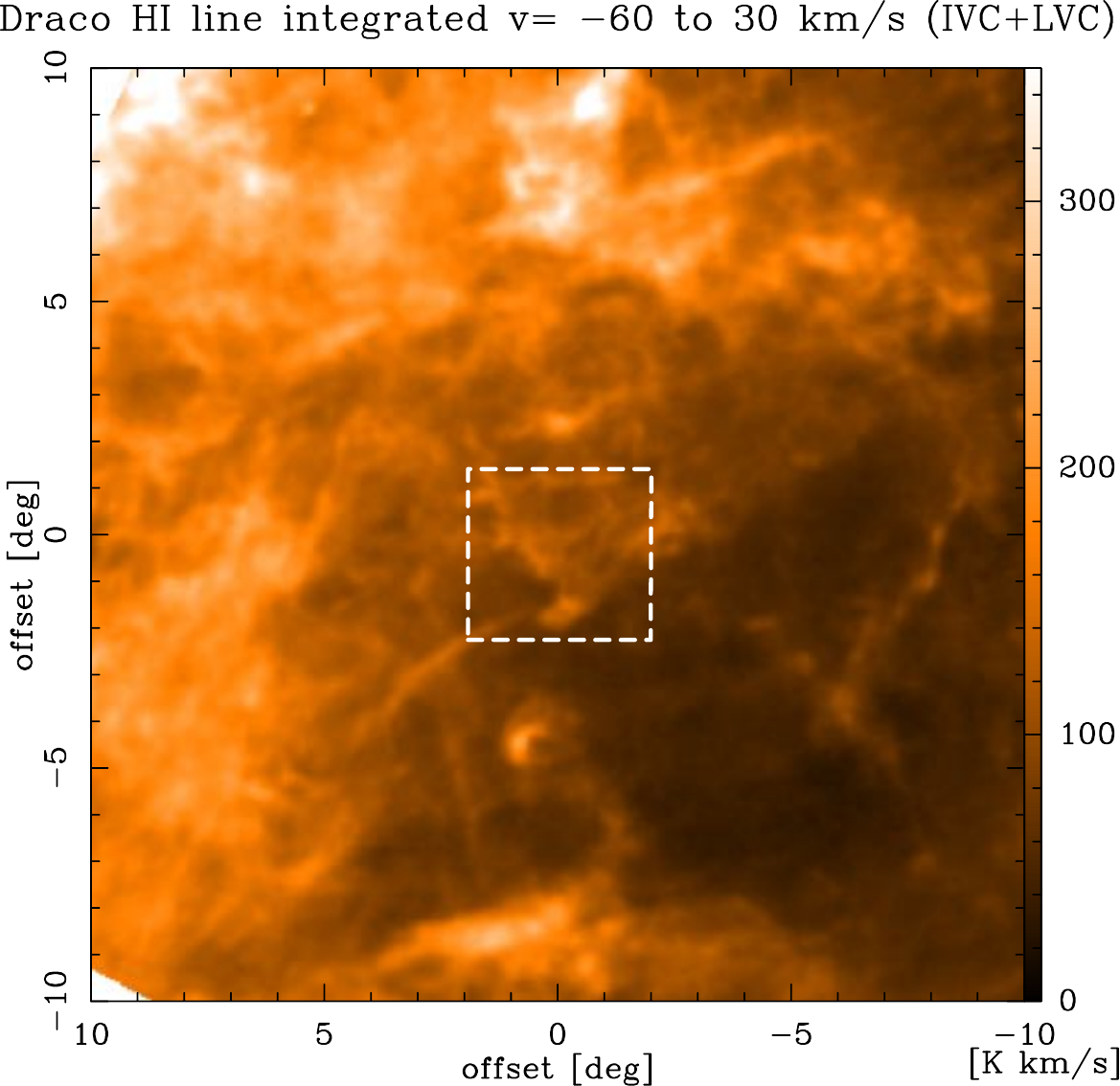}
\caption{FIR and \HI\ maps of Draco. {\bf Left:} Combined {\sl
    Herschel} and Planck 250 $\mu$m map of the Draco cloud and its
  environment. The offsets in degrees correspond to the center position
  of RA(2000)=16$^h$47$^m$57$^s$, Dec(2000)=61$^\circ$45$'$16$''$
  ($l$=91.829, $b$=38.156). The angular resolution of the {\sl
    Herschel} map (in the image center) is 36$''$ and the one of the
  Planck map is $\sim$5$'$. The white dashed square indicates the area
  for which we show the dust column density map. {\bf Right}: Velocity
  integrated \HI\ map from the Effelsberg \HI\ survey (EBHIS) at 10$'$
  resolution.  The velocity range covers the IVC and LVC. Maps of the
  other velocity ranges are shown in Appendix~\ref{appendix:maps}.}
\label{draco-250}
\end{figure*}


\begin{figure*}
\centering
\includegraphics[width=7cm,angle=0]{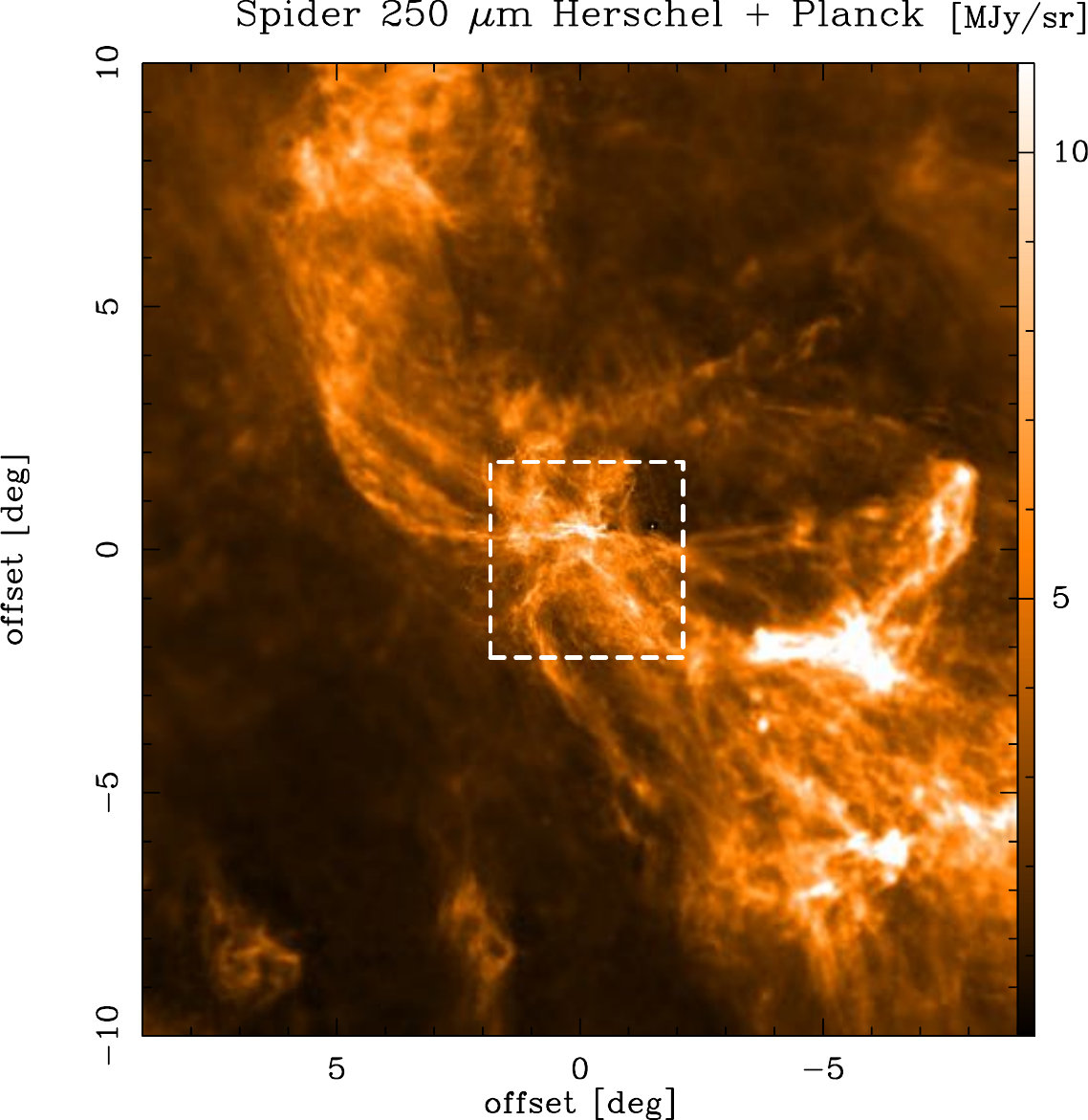}
\hspace{0.5cm}
\includegraphics[width=7.2cm,angle=0]{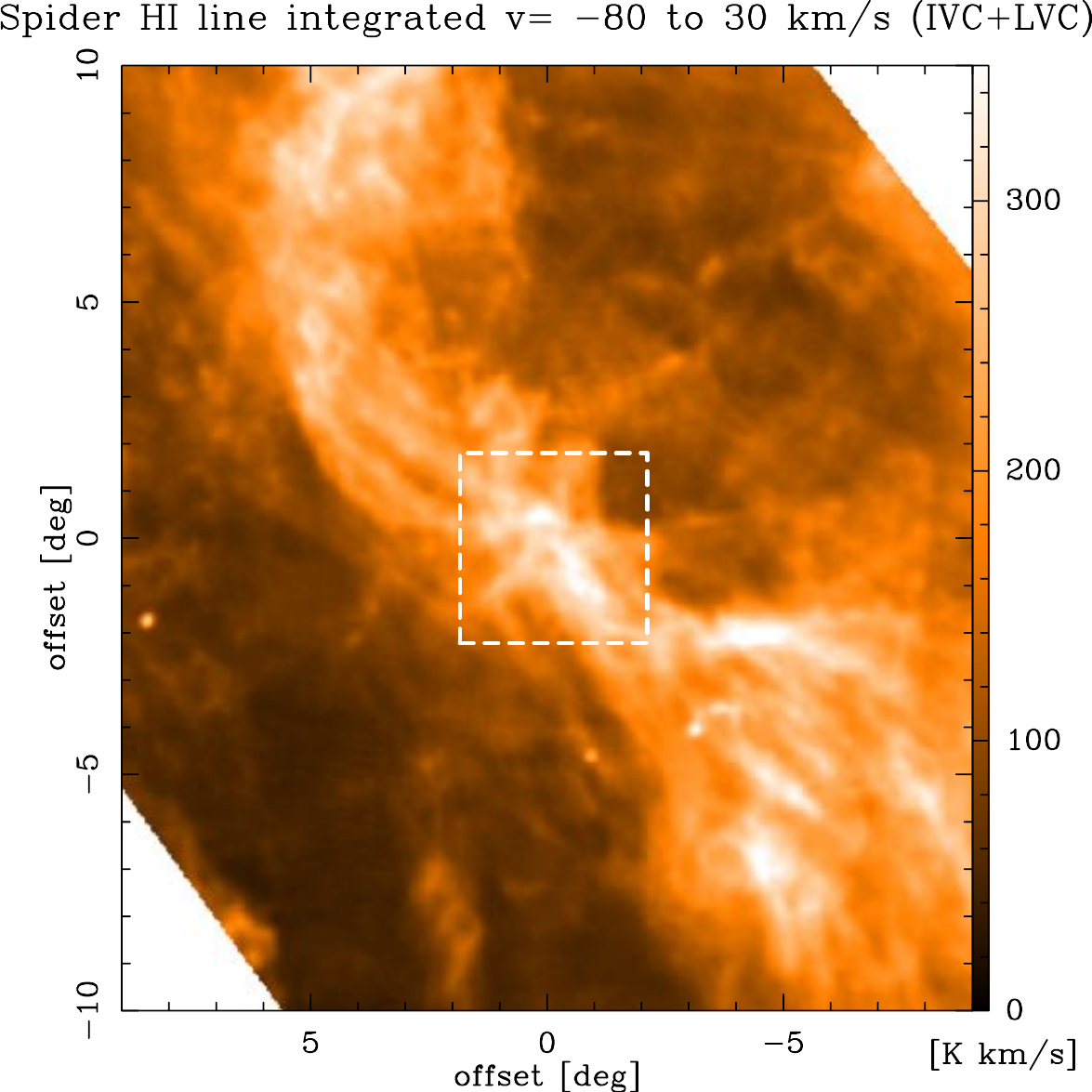}
\caption{Same as Figure~\ref{draco-250} but for Spider. The offsets
  correspond to the center position of RA(2000)=10$^h$37$^m$58$^s$,
  Dec(2000)=72$^\circ$59$'$42$''$ ($l$=135.200, $b$=40.800).}
\label{spider-250}
\end{figure*}

\section{The observational cloud sample} 
\label{sec:sources}
Draco and Spider are both part of the interstellar cirrus,
associated with diffuse \HI\ clouds \citep{Heiles1974}. Draco
constitutes a relatively isolated feature, containing gas concentrated
within local, intermediate, and high-velocity clouds, referred to as
LVC, IVC, and HVC, respectively. The distances to the IVC and LVC were
established to be within the range of 463 to 618 pc by
\citet{Gladders1998} through the utilization of sodium doublet
absorption. In a recent study, \citet{Zucker2020} employed Gaia DR2
parallax measurements to deduce a distance of 481$\pm$50 pc, a value
we will adopt in this paper.
Spider and Polaris are situated within the North Celestial \HI\ Loop
\citep{Meyerdierks1991}. Using Gaia data, \citet{Zucker2020}
determined distances ranging from 341 to 472 pc (with uncertainties
between 10 and 40 pc) for three locations close to the Polaris
cloud. It is important to note that these locations are distant from
the area of the \CII\ observation, which primarily focuses on the
brightest region evident in dust and CO emissions. An additional
Gaia-based study by \citet{Yan2019} reports a distance of 489
pc. These distances surpass earlier estimates. \citet{Heithausen1990}
discussed various potential distances to Polaris and proposed an upper
limit of 240 pc, while \citet{Falgarone2009} employed a distance of
150 pc. In order to maintain consistency with our prior work on dust
column density probability distribution functions
\citep{Schneider2022}, the results of which will be employed herein,
we opt for adopting a distance of 489 pc.  For Spider,
\citet{Zucker2020} determined a value of 369$\pm$18 pc (we will use
here) for the distance which places this region further away than
earlier estimated of around 100 pc \citep{deVries1987}.
Musca is located in the southern Chamaleon-Musca cloud
complex. Distances between 140 and 225 pc are reported
\citep{Franco1991,Knude1998,Bonne2020a}. We here adopt a value of 150
pc to be consistent with \citep{Schneider2022}.  It is not
straightforward to give an error range of this value, see Appendix A
in \citet{Bonne2020a}.  The reddening of stars due to cloud material
starts at 90 pc and reaches its peak at $\approx$180 pc. It is,
however, difficult to assess at which distance the gas of the Musca
filament is located.

Figure~\ref{draco-250} provides an overview of the {\sl Herschel} 250
$\mu$m map of Draco, outlined by a white dashed square, situated
within a larger scale Planck map at the same wavelength (left
panel). Additionally, the right panel depicts \HI\ emission across the
intermediate and low velocity range spanning from $-$60 to 30 km
s$^{-1}$. Notably, the velocity structure within the \HI\ data becomes
even more intricate when dissecting the velocity ranges. Refer to
Figure~\ref{app-draco-hi} in Appendix A for a depiction of faint gas
emission at local velocities ranging between $-$10 and 30 km s$^{-1}$,
a prominent component representing intermediate velocities spanning
from $-$30 to $-$10 \kms, and focused emission at high velocities
ranging from $-$200 to $-$100 km s$^{-1}$.  Figure~\ref{spider-250}
shows the combined {\sl Herschel} and Planck 250 $\mu$m map for the
Spider region where distinct LVC ($-$10 to 30 km s$^{-1}$) and IVC
($-$80 to $-$10 km s$^{-1}$) components are observable, although no
HVC is apparent. It is important to highlight that Draco stands out as
the only region dominated by \HI\ emission of the IVC, whereas Spider
solely exhibits a weak IVC component. Furthermore, Polaris and Musca
exclusively feature gas at local velocities, derived from CO
observations, which correspond to \HI\ emission velocities of $-$3.7
\kms\ and 3 km s$^{-1}$, respectively.

\begin{table}  
  \caption{Observational parameters for the \CII\ data.} \label{obs:table}   
  \begin{center}  
\begin{tabular}{lcc|c|l}  
\hline \hline   
Cloud        & $\alpha_{J2000}$  & $\delta_{J2000}$        & rms  & S/N  \\  
             &                  &                        & [K]  & \\
\hline
\multicolumn{3}{l}{ {\bf Draco}}  &  & \\   
\hline
Draco Front 1 & 16$^h$58$^m$18$^s$ & 61$^\circ$30$'$10$''$  &  0.0445  & 3.4 \\ 
Draco Front 2 & 16$^h$50$^m$54$^s$ & 60$^\circ$54$'$37$''$  &  0.0270  & 3.2 \\ 
Draco Nose 1  & 16$^h$49$^m$06$^s$ & 59$^\circ$55$'$58$''$  &  0.0251  & 4.4 \\ 
Draco Nose 2  & 16$^h$46$^m$29$^s$ & 60$^\circ$19$'$31$''$  &  0.0344  & 4.7 \\     
Draco IVC    & 16$^h$32$^m$13$^s$ & 61$^\circ$39$'$10$''$  &  0.0487  & - \\  
\hline   
\multicolumn{3}{l}{ {\bf Spider}} &  &  \\   
\hline
Spider 1     & 10$^h$40$^m$59$^s$ & 73$^\circ$22$'$21$''$  &  0.0445  & -  \\      
Spider 2     & 10$^h$33$^m$58$^s$ & 73$^\circ$56$'$12$''$  &  0.0647  & -   \\  
\hline   
{\bf Polaris} & 01$^h$59$^m$32$^s$ & 87$^\circ$39$'$41$''$ &  0.0317 & -  \\  
\hline  
{\bf Musca}   & 12$^h$24$^m$41.6$^s$ & -71$^\circ$46$'$41$''$  &  0.0278 & - \\    
\hline  
\end{tabular}  
\end{center}  
\vskip0.1cm \tablefoot{The coordinates specify the positions of each
  setting of the seven-pixel \CII\ array. For the rms determination, the seven
  positions from the single array were averaged and resampled to a
  velocity resolution of 0.6 km s$^{-1}$. The signal-to-noise (S/N)
  ratio (peak main beam brightness temperature over rms in one channel) 
  is given in the last column if the \CII\ line was detected. }
\end{table}

Draco (IVC G091.0+38.0, MBM41) is located at $b$=38.4$^\circ$ at a
height of around 298 pc above the Galactic plane, considering a 
distance of 481 pc. There are no OB stars in the immediate environment, but 
there is noteworthy 160 $\mu$m flux in the {\sl Herschel} map.
Emission at 70 $\mu$m is at the noise level. Assuming that the dust is
only heated by external FUV radiation of stars, the 160 $\mu$m flux can be
translated into a FUV field (see Sect.~\ref{subsec:fuv}) and we derive
a field of 3.6 G$_{\rm o}$ averaged over the 4 positions (the average
FUV field is 1.6 G$_\circ$ from a census of the stars).

It has been recognized for quite some time \citep{Mebold1985} that the
IVC descends toward the Galactic plane with a significant velocity and
experiences deceleration due to its interaction with the surrounding
warm neutral medium. Notably, observations have revealed high FIR
emissivities and elevated CO abundances \citep{Herbstmeier1993}, which
were interpreted as arising from a direct interaction with the
HVC. This high-velocity gas could originate from the infall of
extragalactic material. However, recent \HI\ surveys
\citep{Westmeier2018} have identified a velocity gap between the IVC
and the HVC in Draco \ref{app-pv}, rendering a collision scenario less
plausible. Furthermore, the Draco region prominently exhibits
emissions at intermediate velocities. A proposed explanation for
Draco's origin involves a Galactic fountain process, where material
from the Galactic disk is lifted above the plane and subsequently
returns at high velocities to the disk (as elaborated in
\citet{Lenz2015} and references therein). For a more comprehensive
discourse on the underlying physics and potential sources of IVCs,
readers are encouraged to explore \citet{Putman2012},
\citet{Roehser2014}, \citet{Roehser2016a}, \citet{Roehser2016b},
\citet{Kerp2016}, and the accompanying references.

By investigating probability distribution functions of \HI\ (N$_{\rm
  HI}$-PDF) and of {\sl Herschel} derived total hydrogen column
density (N-PDFs), \citet{Schneider2022} showed that the N-PDF of Draco
has the form of a double-log-normal. The authors propose that one
log-normal arises from atomic gas and the other one from molecular
gas. The \HI-to-H$_2$ transition is defined where the two log-normal
dust N-PDFs have equal contributions and takes place at \av=0.33
(N(H)=6.2 10$^{20}$ cm$^{-2}$).  Importantly, the absence of a
power-law tail in the distribution implies that self-gravity does not
currently play a significant role in the region.  Nevertheless,
certain regions within Draco have exhibited detections of \CI, CO, and
other molecules with critical densities exceeding $10^3$ cm$^{-3}$
\citep{Mebold1985,Herbstmeier1993} and are thus mostly
molecular. Furthermore, \citet{Deschenes2017}, utilizing {\sl
  Herschel} dust observations, inferred that the molecular gas
primarily comprises small ($\sim$0.1 pc), dense ($n\sim$1000
cm$^{-3}$), and cold (T$\sim$10-20 K) clumps.  Despite the presence of
such molecular structures, no indications of active star formation
have been observed within the cloud, as evidenced by the lack of
detected pre- or proto-stellar cores.

In contrast to Draco, the Spider region has received significantly
less attention.  \citet{Barriault2010a,Barriault2010b} presented maps
of \HI, OH, and CO emission and concluded that around 20\% of the gas
exists in the molecular phase. CO, serving as a tracer of CO-bright
H$_2$, is found where two \HI\ velocity components merge into one
component or where there is a velocity shear. \citet{Barriault2010b}
derived an upper limit for the volume density of 2$\times$10$^3$
cm$^{-3}$.  The {\sl Herschel} dust and Effelsberg \HI\ emission
distributions, shown in in Fig.~\ref{spider-250}, support
qualitatively this idea because the densest parts of the Spider cloud
are located in the center of various flows. As already pointed out, in
contrast to Draco, \HI\ emission in Spider is dominated by the LVC and
not the IVC (see Fig.~\ref{app-spider-hi}) and has no HVC
component. Similarly to Draco, we derive the FUV field from the 160
$\mu$m flux, assuming radiative excitation, yielding a maximum value
of 2.9 G$_{\rm o}$.

The Polaris flare exhibits abundant extended, diffuse emission
previously observed through IRAS at 100 $\mu$m \citep{Low1984}, and
now unveiled in large detail using {\sl Herschel}
\citep{Deschenes2010}. The most densely concentrated region, MCLD
123.5+24.9, referred to as the 'saxophone,' has been scrutinized
across various CO lines \citep{Heithausen1990, Falgarone1998}, as well
as in \CI\ \citep{Bensch2003}, housing a handful of prestellar
cores. The peak column density at the core positions ranges from 6 to
13 $\times$ 10$^{21}$ cm$^{-2}$ \citep{Derek2010}. The density at this
position, which coincides with our \CII\ position, is derived to be
between $2-5\times$10$^4$ cm$^{-3}$ from CO, HCN, and {\sl Herschel}
FIR data \citep{Grossmann1992,Heithausen1995,Derek2010}.  The incident
FUV field at the cloud's location was estimated to be approximately
one G$_{\rm o}$ \citep{Bensch2003}. By leveraging the methodology that
converts the 160 $\mu$m flux to a FUV field, we ascertain a field
strength of around $\sim$1.5 G$_{\rm o}$ at the precise coordinates of
the \CII\ observation site.
In an earlier investigation, \citet{Heithausen1990} approximated that
roughly 40\% of the hydrogen in the Polaris flare is
molecular. Consequently, it follows that this cloud is more advanced
in its evolutionary stage compared to Draco and Spider, supported by
its N-PDF \citep{Schneider2013, Schneider2022}.

Musca is located in the south, embedded in the extended Chameleon
complex. It presents itself as a prominent filamentary structure
extending over 6 pc, as illustrated in previous works
\citep{Kainulainen2009,Kainulainen2016,Cox2016}, yet exhibits minimal
star-formation activity. Notably, a sole protostar has been identified
in the northernmost part of the filament, while the cloud remains
relatively unaffected by protostellar feedback.  The external FUV
field is approximated to be 3.4 G$_\circ$ as an upper limit (without
considering extinction), as deduced from an analysis of neighboring
stars \citep{Bonne2020a,Bonne2020b}. Alternative estimates place the
FUV field at 5.8 G$_\circ$ based on the 160 $\mu$m flux and roughly 10
G$_\circ$ according to the Musca map featured in
\citet{Xia2022}. Within the dense crest region, peak column densities
surpass those observed in Polaris, reaching N$\sim$10$^{22}$ cm$^{-2}$
\citep{Cox2016} at the position of the protostar.  In all other
locations, the column densities are smaller than $8\times$10$^{21}$
cm$^{-2}$ and the (column)-densities are also notably lower at the
exact location of the \CII\ observation. The volume density at this
position is at least $7\times$10$^3$ cm$^{-3}$, based on CO and
\CI\ observations presented in \citet{Bonne2020b}. They performed a
non-LTE analysis of the observed tracers and obtained this density for
the warm gas layer slightly outside of the denser (up to 10$^4$
cm$^{-3}$) Musca ridge.


\section{Observations} \label{sec:obs}

\subsection{SOFIA} 

%

Draco was observed during Cycle-5, both within guaranteed time and
open time allocations, under the program number 05\_0208, with
N. Schneider as the principal investigator (PI). A total of five
positions were targeted for observation, with their coordinates
provided in Table~\ref{obs:table}, and their locations given in
Fig.~\ref{fig:map-draco}. These observations took place over the
course of three flights utilizing the Stratospheric Observatory for
Infrared Astronomy (SOFIA), operating from Palmdale, California.
Among the selected positions, two were situated at peaks in dust
column density in the southern region, designated as Nose 1 and
Nose 2. These locations had previously exhibited detections of CO
(refer to Fig.~\ref{fig:spectra-draco} for our CO spectra) and other
molecular lines. Additionally, two positions were chosen within the
eastern region, labeled as Front 1 and Front 2. It is noteworthy that
all these positions display prominent \HI\ emission within the IVC and
LVC velocity ranges.  In contrast, the position denoted as IVC is
positioned farther away from the regions of high column density. This
position is exclusively discernible within the IVC and HVC velocity
range. \\ 
The Front 1 position was observed on November 8, 2016, using the GREAT
instrument \citep{Heyminck2012}. The 7-pixel GREAT/LFA array was tuned
to the \CII\ 158 $\mu$m line, and the single-pixel L1 channel was
tuned to the \NII\ 1.461 THz line. This observation was repeated, and
another position (Nose 1) was added on February 3, 2017, employing
upGREAT \citep{Risacher2018}. The 2$\times$7 pixels LFA array was
again tuned to the \CII\ 158 $\mu$m line, and the L1 channel on the
\NII\ 1.461 THz line. Three further positions (Nose 2, Front 2, and
IVC) were observed on February 14, 15, 16, and 17, 2017, in the same
upGREAT/LFA setup, but tuning the L1 channel to the CO 11$\to$10 line
at 1.1 THz. All observations were carried out in total power mode, and
different emission-free positions with offsets of typically
10$'$-15$'$ to the center positions were used. The total observing
time for each position (ON+OFF) was typically one hour. Beam
efficiencies used in this paper were determined using Mars as a
calibrator for each pixel; the average is 0.64 for the LFA and 0.65
for L1. Third-order spectral baselines were applied to the LFA/L1
spectra and then averaged with a fixed velocity axis and 1/rms$^2$
weighting, smoothed to a channel width of 0.6 km s$^{-1}$. The main
beam sizes are 19$''$ at 1.1 THz, 17$''$ at 1.4 THz, and 14.1$''$ at 158 $\mu$m,
respectively.

Spider was observed during Cycle-6 under the program number 06\_0153
(PI N. Schneider). The observations were carried out on May 23, 2018,
from Palmdale, California, using upGREAT on SOFIA. The upGREAT/LFA
array (2$\times$7 pixels) was tuned to the \CII\ 158 $\mu$m line, and
the HFA array (7 pixels) was tuned to the \OI\ 63 $\mu$m line. Two
pointings were performed in total power mode, centering on positions
Spider 1 at RA(2000) = 10$^h$40$^m$59$^s$, Dec(2000) =
73$^\circ$22$'$21$''$, and Spider 2 at RA(2000) = 10$^h$33$^m$58$^s$
(Fig.~\ref{fig:map-spider}), Dec(2000) = 73$^\circ$56$'$12$''$. The
reference positions were located at an offset of 15$'$ east of the
center positions. Spider 1 is identical with 'S3' in
\citet{Barriault2010a} and represents the peak of IR emission at 100
$\mu$m.
Spider 2 coincides with position 'S6' in \citet{Barriault2010a} and
has peak emission in the atomic phase.  The total observing time for
each position (ON+OFF) was typically one hour.  The determination of
beam efficiencies was accomplished by utilizing Mars as a calibrator
for each pixel. The calculated averages are 0.64 for the Low-Frequency
Array (LFA) and 0.66 for the High-Frequency Array (HFA).  For the
LFA/HFA spectra, baseline corrections of the first and third orders
were applied. The corrected spectra were then subjected to averaging,
employing a fixed velocity axis and a weighting scheme of
1/rms$^2$. Further refinement was achieved by smoothing to a channel
width of 0.6 km s$^{-1}$.  The physical dimensions of the main beams
are 6.3$''$ at a wavelength of 63 $\mu$m and 14.1$''$ at 158 $\mu$m,
respectively.



We also use archival \CII\ data from a SOFIA PI-program on Polaris
(Cycle-5, 75\_0020 PI W. Reach) and Musca (Cycle-6, 06\_0177, PI
S. Bontemps).

In the Polaris flare a single position-switch pointing at
RA(2000)=01$^h$59$^m$32.0$^s$, Dec(2000)=87$^\circ$39$'$41.0$''$ was
performed 2017 June 14 from Palmdale, California, in the LFA/HFA
configuration with the \CII\ 158 $\mu$m line and the \OI\ 63 $\mu$m
line, respectively. The reduced data on a main beam brightness
temperature scale were taken from the SOFIA
archive.\footnote{\href{https://irsa.ipac.caltech.edu/applications/sofia}{https://irsa.ipac.caltech.edu/applications/sofia}}
The position corresponds to the 'core 4' position in \citet{Derek2010}
within the MCLD 123.5+24.9 region.

Musca was observed 2018 June from Christchurch, New Zealand.  A single
pointing in position-switch at RA(2000)=12$^h$24$^m$41.6$^s$,
Dec(2000)=-71$^\circ$46$'$41.0$''$ was carried out in the same
\CII\ and \OI\ configuration as for Polaris. The total duration of the
observations was 70 min. No line was detected in \CII\ or \OI\ in the
individual array spectra (see Fig.~A.1 in \citet{Bonne2020a}).

\subsection{(Sub)Millimeter line data} 

For Draco and Spider, we utilized unpublished CO data acquired at the
IRAM 30m telescope in 2017 and 2018 for two projects (002-17, led by
PI Q. Salome; 003-18, led by PI J. Kerp). \\ In August 2017, we
conducted small-scale maps in the $^{12}$CO and $^{13}$CO 1$\to$0 and
2$\to$1 lines, centered on the five SOFIA positions in Draco,
employing the EMIR E090 and E230 receivers in frequency-switching
mode. In this study, we leveraged $^{13}$CO 2$\to$1 observations from
all positions and all CO data for the Front 2 and Nose 2 positions
from this program, as they were excluded from the 2018 observing
campaign. All data were smoothed to a velocity resolution of 0.5 km
s$^{-1}$ for the 1$\to$0 line observations and 0.25 km s$^{-1}$ for
the 2$\to$1 line to match the 2018 observations.  \\
In July 2018, the observation focus shifted to other regions in
Draco. Nevertheless, we conducted small-scale maps around the Front 1
position (110$''\times$110$''$) and the Nose 1 position
(400$''\times$400$''$). For Spider, a map of dimensions
140$''\times$140$''$ was obtained around the Spider 1 position, while
Spider 2 involved a single extended integration. All observations were
performed in total power mode with a reference position situated 33$'$
east of the map center. We used the EMIR E090 and E230 receivers in
parallel, tuned to the $^{12}$CO 1$\to$0 and 2$\to$1 lines at 115.271
GHz and 230.538 GHz, respectively, ensuring that the $^{13}$CO 1$\to$0
line fell within the bandpass. As the backend, we utilized the Fourier
Transform Spectrometer (FTS) in a configuration that yielded velocity
resolutions of 0.51 and 0.53 km s$^{-1}$ for the $^{12}$CO and
$^{13}$CO 1$\to$0 data, and 0.25 km s$^{-1}$ for the $^{12}$CO 2$\to$1
data.  The beamsize of the CO 2$\to$1 data is 11$''$ and of the CO
1$\to$0 data 22$''$.  In this study, we incorporated all CO data from
the Front 1, Nose 1, and IVC positions, as well as the CO data from
Spider, which was not covered in 2017. The Front 1 and Nose 1
positions were also observed in 2017, and the spectra exhibit
consistent line positions and intensities. All data are presented on a
main beam brightness temperature scale (using main beam
efficiencies\footnote{\href{http://www.iram.es/IRAMES/mainWiki/Iram30mEfficiencies}{http://www.iram.es/IRAMES/mainWiki/Iram30mEfficiencies}}
of 0.78 at 115 GHz and 0.59 at 230 GHz). \\
For Polaris, we use data from the IRAM key-project "Small-scale
structure of pre-star forming regions" (PIs E. Falgarone,
J. Stutzki). The data were made available via the Centre de donn\'ees
astronomique Strasbourg (CDS). We here only employ spatially smoothed
isotopomeric CO 2$\to$1 and 1$\to$0 spectra extracted for the
\CII\ position (the maps are all large enough to allow for smoothing
to 70$''$).  For more details on the observations see
\citet{Falgarone1998}. For Musca, we use CO data presented in
\citet{Bonne2020a,Bonne2020b} that stem from various CO observing runs
at the APEX telescope. For technical details we refer to the relevant
papers given above.  In order to allow for a comparison to the
array-averaged \CII\ data, we smoothed all CO spectra for all sources
to a beam of 70$''$.

For one position in Draco (Nose 1), we use integrated intensities of
the 1$\to$0 line of atomic carbon (\CI) at 490 GHz, given in Table~1
in \citet{Heithausen2001} that stem from observations performed with
the Heinrich-Hertz telescope (HHT) located on Mount Graham in Arizona.
Four positions in Draco were covered, with one observations
(0$''$,0$''$) very close to our Nose 1 position and the others in
steps of 30$''$ offset in x-direction. The beamsize of the HHT at 490
GHz is 16$''$ and thus smaller than what we use here to compare line
intensities and ratios (70$''$). However, the \CI\ line intensity does
not change significantly over the range of 0 to 90$''$ so that we can
assume to first order a beam filling of unity. This will be important
for the PDR modeling.

\subsection{Complementary data sets} 
We use publicly
available\footnote{\href{http://www.cita.utoronto.ca/DHIGLS}{http://www.cita.utoronto.ca/DHIGLS}}
observations of the \HI\ 21 cm line (1420 MHz) for the Draco and
Spider regions from the DRAO Synthesis Telescope and the Green Bank
Telescope (GBT) \citep{Blagrave2017}.  The data have an angular
resolution of 1$'$ and a velocity resolution of 1.32 km s$^{-1}$ for a
channel spacing of 0.824 km s$^{-1}$ and comprise a total velocity
range of $-$103 to 33 km s$^{-1}$ for Draco and $-$164 to 44 km
s$^{-1}$ for Spider.

Additionally, we make use of the all-sky \HI\ data from the
Effelsberg-Bonn HI survey \citep{Winkel2016} at an angular resolution
of $\sim$10$'$. The EBHIS
survey\footnote{\href{https://astro.uni-bonn.de/~jkerp/index.php?page=EBHISproject}{https://astro.uni-bonn.de/~jkerp/index.php?page=EBHISproject}.}
is a joined project of the AIfA and the MPIfR to image the neutral
hydrogen content of the Milky Way galaxy and trace extragalactic
sources. The spectra have a channel width of 1.3 km s$^{-1}$, the rms
noise is less than 90 mK, and the data is corrected for stray
radiation \citep{Roehser2014}.  We obtained fits data cubes from the
CDS.

\begin{figure*} 
  \centering
\includegraphics[width=15cm,angle=0]{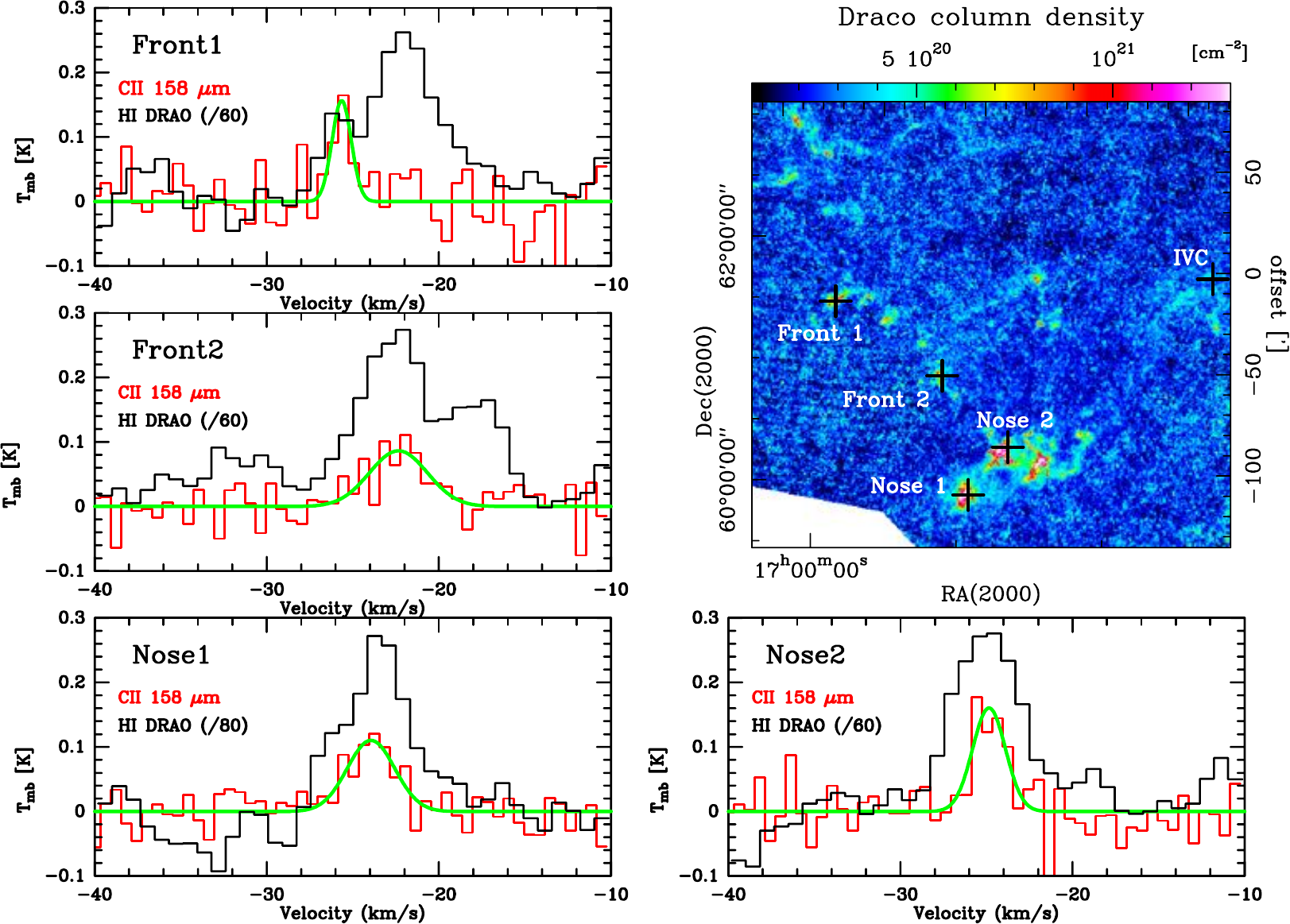}
\caption{Spectra and column density map of Draco. Top-right panel:
  {\sl Herschel} hydrogen column density map derived from dust in
  color in which the observed \CII\ positions are indicated with black
  crosses.  In the panels around, spectra of the \CII\ 158 $\mu$m line
  and the \HI\ 21 cm line (DRAO), both at $\sim$1$'$ resolution, are
  displayed.  We note that the \HI\ line was reduced for better
  visibility. The green curve indicates a single Gaussian fit to the
  \CII\ line.}
\label{fig:map-draco}
\end{figure*}

\begin{figure*} 
\centering
\includegraphics[width=7.5cm,angle=0]{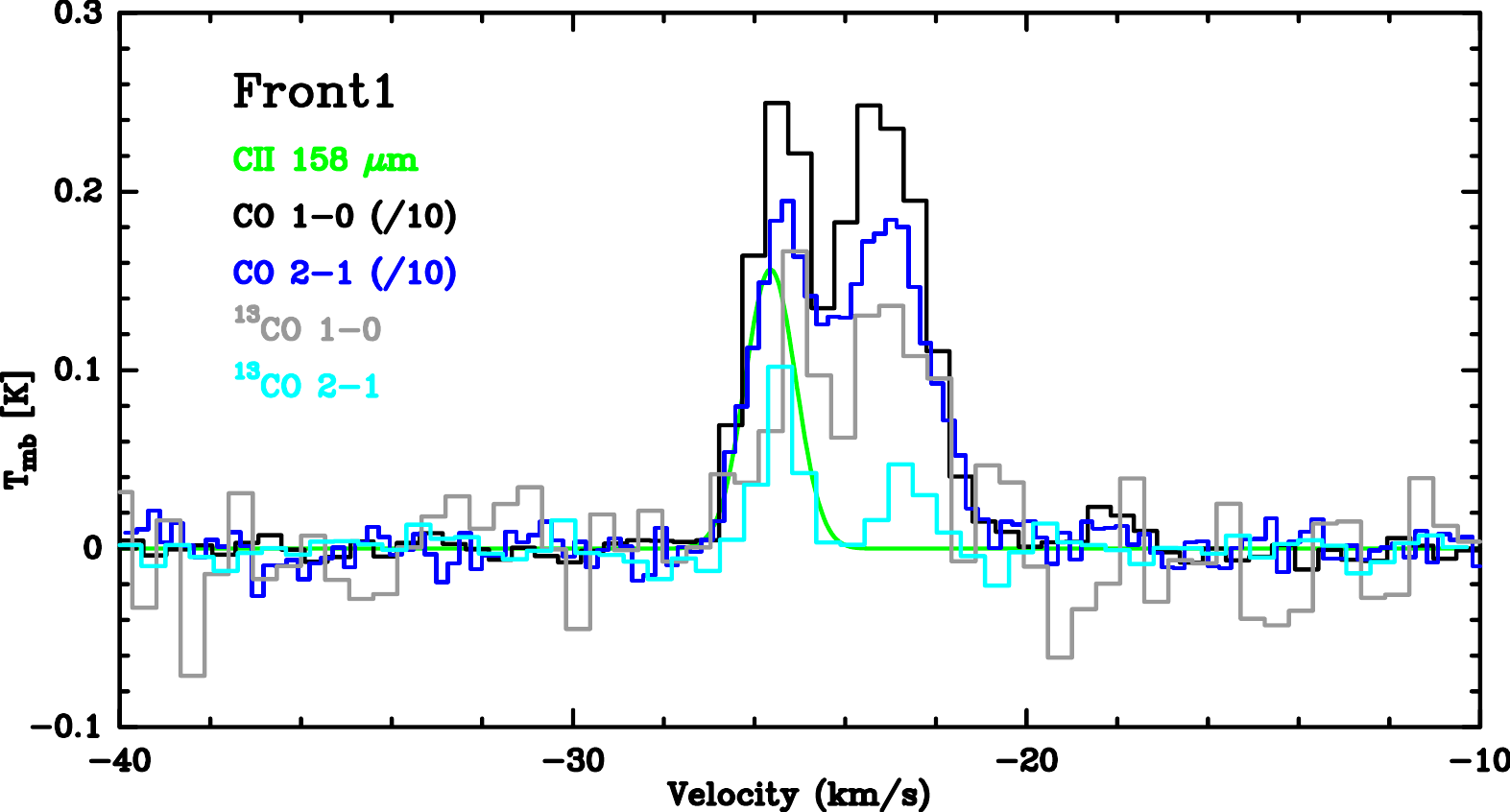}
\includegraphics[width=7.5cm,angle=0]{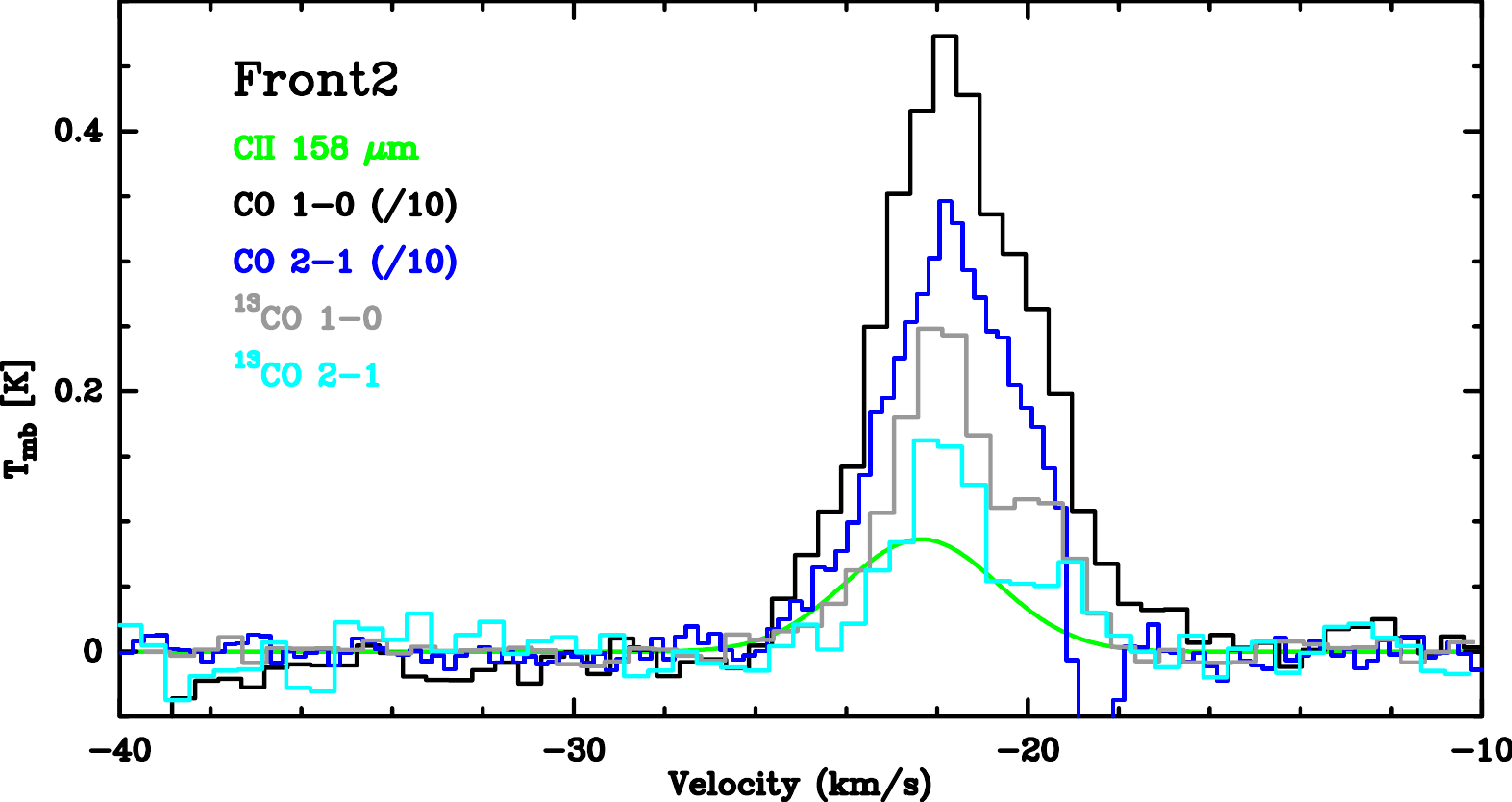}
\includegraphics[width=7.5cm,angle=0]{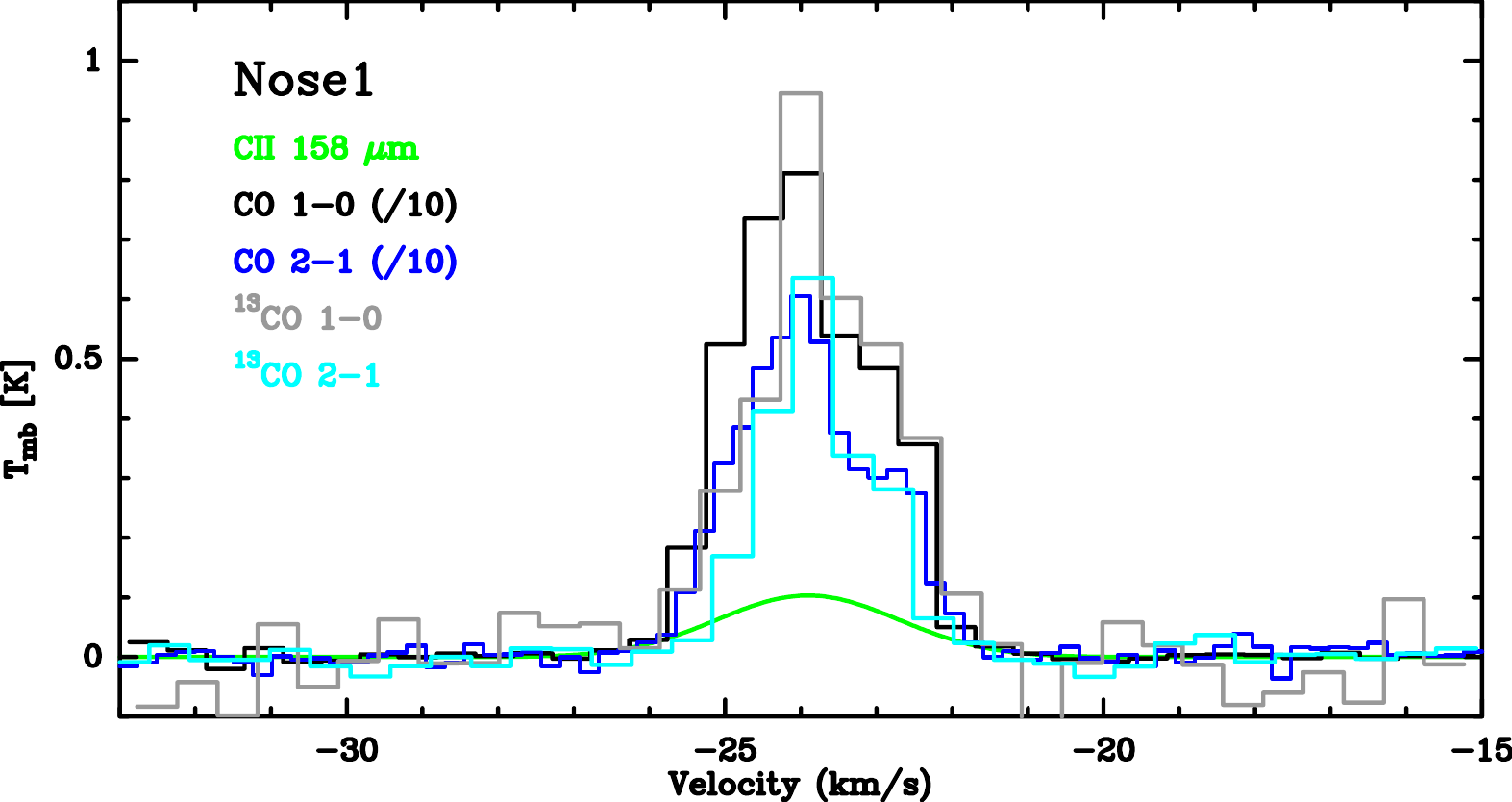}
\includegraphics[width=7.5cm,angle=0]{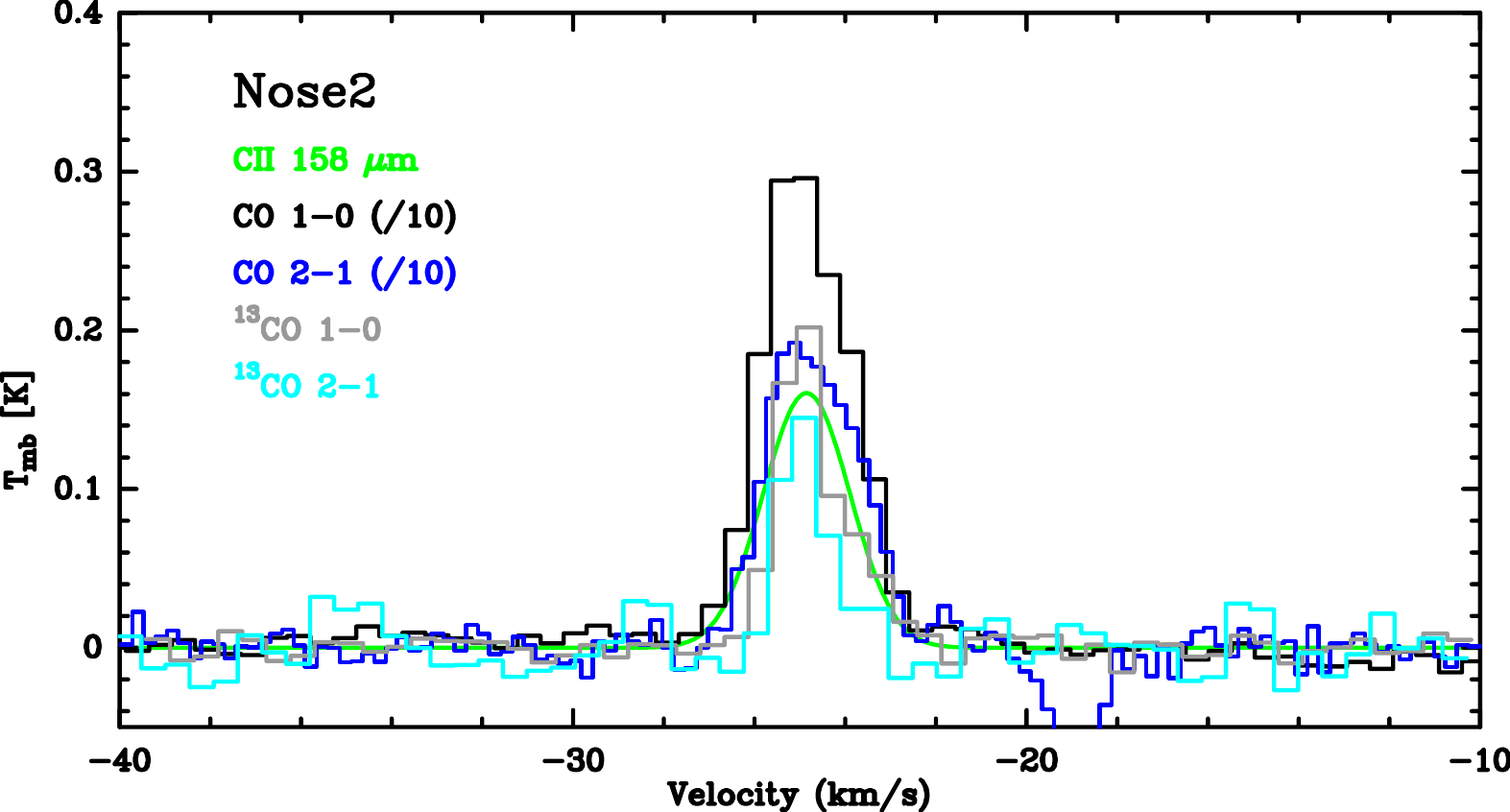}
\caption{Spectra of \CII\ and CO emission at the 4 positions. For
  better visibility of the \CII\ spectra, only the Gaussian fit is
  shown (in green) and  the $^{12}$CO line intensities were
  reduced by a factor of 10.}
\label{fig:spectra-draco}
\end{figure*}


\begin{table*}    
\caption{Physical parameters of Draco for the five observed positions.} \label{draco-obs1}
\begin{center}  
\begin{tabular}{lcccc|cc|cc}
\hline \hline   
Position & \CII\     & \CII\ & \CII\     & \CII\          & N(H) & T$_d$ & FUV        & FUV  \\
         & $T_{mb}$   & v     & FWHM      & I              &      &      & 160 $\mu$m & stars\\
&  [K]    & [km s$^{-1}$] & [km s$^{-1}$] & [K km s$^{-1}$] & [10$^{20}$ cm$^{-2}$] & [K] & [G$_\circ$] & [G$_\circ$] \\
\hline
Front 1   &   0.16   & -25.6 & 1.32 & 0.22 &  8.48 & 13.4 & 3.7 & $1.55^{1.59}_{1.51}$\\
Front 2   &   0.09   & -22.4 & 4.03 & 0.37 &  5.80 & 13.8 & 3.1 & $1.56^{1.60}_{1.52}$\\
Nose 1    &   0.11   & -23.9 & 3.26 & 0.38 & 12.70 & 13.1 & 4.3 & $1.57^{1.61}_{1.52}$\\
Nose 2    &   0.16   & -24.9 & 2.02 & 0.36 & 13.17 & 12.9 & 3.2 & $1.57^{1.61}_{1.53}$\\
IVC      &    -     &  -    &  -   &  -   &  3.21 & 13.3 & 2.2 & $1.58^{1.61}_{1.54}$\\
\hline
\end{tabular}
\end{center}
\vskip0.1cm \tablefoot{The name of the position is given in the first
  column, followed by the \CII\ main beam brightness temperature, line
  velocity, FWHM, and line integrated intensity. These values were
  determined from a single Gaussian line fit.  Column 6 and 7 give the
  hydrogen column density and temperature from {\sl Herschel}. Column
  8 displays the FUV field in Habing units, determined from the 160
  $\mu$m flux, and column 9 the one from the census of the stars
  (Sec.~\ref{subsec:fuv}). The FUV values are given for the
  heliocentric distance D of the source. The upper and lower values
  indicate the field considering the uncertainty in the distances
  (D+$\Delta$D and D-$\Delta$D).}
\end{table*}

\begin{table*}  
\caption{Draco's CO line parameters determined from a Gaussian line fit with two velocity components.} \label{draco-obs2}   
  \begin{center}    
    \begin{tabular}{c|cccc|cccc|cccc|cccc}
      \hline\hline 
 & \multicolumn{4}{l}{$^{13}$CO 1$\to$0} & \multicolumn{4}{|l}{$^{12}$CO 1$\to$0} & \multicolumn{4}{|l}{$^{13}$CO 2$\to$1} & \multicolumn{4}{|l}{$^{12}$CO 2$\to$1}  \\    
 & T$_{mb}$ & v & $\Delta$v  & I  & T$_{mb}$ & v & $\Delta$v & I & T$_{mb}$ & v & $\Delta$v & I & T$_{mb}$ & v & $\Delta$v & I \\
 & {\tiny [K]}     & {\tiny [$\frac{{\rm km}}{{\rm s}}$]} & {\tiny [$\frac{{\rm km}}{{\rm s}}$]} &  {\tiny [$\frac{{\rm K km}}{{\rm s}}$]} &  {\tiny [K]} & {\tiny [$\frac{{\rm km}}{{\rm s}}$]} & {\tiny [$\frac{{\rm km}}{{\rm s}}$]}  & {\tiny [$\frac{{\rm K km}}{{\rm s}}$]} &  {\tiny [K]} & {\tiny [$\frac{{\rm km}}{{\rm s}}$]} & {\tiny [$\frac{{\rm km}}{{\rm s}}$]} &  {\tiny [$\frac{{\rm K km}}{{\rm s}}$]} & {\tiny [K]} &  {\tiny [$\frac{{\rm km}}{{\rm s}}$]} & {\tiny [$\frac{{\rm km}}{{\rm s}}$]} & {\tiny [$\frac{{\rm K km}}{{\rm s}}$]} \\
%
\hline
Front 1 a & 0.16 & -25.1 & 1.01 & 0.17 & 2.44 & -25.4 & 1.43 & 3.72  & 0.10 & -25.4 & 0.91 & 0.10 & 1.75 & -25.4 & 1.54 & 2.87  \\
Front 1 b & 0.14 & -22.9 & 2.31 & 0.35 & 2.52 & -23.1 & 2.15 & 5.77  & 0.05 & -22.7 & 1.09 & 0.06 & 1.82 & -23.1 & 2.28 & 4.42  \\
{\tiny -29.5,-20} & \multicolumn{4}{l}{I$_{int}$ = 0.54 K km s$^{-1}$} & \multicolumn{4}{|l}{I$_{int}$ = 9.5 K km s$^{-1}$}
& \multicolumn{4}{|l}{I$_{int}$ = 0.14 K km s$^{-1}$} & \multicolumn{4}{|l}{I$_{int}$ = 7.3 K km s$^{-1}$}  \\    
\hline
Front 2 a & 0.24 & -22.0 & 2.33 & 0.60 & 3.73 & -21.8 & 3.47 & 13.78 & 0.16 & -21.9 & 2.09 & 0.36 & 3.14 & -22.1 & 2.72 & 9.09  \\
Front 2 b & 0.10 & -19.5 & 1.39 & 0.15 & 0.90 & -20.3 & 4.23 & 4.04  & 0.06 & -19.2 & 1.34 & 0.09 & 1.53 & -20.2 & 1.39 & 2.26  \\
{\tiny -27,-18} & \multicolumn{4}{l}{I$_{int}$ = 0.82 K km s$^{-1}$}
& \multicolumn{4}{|l}{I$_{int}$ = 18.0 K km s$^{-1}$} & \multicolumn{4}{|l}{I$_{int}$ = 0.47 K km s$^{-1}$} & \multicolumn{4}{|l}{I$_{int}$ = 11.1 K km s$^{-1}$}  \\    
\hline
Nose 1 a  & 0.85 & -24.0 & 1.50 & 1.35 & 8.22 & -24.2 & 1.80 & 15.78 & 0.61 & -24.0 & 1.33 & 0.87 & 5.69 & -24.1 & 1.81 & 11.00 \\
Nose 1 b  & 0.38 & -22.6 & 0.96 & 0.39 & 3.71 & -22.7 & 0.78 &  3.07 & 0.24 & -22.6 & 0.53 & 0.14 & 2.25 & -22.6 & 0.76 & 1.82 \\
{\tiny -27,-20} & \multicolumn{4}{l}{I$_{int}$ = 1.78 K km s$^{-1}$} & \multicolumn{4}{|l}{I$_{int}$ = 19.0 K km s$^{-1}$}
& \multicolumn{4}{|l}{I$_{int}$ = 1.04 K km s$^{-1}$} & \multicolumn{4}{|l}{I$_{int}$ = 12.9 K km s$^{-1}$}  \\    
\hline
Nose 2 a  & 0.21 & -25.0 & 1.17 & 0.26 & 3.02 & -25.2 & 1.74 &  5.60 & 0.14 & -25.1 & 0.94 & 0.14 & 1.92 & -24.9 & 2.27 & 4.63 \\
Nose 2 b  & 0.06 & -23.6 & 1.27 & 0.08 & 1.22 & -23.8 & 1.32 &  1.71 & 0.04 & -24.3 & 1.42 & 0.06 & 0.36 & -23.3 & 2.81 & 1.08 \\
{\tiny -28,-20} & \multicolumn{4}{l}{I$_{int}$ = 0.35 K km s$^{-1}$} & \multicolumn{4}{|l}{I$_{int}$ = 7.6 K km s$^{-1}$}
& \multicolumn{4}{|l}{I$_{int}$ = 0.18 K km s$^{-1}$} & \multicolumn{4}{|l}{I$_{int}$ = 4.9 K km s$^{-1}$}  \\    
\end{tabular}
\end{center}
  \vskip0.1cm \tablefoot{I$_{int}$ gives the line integrated intensity
    in the total velocity range in km s$^{-1}$ indicated in the first
    column. The typical errors for the fitting are 0.2-0.5 K km
    s${-1}$ for the line integrated intensity in $^{12}$CO, 0.01 to
    0.08 K km s$^{-1}$ for $^{13}$CO, $\sim$0.06 km s$^{-1}$ for the
    line position and $\sim$0.05 km s$^{-1}$ for the line width for
    all CO lines, respectively. }
\end{table*}

\section{Results} \label{sec:results}

\subsection{\CII, CO, and \HI\ data: Line intensities and profiles}
The following sections discuss the line observations for all all sources.

\subsubsection{Draco} \label{sec:draco-results}

Figure~\ref{fig:map-draco} displays the total hydrogen column density
map at 36$''$ of Draco, derived from {\sl Herschel} dust flux maps
\citep{Schneider2022}. The positions observed in \CII\ with SOFIA and
CO with the IRAM 30m telescope are indicated. The \CII\ spectra,
averaged over the entire array for each position, resulting in an
effective angular resolution of 70$''$, are shown in the surrounding
panels, along with \HI\ spectra from the DRAO survey.
Figure~\ref{fig:spectra-draco} presents the \CII\ spectra alongside
various CO isotopologues and transitions ($^{12}$CO and $^{13}$CO
1$\to$0 and 2$\to$1), all at an angular resolution of 70$''$. The IVC
position is not displayed due to the absence of a \CII\ detection.
Table~\ref{draco-obs1} provides the observed \CII\ main beam
brightness temperature, line position and width, and line-integrated
intensity resulting from a single Gaussian line fit. Additionally, it
includes the total hydrogen column density and dust temperature from
{\sl Herschel}, both measured within a 70$''$
beam. Table~\ref{draco-obs2} lists the main beam brightness
temperatures, line positions and widths, and line-integrated
temperatures for the $^{12}$CO and $^{13}$CO 2$\to$1 and 1$\to$0
lines, derived from Gaussian fits with two components.  The \CII\ line
was detected at a level of 0.1-0.2 K, corresponding to a
Signal-to-Noise (S/N) ratio of 3-5 (Table~\ref{obs:table}), within
velocities ranging from $-$22 to $-$26 km s$^{-1}$ at four out of the
five positions. Only the IVC position did not exhibit a \CII\ line
above the noise threshold. Neither the \NII\ nor the \OI\ line or the
CO 11$\to$10 line were detected at any position.

\begin{figure*} 
\centering
\includegraphics[width=15cm,angle=0]{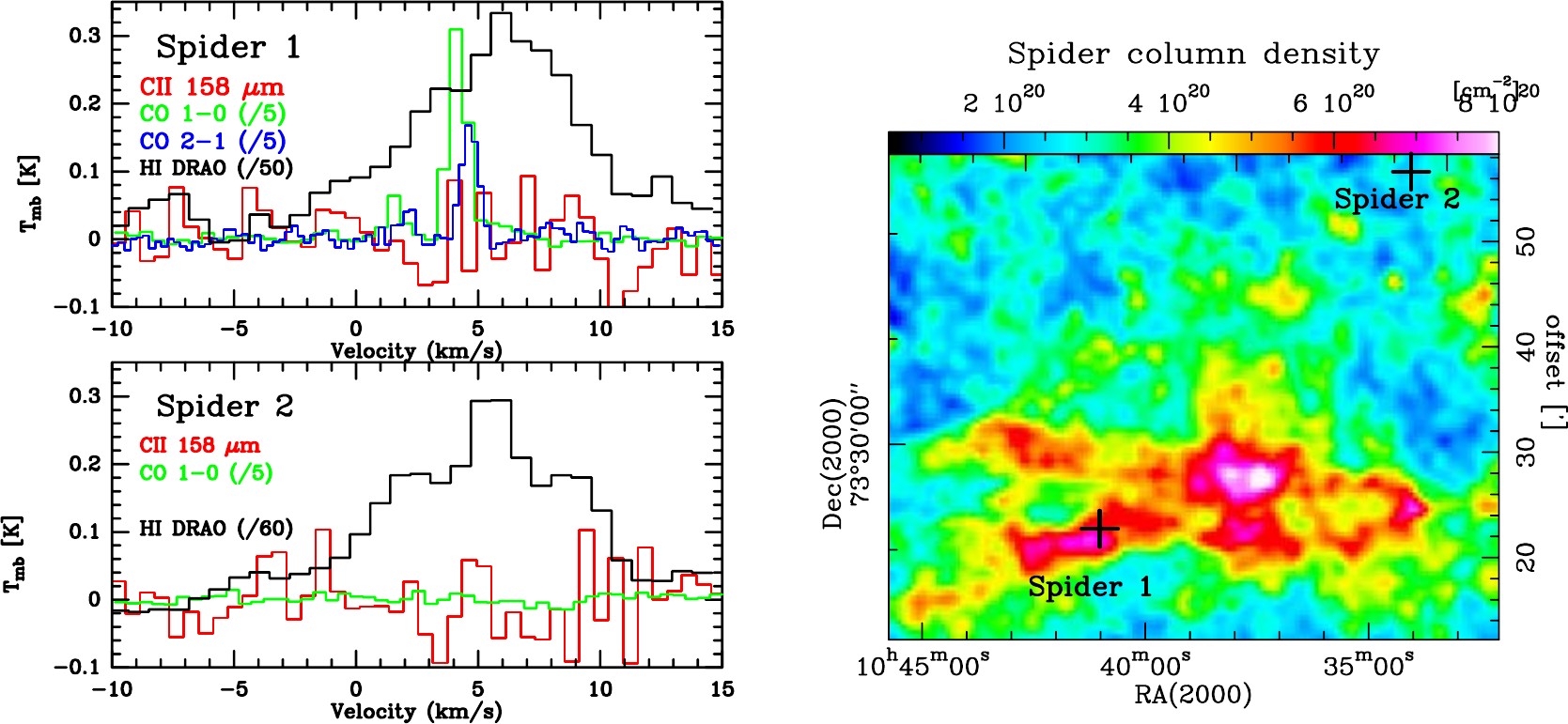}
\caption{Spectra and column density map of Spider. Top-right panel:
  {\sl Herschel} hydrogen column density map derived from dust in
  color in which the observed \CII\ positions are indicated with black
  crosses.  Left panels: Spectra of the \CII\ 158 $\mu$m, CO, and
  \HI\ (DRAO) lines, all at $\sim$1$'$ resolution. We note that some
  line intensities were reduced for better visibility. }
\label{fig:map-spider}
\end{figure*}


\begin{table*}
\caption{Spider observational \CII\ and CO data.} \label{spider-obs}
\begin{center}
  \begin{tabular}{c|c|cccc|cccc|ccc}
    \hline\hline
 & CII  &  \multicolumn{4}{|l}{$^{12}$CO 1$\to$0} & \multicolumn{4}{|l}{$^{12}$CO 2$\to$1} & \multicolumn{1}{|c}{N(H)} & T$_d$    \\
 & I  & T$_{mb}$ & v   & $\Delta$v  & I & T$_{mb}$  & v  & $\Delta$v & I &  &   \\
& [$\frac{{\rm K km}}{{\rm s}}$] & [K] & [$\frac{{\rm km}}{{\rm s}}$] & [$\frac{{\rm km}}{{\rm s}}$] & [$\frac{{\rm K km}}{{\rm s}}$] & [K] & [$\frac{{\rm km}}{{\rm s}}$] & [$\frac{{\rm km}}{{\rm s}}$] & [$\frac{{\rm K km}}{{\rm s}}$] & [10$^{20}$ cm$^{-2}$] & [K] \\ 
    \hline
Spider 1 a & $<$0.06  & 1.57  & 4.18 & 0.92 & 1.54 & 0.85  & 4.67 & 0.81 & 0.73 & 6.30 & 17.7 \\
Spider 1 b & $<$0.06  & 0.31  & 1.59 & 0.77 & 0.26 & 0.22  & 2.12 & 0.65 & 0.16 & 6.30 & 17.7 \\
Spider 2   & $<$0.06  &  -    &  -    &  -  &  -   &  -    &  -   &  -   &  -    & 2.4 & 16.8 \\
\end{tabular}
\end{center}
\vskip0.1cm \tablefoot{The CO line parameters were determined from a
  Gaussian line fit with two components. The error (for both CO lines)
  of the line integrated intensity is 0.035 K km s$^{-1}$ and 0.025 km
  s$^{-1}$ for the linewidth. There was no CO detection at the Spider
  2 position. The last two columns give the hydrogen column density
  and temperature.}
\end{table*}

While the CO and \HI\ spectra exhibit a complex line shape
characterized by at least two distinct components, the \CII\ line
corresponds to one of these velocity components (Front 1a at $-$25.6
km s$^{-1}$, Front 2a at $-$22.4 km s$^{-1}$), Nose 1 at $-$23.9 km
s$^{-1}$, and Nose 2 at $-$24.9 km s$^{-1}$).  The $-$22 km s$^{-1}$
component is evident in both CO and \HI\ spectra for both Front 1 and
Front 2 positions. However, the Front 2 position displays an
additional component ranging from $-$18 to $-$20 km s$^{-1}$ in both
\HI\ and CO spectra. In contrast, no significant velocity difference
is discernible between the two nose positions.  Overall, the \HI\ line
exhibits larger width compared to the \CII\ and CO lines. The distinct
components of the CO lines are unmistakably distinct velocity features
and are not attributed to self-absorption effects. This assertion is
supported by the observation that the optically thin $^{13}$CO lines
present the same two-component line profile.

\subsubsection{Spider} \label{sec:spider-results}

Figure~\ref{fig:map-spider} presents a map of the total hydrogen
column density in Spider at a resolution of 36$''$,
utilizing {\sl Herschel} dust observations. This methodology is the
same as the one employed for Draco, as described in
\citet{Schneider2022}. The plot identifies the positions observed for
\CII\ using SOFIA and for CO using the IRAM 30m telescope. The left
panels display the CO and \HI\ DRAO spectra, while
Table~\ref{spider-obs} provides a summary of observations from SOFIA,
{\sl Herschel}, and IRAM 30m.  No emission in CO or \CII\ within the
IVC or LVC velocity ranges is detected at the Spider 2 position. The
column density is notably low, measuring 2.4 $\times$ 10$^{20}$
cm$^{-2}$ (within a $\sim$1$'$ beam). In Spider 1, narrow
$^{12}$CO 1$\to$0 and 2$\to$1 lines in the LVC were observed. A
prominent velocity component at 4.2 km s$^{-1}$ is evident, with a
corresponding main beam brightness of T$_{mb}$=1.6 K for $^{12}$CO 1$\to$0,
along with a weaker component at 1.6 km s$^{-1}$ with T$_{mb}$=0.3 K
for $^{12}$CO 1$\to$0. It is worth noting that the emission between
$\sim$1 and 4 km s$^{-1}$ corresponds to just one component of the
broader LVC \HI\ line, which spans significantly wider velocities
(refer to Fig.~\ref{app-spider-hi}).  The {\sl Herschel} data yields a
column density value of N(H) = 6.3 $\times$ 10$^{20}$ cm$^{-2}$ at the
Spider 1 position. This aligns with a peak in IR excess emission, where
\citet{Barriault2010a} derived a column density of N(H) = 4.05
$\times$ 10$^{20}$ cm$^{-2}$. The authors highlight that this excess
suggests a potential decrease in the quantity of small grains and/or a
possible reduction in the temperature of larger grains
\citep{Abergel1996}. Such conditions may provide a favorable
environment for H$_2$ formation, as a lower temperature is conducive
to this process.

\begin{figure*} 
\centering
\includegraphics[width=14cm,angle=0]{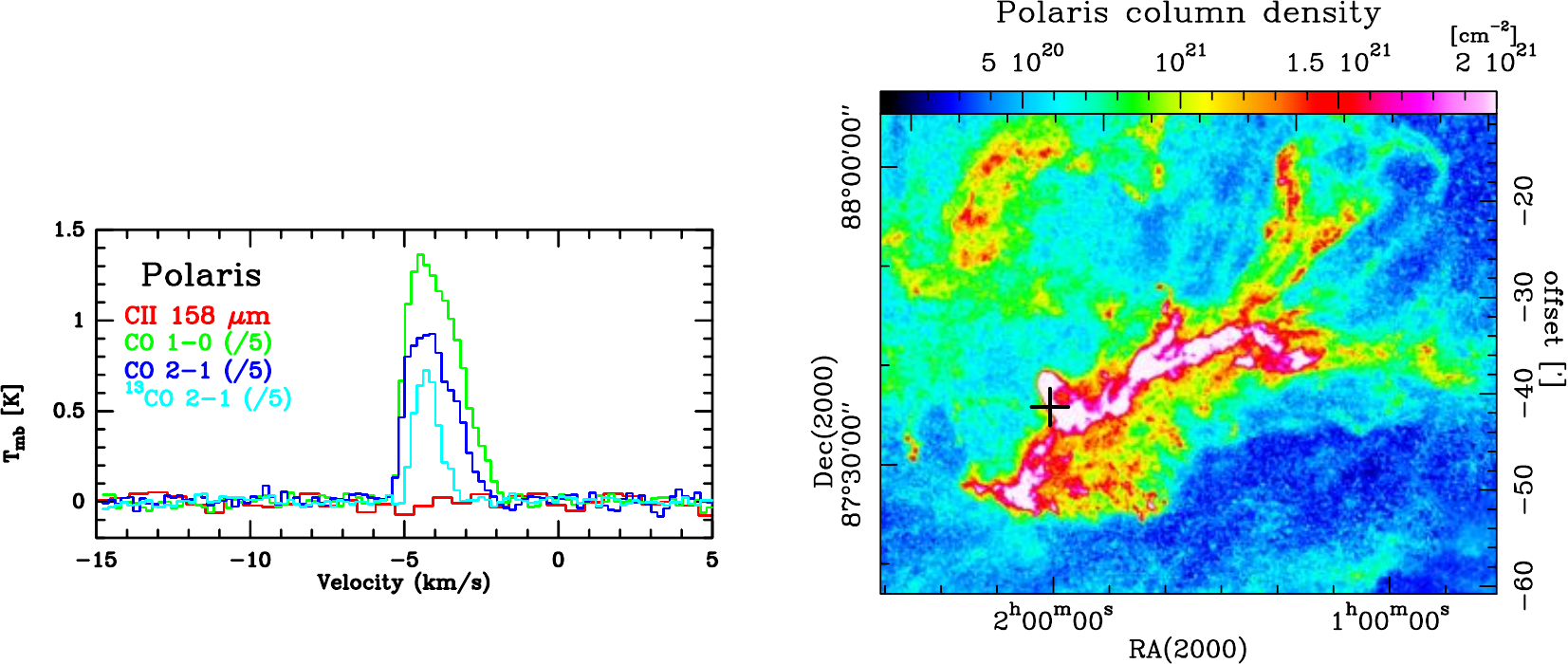}
\caption{Spectra and column density map of Polaris. Right panel:
  Cutout of the {\sl Herschel} hydrogen column density map derived
  from dust of Polaris in color in which the observed \CII\ position
  indicated with black crosses.  Left panel: Spectra of the \CII\ 158
  $\mu$m and CO lines, all at $\sim$1$'$ resolution. We note that some
  line intensities were reduced for better visibility.  }
\label{fig:map-polaris}
\end{figure*}

\begin{figure*} 
\centering
\includegraphics[width=14cm,angle=0]{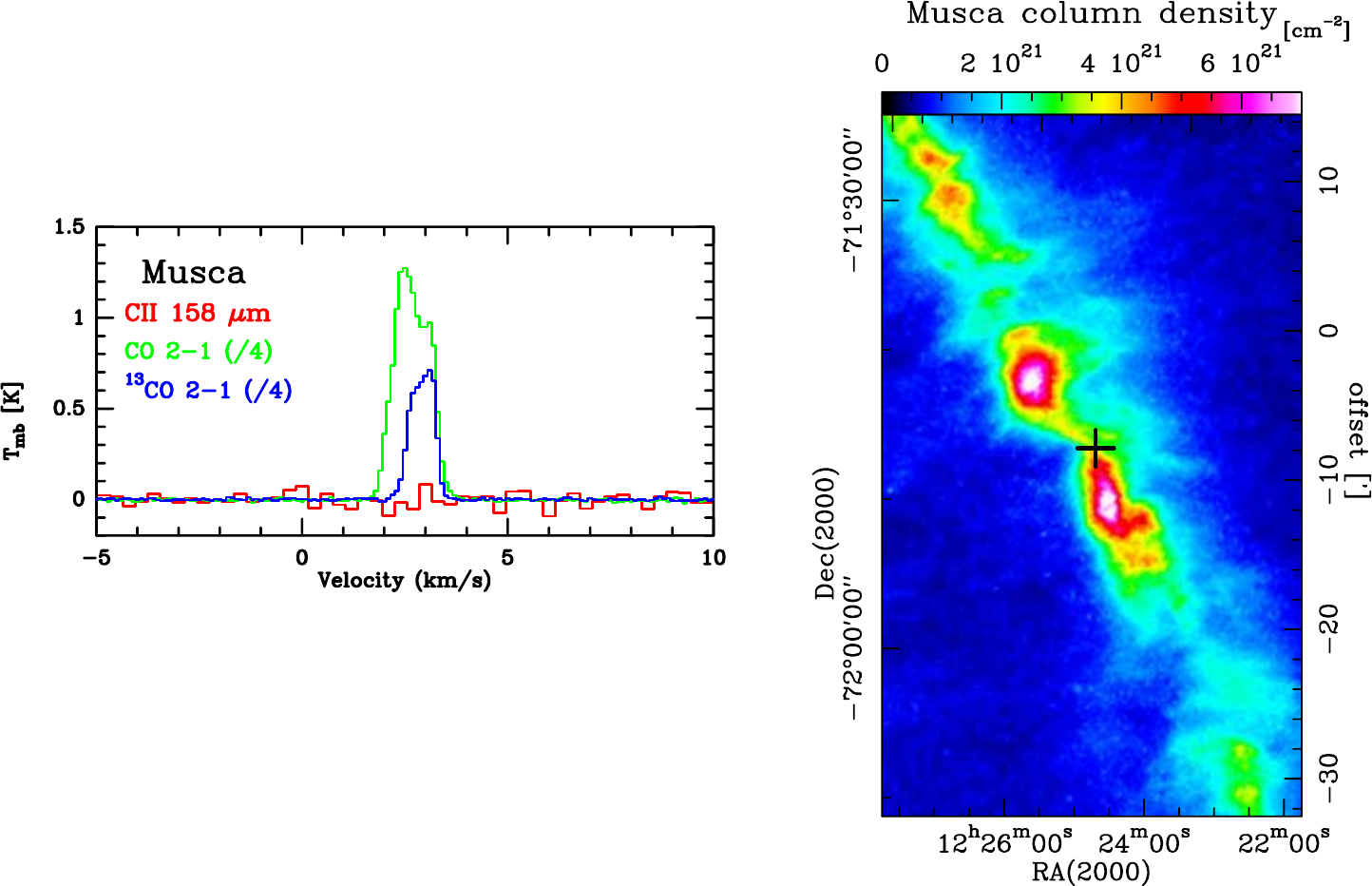}
\caption{Spectra and column density map of Musca. Top-right panel:
  Cutout of the {\sl Herschel} hydrogen column density map derived
  from dust of Musca in color in which the observed \CII\ position
  indicated with black crosses.  Left panel: Spectra of the \CII\ 158
  $\mu$m and CO lines, all at $\sim$1$'$ resolution. We note that some
  line intensities were reduced for better visibility.  }
\label{fig:map-musca}
\end{figure*}

\begin{table*}
\caption{Polaris and Musca observational \CII\ and CO data.} \label{polaris-musca-obs}
\begin{center}
  \begin{tabular}{c|c|cccc|cccc|ccc}
    \hline\hline
 & CII  &  \multicolumn{4}{|l}{$^{12}$CO 1$\to$0} & \multicolumn{4}{|l}{$^{12}$CO 2$\to$1} & \multicolumn{1}{|c}{N(H)} & T$_d$    \\
 & I  & T$_{mb}$ & v   & $\Delta$v  & I & T$_{mb}$  & v  & $\Delta$v & I &  &   \\
    & [$\frac{{\rm km}}{{\rm s}}$] & [K] & [$\frac{{\rm km}}{{\rm s}}$] & [$\frac{{\rm km}}{{\rm s}}$] & [$\frac{{\rm K km}}{{\rm s}}$] & [K] & [$\frac{{\rm km}}{{\rm s}}$] & [$\frac{{\rm km}}{{\rm s}}$] & [$\frac{{\rm K km}}{{\rm s}}$] & [10$^{20}$ cm$^{-2}$] & [K] \\ 
    \hline
Polaris & $<$0.03  & 5.46 & -4.06 & 1.92 & 11.1 & 3.82  & -4.16 & 1.75 &  7.1 & 32.2  & 13.8 \\
Musca   & $<$0.03  &  -    &   -  &  -   &  -   & 5.11  &  2.65 & 1.02 & 5.58 & 34.3  & 14.0 \\
\end{tabular}
\end{center}
\vskip0.1cm \tablefoot{The CO line parameters were determined from a single 
  Gaussian line fit. The error of the line integrated intensity is 0.09 K km 
  s$^{-1}$ for Polaris and 0.01 K km s$^{-1}$ for Musca, respectively. The error for the linewidth is 0.017 km s$^{-1}$ for 
  Polaris and 0.002 km s$^{-1}$ for Musca, respectively. There is no high angular resolution CO 1$\to$0 data for Musca available.
  The last two columns give the dust column density and temperature.}
\end{table*}

\begin{table}
\caption{Distance and FUV field.} \label{tab:flux}
\begin{center}
  \begin{tabular}{l|c|c|c|c}
    \hline\hline
{\tiny Source}  & {\tiny D${^a}$} & {\tiny z$^b$}  & {\tiny FUV from}               &  {\tiny FUV from} \\
                & {\tiny [pc]}    &  {\tiny [pc]}   & {\tiny 160 $\mu$m [G$_\circ$]} &  {\tiny stars [G$_\circ$]} \\
\hline
{\tiny Draco}   & {\tiny 481$\pm$50}  & {\tiny 298} & {\tiny 3.6$^c$} & {\tiny 1.6} \\ 
{\tiny Spider}  & {\tiny 369$\pm$18}  & {\tiny 240} & {\tiny 2.9}      & {\tiny 1.6} \\ 
{\tiny Polaris} & {\tiny 489$\pm $10} & {\tiny 206} & {\tiny 1.5}      & {\tiny 1.3} \\  
{\tiny Musca}   & {\tiny 150}         & {\tiny 24}  & {\tiny 5.8}      & {\tiny 1.4$^d$} \\ 
\end{tabular}
\end{center}
\vskip0.1cm
\tablefoot{FUV field determined at the positions where \CII\ was observed with different methods.\\
$^a$Distance from the sun according to the values given in Sect.~\ref{sec:sources}. \\
$^b$Approximate height above the Galactic plane. \\
$^c$Average value of the 4 positions in Draco where \CII\ was detected. \\ 
$^d$Note that \citet{Bonne2020a} derived a field of 3.4 G$_\circ$ for Musca also from a census of the stars but 
without considering extinction and assuming a 2D geometry, that is, all stars in the plane of the sky at the Musca distance. }
\end{table}

\subsubsection{Polaris and Musca} \label{sec:polaris-musca-results}

Figure~\ref{fig:map-polaris} and \ref{fig:map-musca} display the {\sl
  Herschel} total hydrogen column density maps from dust for Polaris
and Musca \citep{Schneider2022}, respectively, with the observed \CII\ position
indicated. The left panels show various CO lines, all at a resolution
of 70$''$, to indicate the velocity of the molecular gas. For Polaris,
the bulk emission of the cloud is traced by the $^{13}$CO 2$\to$1 line at
v$\approx -4.5$ km s$^{-1}$. The $^{12}$CO 2$\to$1 and 1$\to$0 lines have
an additional component at lower velocities around $-3$ km
s$^{-1}$. These velocity features are present throughout the Polaris
region and were studied in detail by \citet{Hily-Blant2009}. They
concluded that the occurrence of these two velocity components is a
signature of intermittency in turbulent molecular gas.  

In the case of Musca, the main velocity component occurs at v$\approx 3$ km
s$^{-1}$, while the $^{12}$CO line shows an additional component
around $2.5$ km s$^{-1}$. This emission feature was interpreted by
\citet{Bonne2020a} as being caused by the dissipation of turbulence in
a low-velocity shock. The \CII\ line was not detected
above the noise limit. The physical properties of both regions are
summarized in Table~\ref{polaris-musca-obs}.

\section{Analysis} \label{sec:analysis}

In the following, we will determine the FUV field in the sources using
different methods (Sect.~\ref{subsec:fuv}), and then derive the
physical properties of the gas in the PDR with the help of a PDR model
(Sect.~\ref{subsec:pdr}), a non-LTE radiative transfer code
(Sect.~\ref{subsec:radex}), and a shock model
(Sect.~\ref{subsec:shock}). The objective of this exercise is to
distinguish which heating mechanism is responsible for the emission of
the \CII\ and CO lines.

\subsection{Determination of the FUV field} \label{subsec:fuv}

For the determination of the FUV field, we employed several approaches
we summarize in the following.

Firstly, we translated the observed 160 $\mu$m flux into an FUV field,
assuming that only radiation from stars is responsible for heating the
dust, which is then fully re-radiated at FIR wavelengths. The FUV
field was estimated using the correlation
\citep{Kramer2008,Roccatagliata2013,Schneider2016}: $F_{\text{FUV}}
      [G_\circ] = (4 \pi/1.6) \, I_{\text{FIR}} \, \times 1000$ where
      the 160 $\mu$m intensity $I_{\text{FIR}}$ is given in units of
      $10^{-17}$ erg cm$^{-2}$ s$^{-1}$ sr$^{-1}$.

Secondly, we used the continuous model proposed by
\citet{Parravano2003} to estimate the FUV field impinging on our
sources. They assumed a statistically homogeneous distribution of OB
stars in the Galactic plane with a scale height of 85~pc for the OB
stars and typical dust properties for the FUV absorption and
scattering (see below). Despite the local and temporal variations of
these properties, they showed that the model reproduces the typical
radiation field in the solar neighbourhood with a median value of
about 1.6 G$_\circ$.  We treated the distribution of the OB stars as
a continuous UV source and used the same properties for the dust
($\kappa_\mathrm{absorption, FUV} = 8.0\times 10^{-22}$~cm$^2$/H-atom,
$\kappa_\mathrm{scattering, FUV} = 7.5\times 10^{-22}$~cm$^2$/H-atom,
mean scattering angle $g \ga 0.75$ \citep[App. B
  from][]{Parravano2003} and the vertical gas density distribution
(see their Eq.~22 and also our Eq.~\ref{eq:hydrogen_column}) of the
ISM\footnote{Newer numbers for the vertically
integrated column from \citet{Marasco2017} are only higher by 6\,\%.}
to compute the flux above the Galactic plane. In this one-dimensional
configuration no geometrical dilution occurs but only dust extinction
and scattering. The effective optical depth for the FUV radiation is
then $\tau_\mathrm{FUV}=N(\mathrm{H})\times
[\kappa_\mathrm{absorption, FUV}+ (1-R)\kappa_\mathrm{scattering,
    FUV}]$, where $0.9 < R < 1$ depending on the number of scattering
events. That means that practically only absorption acts as an
effective extinction. For Musca falling within the disk of OB stars,
the intensity is then only slightly smaller than in the Galactic
midplane, but even for the highest latitude cloud, Draco, the
radiation field is only reduced by a factor of 0.8 due to the
extinction by the dust column above the Galactic plane of
$N(\mathrm{H})=2.6\times 10^{20}$~cm$^{-2}$. In such a continuous
model, all our clouds thus experience an external radiation field
between 1.2 and 1.6 G$_\circ$.
\begin{figure*}[htp] 
\centering
\includegraphics[width=.8\textwidth]{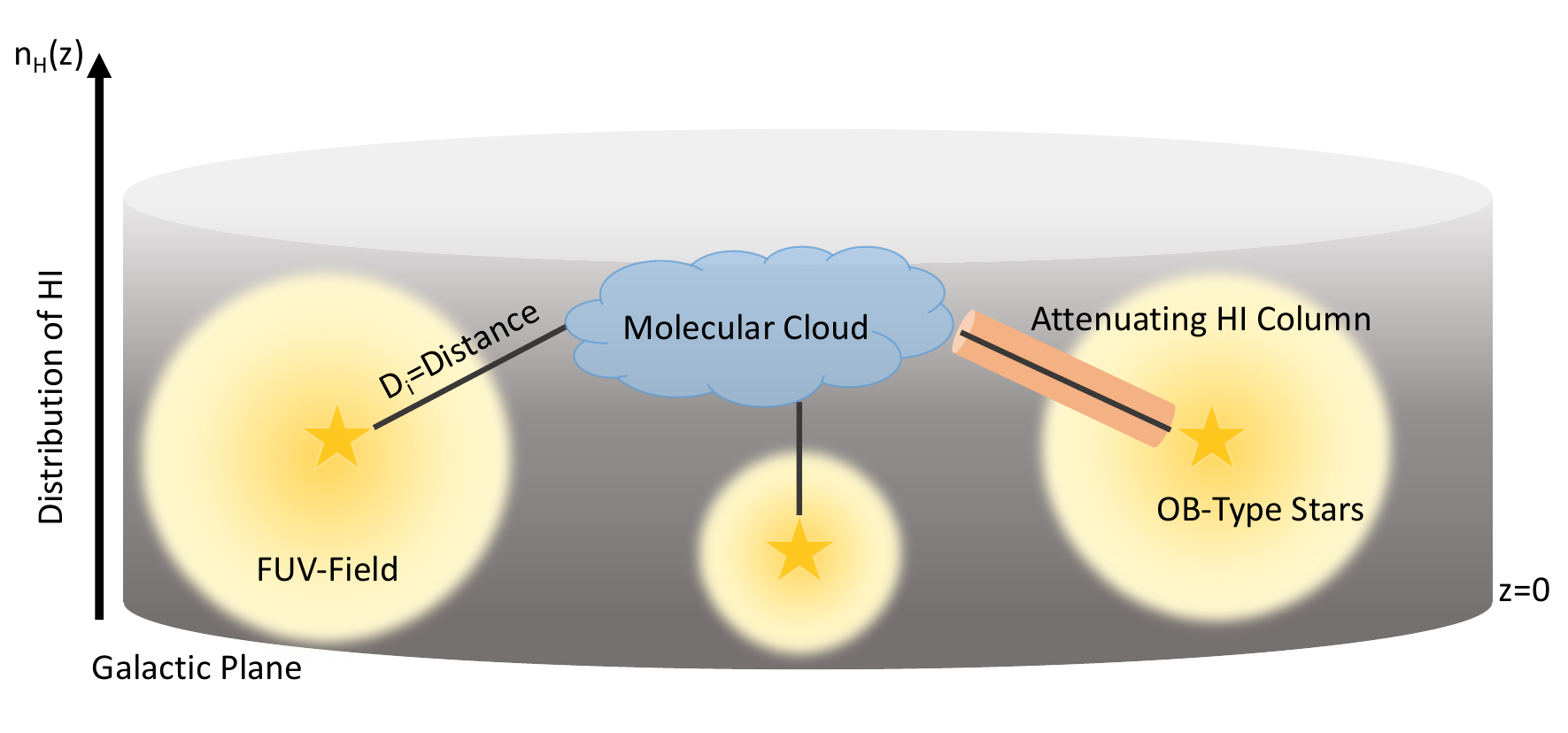}
\caption{Schematic illustration of the method to derive the FUV field
  at each source. The gray cylinder indicates the hydrogen
  distribution with respect to the galactic height z. The yellow stars
  indicate the 3D stellar distribution in the Milky Way. The solid
  black lines originating from the stars indicate the distance from
  the star to the source. The yellow halo around the star illustrates
  the FUV field generated by each star. The orange cylinder placed
  along the path between the star and the source shows schematically
  the hydrogen column which attenuates the FUV field.}
\label{fig:gaia_fuv}
\end{figure*}

In a third approach we dropped the assumption of a homogeneous stellar
distribution since a molecular cloud located in the vicinity to an
OB-cluster can experience a higher FUV field than the one typical for
the solar neighborhood. Thus, we compiled a stellar census from Gaia
DR3 and compared to the information in SIMBAD, the astronomical
database in
Strasbourg \footnote{\href{http://simbad.cds.unistra.fr/simbad}{http://simbad.cds.unistra.fr/simbad}}
and utilized the distances provided in the Introduction (also mostly
Gaia based). Since the data base in Gaia DR3 is much larger than the
one in SIMBAD, we finally only used the stellar census from Gaia. We
note, however, a discrepancy we found for one star in Draco that was
declared as a B-star in SIMBAD via the original HD star catalog
\citep[see compilation in][]{Cannon1993}. In the more recent Apogee
survey, \citet{Jonsson2020} determined a reliable temperature fit of
6800 K, which puts it at a spectral type F2V. This is in agreement
with the Gaia parallax, with sets its distance to 300 pc, which is
actually not the one considered for Draco \citep[480 pc
  following][]{Zucker2020}. We checked in addition the Apogee spectrum
and it shows that the metallicity is subsolar. In summary, everything
agrees with the star being an old F2V star at a distance of 300 pc.
The FUV field was then calculated by taking all stars into account
from the latest Gaia DR3 release \citep{Creevey2023,Drimmel2023},
which provides the necessary information such as the stellar position,
distance, temperature and luminosity. Since most stars are not hot
enough to contribute significantly to the FUV field, we only considered 
stars with a effective temperature $T_{\rm eff}$ above $10\,000\,{\rm
  K}$, thus all spectral types O and B, resulting in $1\,192\,351$
stars\footnote{We initially also considered cooler supergiants
($T_{\rm eff} <10\,000\,{\rm K}$) with a radius larger than 30
$R_\odot$ because they have large luminosities but found that their
contribution to the overall radiation field is small.}. This number is
a lower limit since highly embedded stars were not detected. However,
such embedded stars do not contribute significantly to the FUV field
at our sources since their radiation is extinct in all directions. On
the other hand, there can be a substantial column between the star and
the observed cloud and therefore a direction dependent extinction.
However, the sources of this study all lie within a radius of
$0.5\,{\rm kpc}$, see Table~\ref{tab:flux}, and it is therefore
unlikely that the source is exposed to a radiation field of an
undetected stellar cluster. The temperature information of each star
is provided by the GSP-Phot (General Stellar Parametrizer from
photometry) module\footnote{See
\href{https://gea.esac.esa.int/archive/documentation/GDR3/}{https://gea.esac.esa.int/archive/documentation/GDR3}
for the Gaia DR3 manual.} and the corresponding luminosity information
was taken from the FLAME (Final Luminosity Age Mass Estimator)
module. If the stellar luminosity was not provided by FLAME we used
the stellar radius and temperature, provided by GSP-Phot to derive the
(bolometric) luminosity $L$:
\begin{equation}
    \frac{L}{L_\odot} = \left(\frac{R}{R_\odot}\right)^2\left(\frac{T_{\rm eff}}{T_{\rm eff,\odot}}\right)^4~,
\end{equation}
\noindent
with the effective temperature of the Sun $T_{\rm eff,\odot} =
5772\,{\rm K}$, the solar radius $R_\odot$ and solar luminosity
$L_\odot$. The distance to each star was derived from the parallax
information. However, if the parallax error is large, more than
half of the parallax, we used instead the distance provided by the
GSP-Phot module.

We determined the FUV luminosity by assuming that the spectral
radiance of each star can be approximated by a black-body curve. The
Planck function was integrated over the FUV range spanning from 910 to
2066 \AA, corresponding to a photon energy range of 6 to 13.6 eV. The
FUV luminosity is then defined through the ratio of the spectral
radiance in the FUV range and the entirety of the black-body spectrum:
\begin{equation}
    L_{\rm FUV} = \frac{\pi \int\limits_{\lambda_{910}}^{\lambda_{2066}} B\left(\lambda, T\right)\,\rm{d}\lambda}{ \sigma T^4} L~,
\end{equation}
with the Stefan-Boltzmann constant $\sigma$. The superposition of the
stellar FUV flux of all stars considered gives the FUV field at every
point in the map:
\begin{equation}
    F_{\rm FUV} = \sum\limits_{i} \frac{L_{\rm{FUV},i}}{4\pi D^2_{i}}e^{-\tau_{\rm FUV, i}}~,
\end{equation}
where $D_i$ represents the distance from the source to each star. The
FUV optical depth is defined as described above and the attenuating
column $N_{\rm H}$ is calculated for each star individually, based on
its position with respect to the source. This differs from the
continuous model, where the column is simply determined by galactic
height of the source. Thus, the hydrogen density $n_{\rm H}(z)$
\citep[Eq.~22 in ][]{Parravano2003} integrated along the path
connecting each star at the galactic height $z_i$ with the source at
the galactic height $z_s$ results in the hydrogen column density:

\begin{align}
    N_{\rm H, i}(z_s, z_i, D_i) & = \frac{D_i}{z_s-z_i}\int_{z_i}^{z_s} n_{\rm H}(z) \, {\rm d}z \notag \\
    & = \frac{D_i}{z_s-z_i}\int_{z_i}^{z_s}0.566\left[ 0.69\exp{\left(-\frac{z}{127\,{\rm pc}}\right)^2} \right. \notag  \\
    & +  0.189\exp{\left(-\frac{z}{318\,{\rm pc}}\right)^2} \notag \\
    &\left. + 0.113\exp{\left(-\frac{|z|}{403\,{\rm pc}} \right)}\, {\rm d}z \right]\,{\rm cm^{-2}}~.  
    \label{eq:hydrogen_column} 
\end{align}

\noindent
Equation~\ref{eq:hydrogen_column} is valid for $z_i\neq z_s$, for
$z_i=z_s$ the equation simplifies to $N_{\rm H, i} = D_i n_{\rm
  H}(z_i)$ (with the distance in cm). While Gaia DR3 data allow for a
more sophisticated 3D treatment of the extinction \citep{Zucker2022},
this goes beyond the scope of this paper. Thus, for the sake of
comparison with the previous approach given by \cite{Parravano2003} we
use the above Eq.~\ref{eq:hydrogen_column}.  The method is
schematically illustrated in Fig.~\ref{fig:gaia_fuv}. The resulting
values for the FUV field are around $1.3-1.6\,{\rm G_\circ}$ and given
in Table~\ref{tab:flux}. They are similar to the ones estimated with
the second method and again close to the typical value in the solar
neighborhood. When sticking to this approximation of the interstellar
extinction, we can only quantify the change of values considering the
error of the distance (Table~\ref{draco-obs1} for Draco, the variation
is very small for the other sources and thus not given) but note that
a larger error is introduced by the uncertain detailed $n({\rm H}$)
distribution in the galaxy.

As a fourth method, we compared our numbers to a study conducted by
\citet{Xia2022}. This study derived the FUV field through the fitting
of a spectral energy distribution (SED) to the {\sl Herschel}
fluxes. Furthermore, a dust radiative transfer analysis was performed
using the DUSTY code \citep{Nenkova2000}. The results are presented
for a selection of sources, which includes Polaris and
Musca. Moreover, the empirical correlation $\log({\rm G}_{\rm o}) =
(0.62 \pm 0.12) \,\log({\rm N(H_2) [cm^{-2}]}) \,- \, (11.56\pm2.87)$
was given in \citet{Xia2022} to derive the FUV field.  In order to
estimate a value for the molecular hydrogen column density in Draco,
we used the N-PDF presented in \citet{Schneider2022}. This N-PDF
consists of two lognormals with one peak at A$_{\rm V}$ = 0.4 that we
attribute to molecular hydrogen \citep{Schneider2022}. With that value
of the column density, we obtained a field of 15.7 G$_{\rm o}$ from the
correlation given in \citet{Xia2022}. For Spider, we adopted a value of
7.3 10$^{19}$ cm$^{-2}$ for the molecular hydrogen column density
N(H$_2$), cited in \citet{Barriault2010a}, and derived a field of 6
G$_{\rm o}$.

Table~\ref{tab:flux} summarizes the values for the 160 $\mu$m flux and
the FUV field for all sources at the observed positions. There is
obviously a discrepancy between the values derived from the {\sl
  Herschel} fluxes and the census of the stars. We will come back to
this point in the discussion.

\begin{figure*} 
\centering
\includegraphics[width=7.5cm,angle=0]{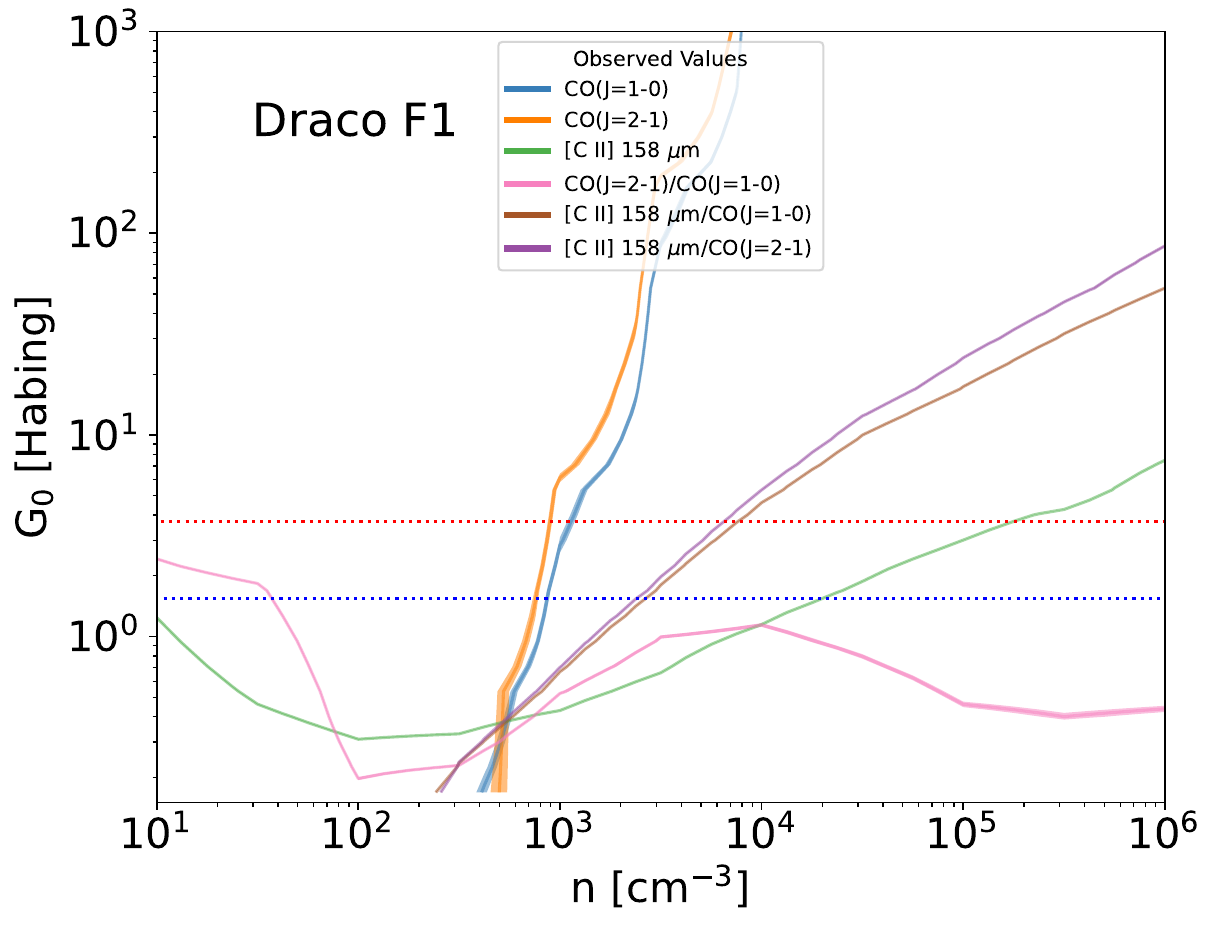}
\includegraphics[width=7.5cm,angle=0]{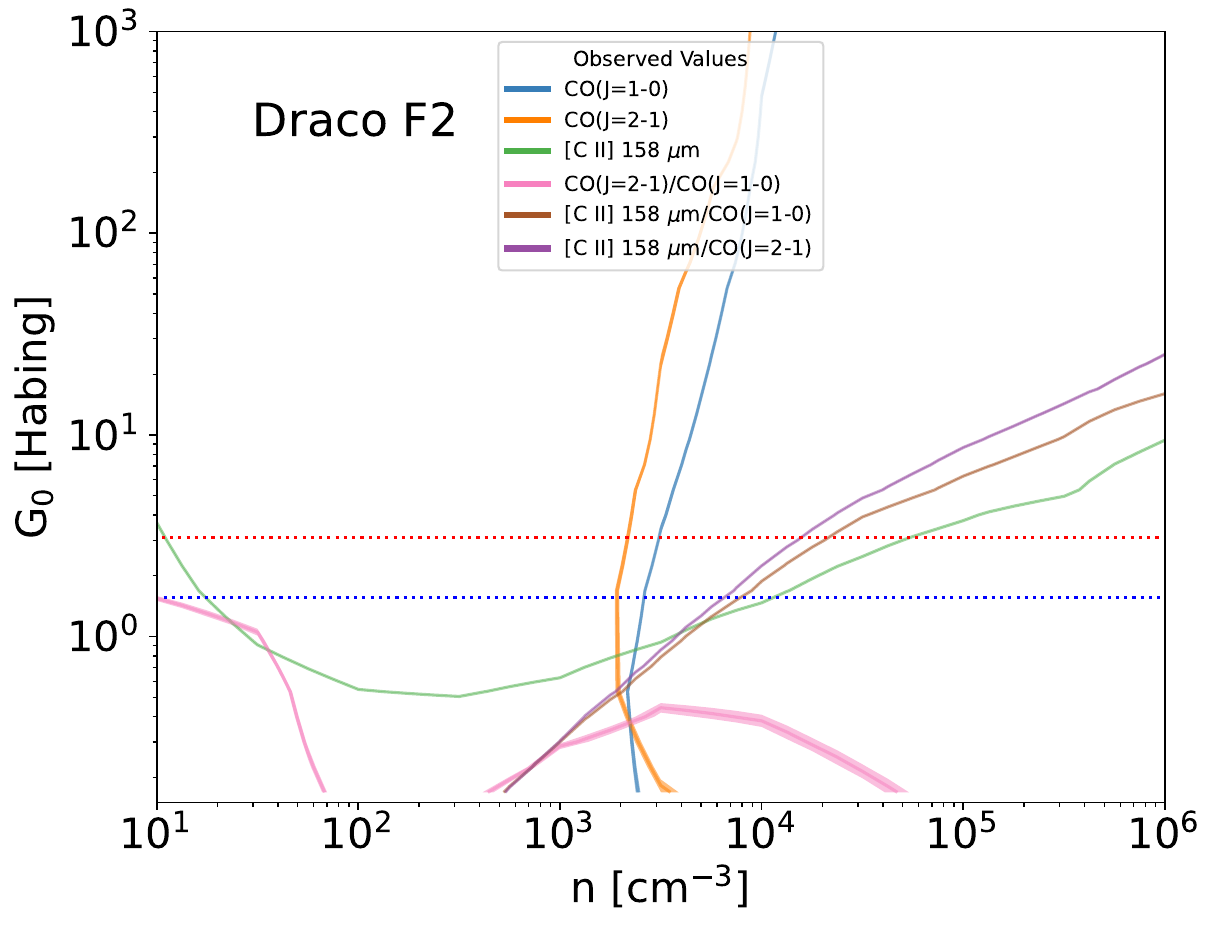}
\includegraphics[width=7.5cm,angle=0]{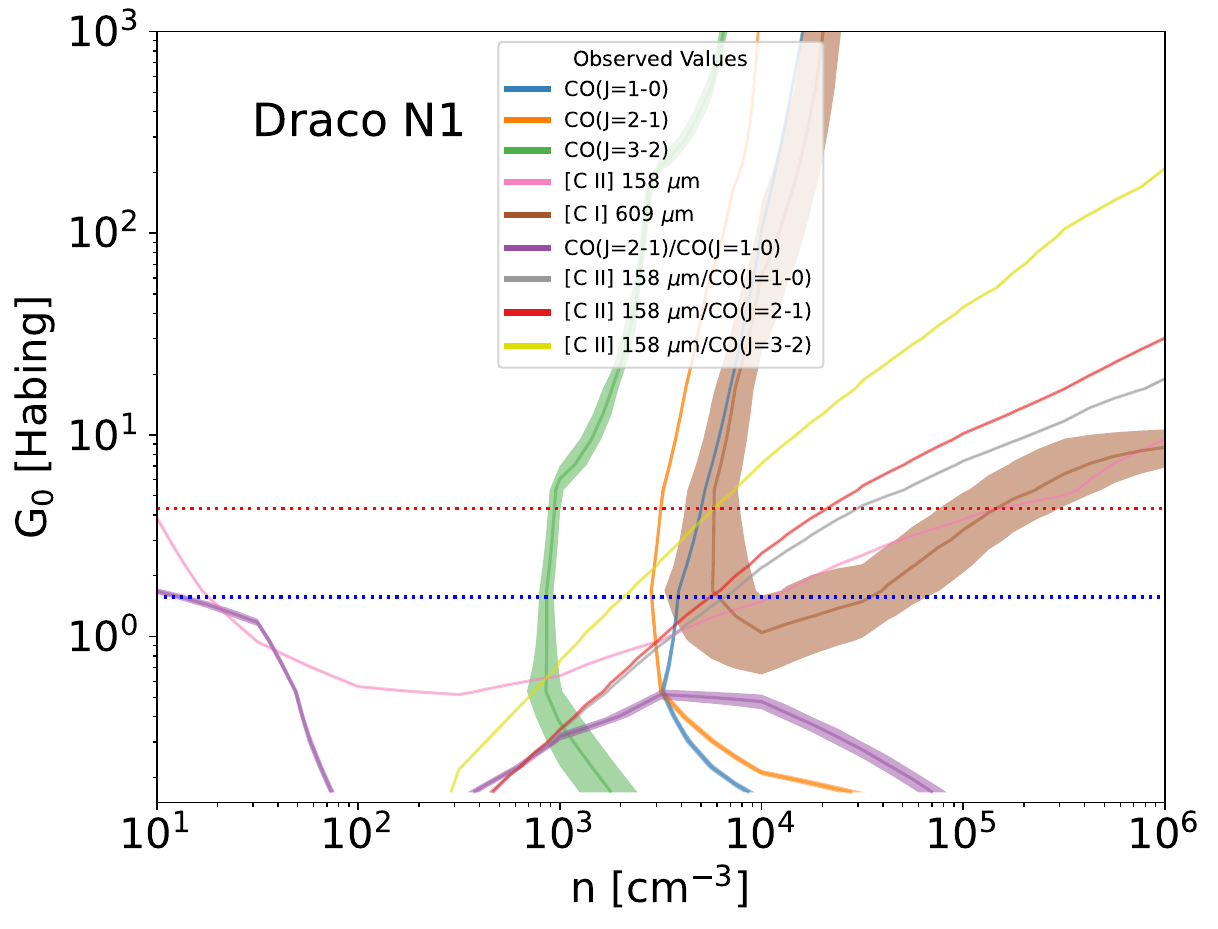}
\includegraphics[width=7.5cm,angle=0]{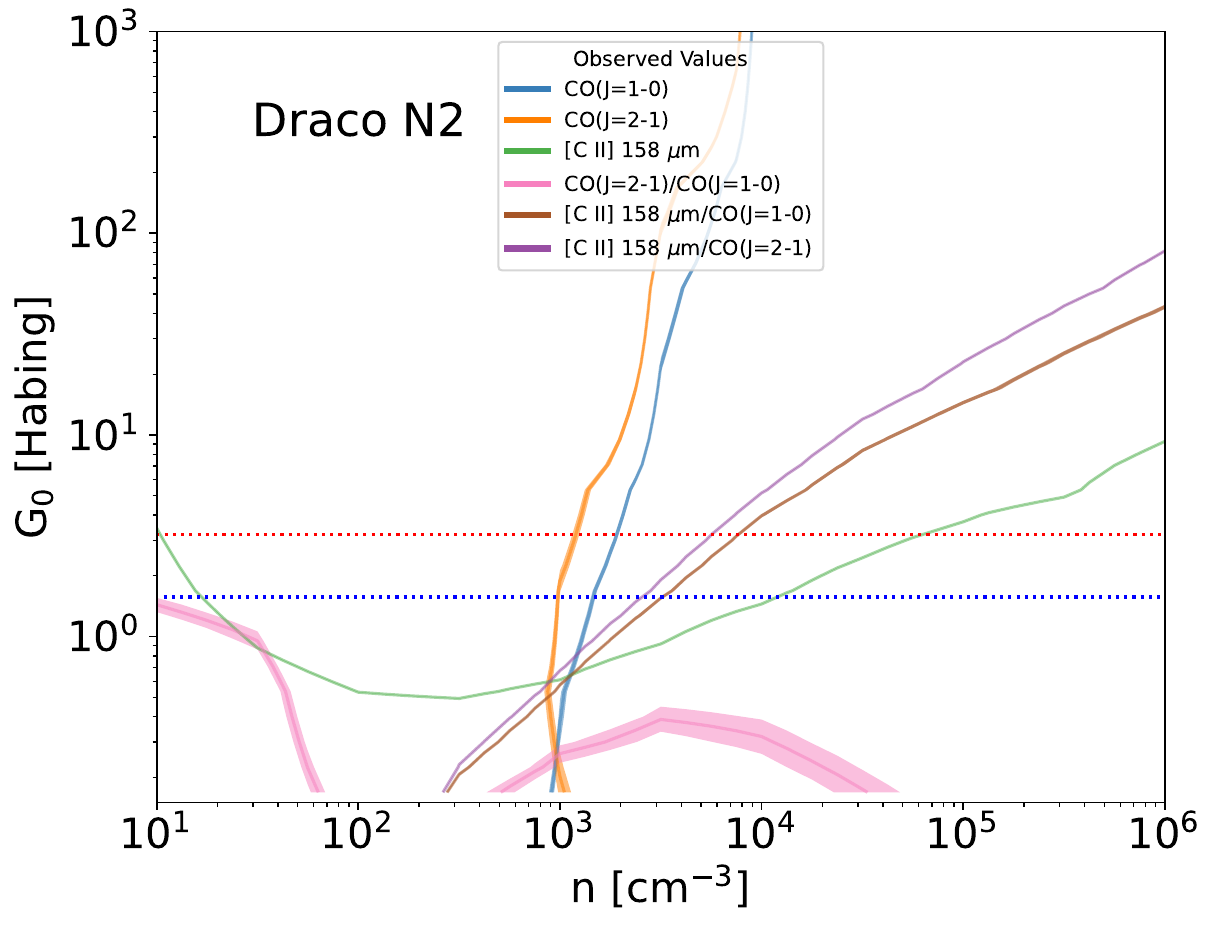}
\caption{PDR modeling results for Draco. Parameter space of hydrogen surface
  density $n$ and FUV field calculated from the KOSMA-$\tau$ model for
  0.1~M$_\odot$ clumps taken from the PDR toolbox for the Draco Front
  1 and Front 2 (left) positions and the Nose 1 and Nose 2 (right)
  positions. The isocontours at different colors show the observed
  line integrated intensities or ratios, including the
  r.m.s. noise. The estimated FUV field for each source from the 160
  $\mu$m flux is indicated by a red dashed line. The FUV field from
  the census of the stars is given as a dashed blue line.  }
\label{fig:pdr-draco}
\end{figure*}

\begin{figure*} 
\centering
\includegraphics[width=6cm,angle=0]{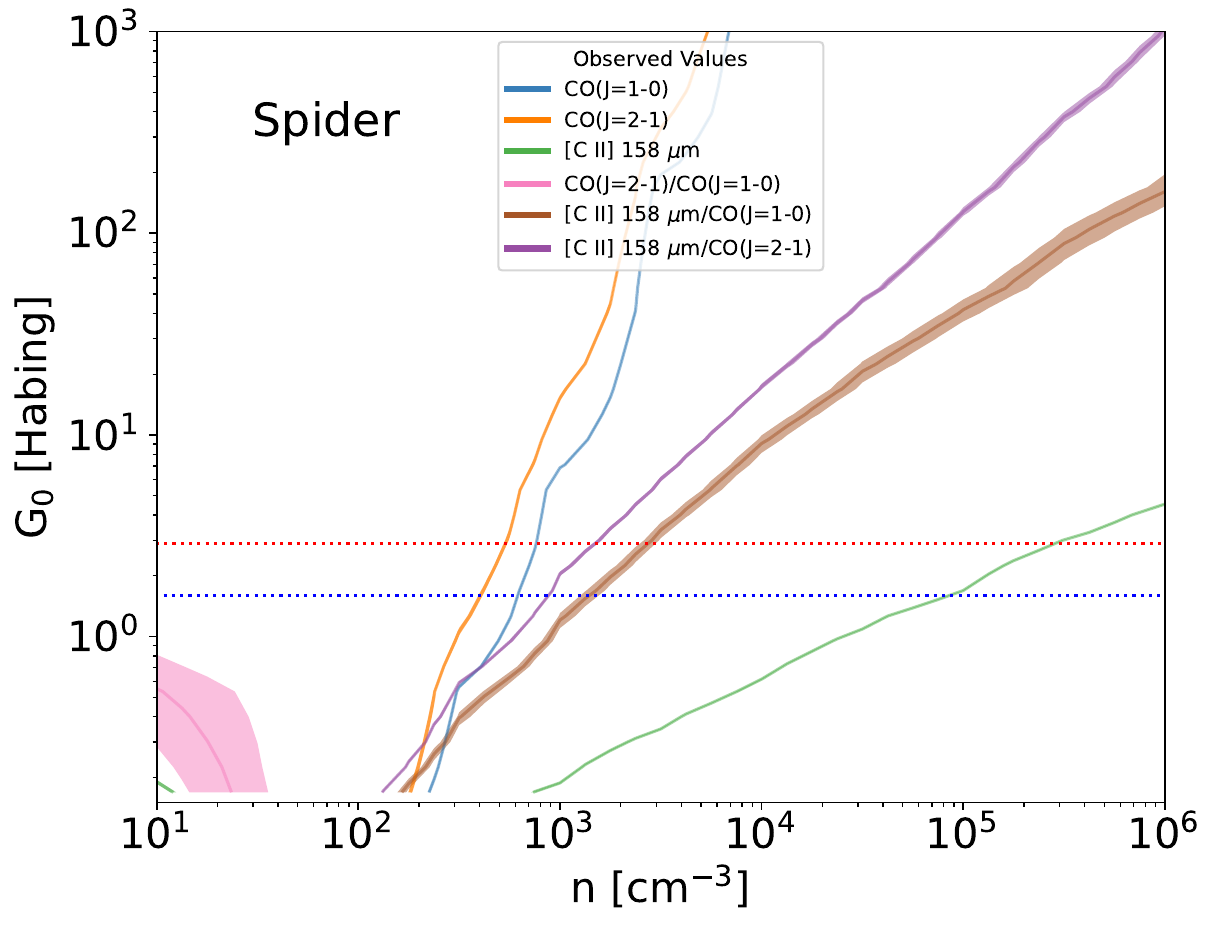}
\includegraphics[width=6cm,angle=0]{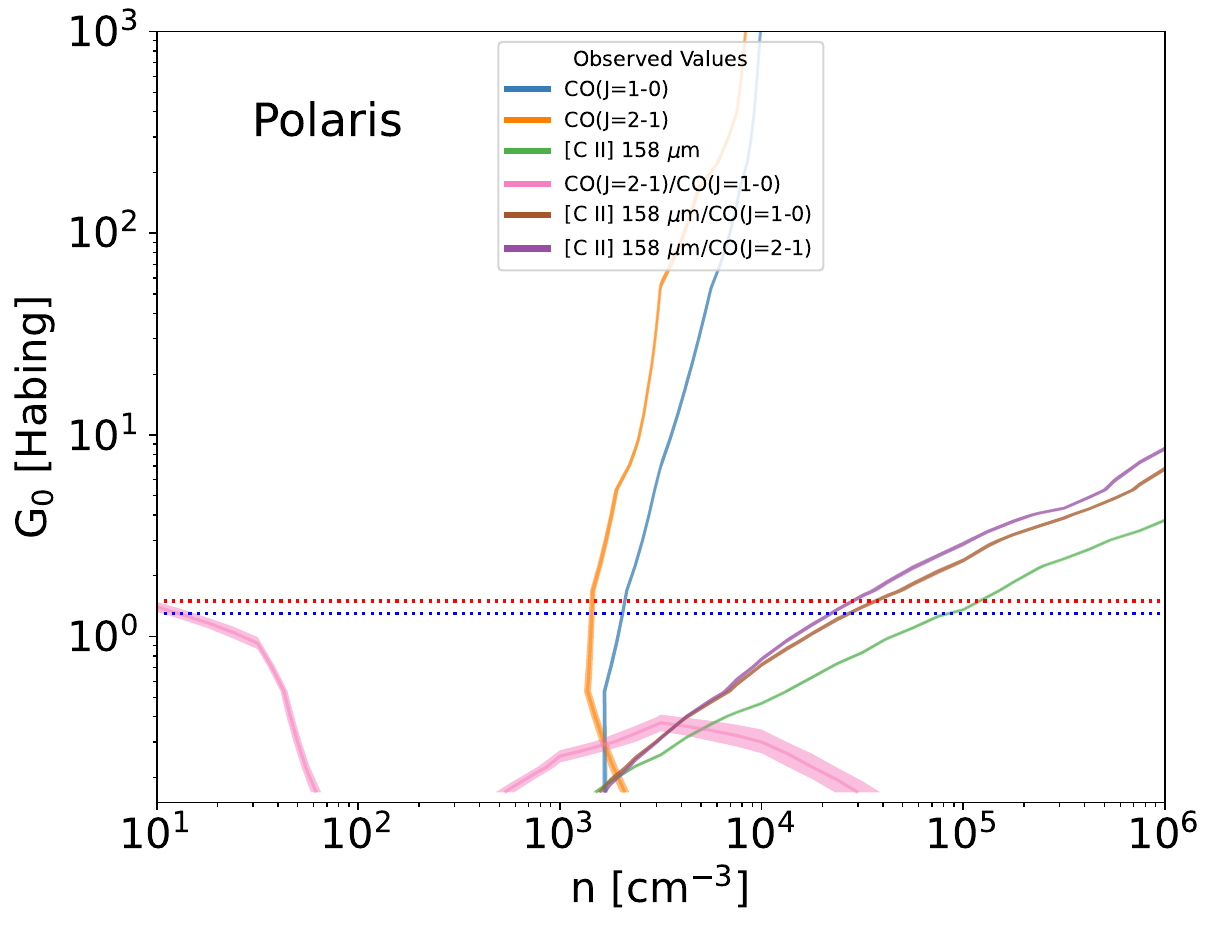}
\includegraphics[width=6cm,angle=0]{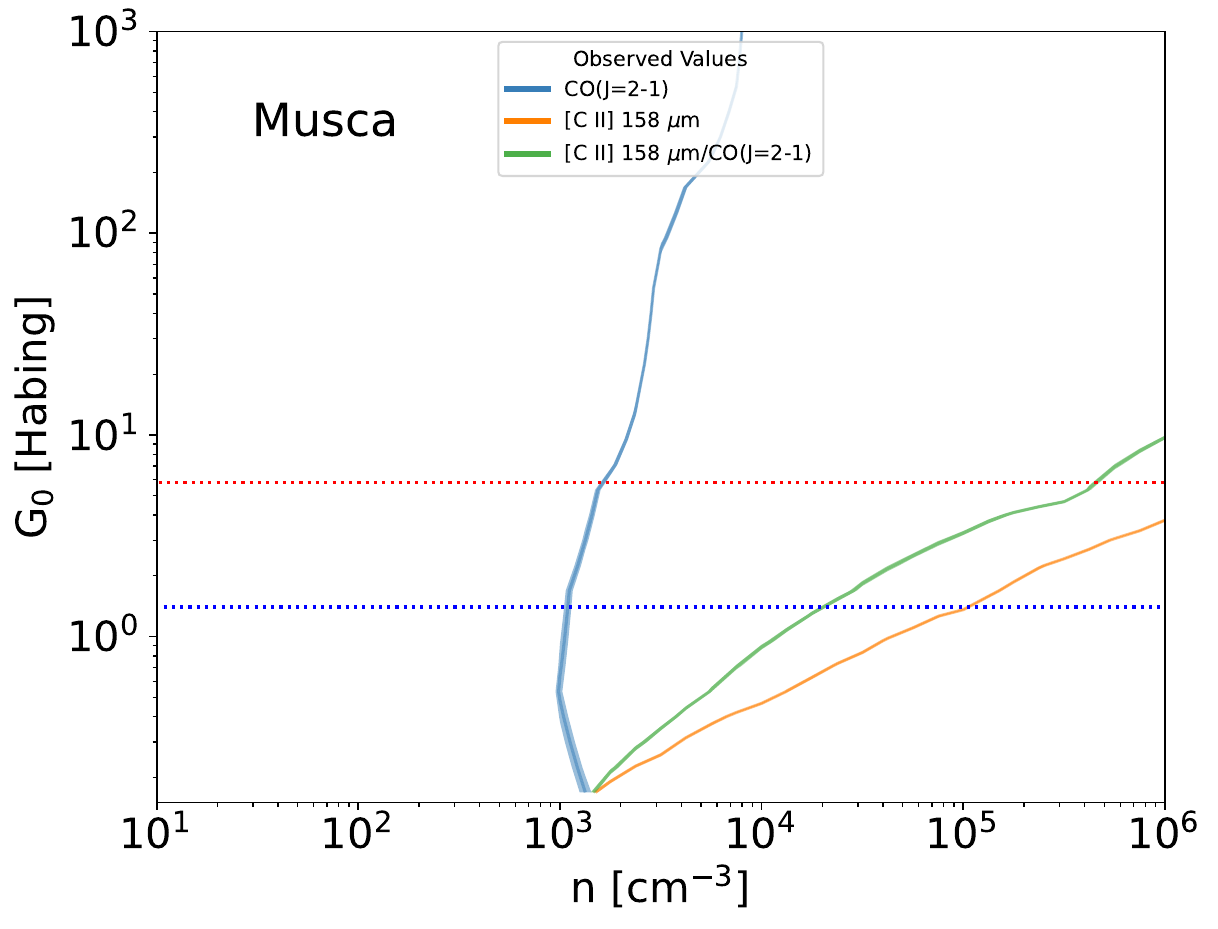}
\caption{PDR modeling results for Spider, Polaris, and
  Musca. Parameter space of hydrogen density $n$ and FUV field
  calculated from the KOSMA-$\tau$ model taken from the PDR toolbox
  for the Spider 1 position (left, 0.1~M$_\odot$ clumps), the Polaris
  single pointing (middle, 1~M$_\odot$ clumps ), and the Musca single
  pointing (right, 0.1~M$_\odot$ clumps).  The isocontours at
  different colors show the observed line integrated intensities or
  ratios, including the r.m.s. noise. The estimated FUV field for each
  source from the 160 $\mu$m flux is indicated by a red dashed
  line. The FUV field from the census of the stars is given as a
  dashed blue line.}
\label{fig:pdr-sources}
\end{figure*}

\subsection{PDR modeling of line emission} \label{subsec:pdr} 

For modeling the observed \CII, CO and \CI\ intensities and ratios, we
used the KOSMA-$\tau$ 1D spherical model \citep{Roellig2022} with an
isotropic radiation field.  The model results are included in the PDR
toolbox\footnote{
\href{https://dustem.astro.umd.edu}{https://dustem.astro.umd.edu}}
\citep{Pound2023} that also delivers results from a plane-parallel PDR
model, the Wolfire-Kaufman models from 2006 and 2020 \citep[see
  references in][]{Pound2023}.  The toolbox presents the models as
grids of model predictions for the intensity or intensity ratio as a
function of the hydrogen nucleus density $n$ and the radiation field
strength FUV. Here, we stick to the KOSMA-$\tau$ model because this
represents a finite configuration that we can adjust to the geometry
of the sources while the plane-parallel models adopted a fixed PDR
depth of A$_{\rm v}$ = 7, too high for the sources in our study.

The KOSMA-$\tau$ model solves the radiative transfer equation with
chemical balance and thermal equilibrium for a clumpy PDR exposed to a
variable FUV radiation field and cosmic rays according to the typical
primary ionization rate of $2\times 10^{-16}$~s$^{-1}$. The model that
we used assumes a dust composition that produces a reddening parameter
R$_{\rm V}$ = 3.1, where R$_{\rm V}$ is the ratio of visual extinction
A$_{\rm V}$ to reddening E(B-V), typical for diffuse clouds
\citep[model 7 from][]{Weingartner2001}.  It assumes a spherical cloud
configuration with a radial density profile approximating a critical
Bonnor-Ebert sphere with a constant central density and a power law
density decay in the outer 80\,\% of the radius. Models are
characterized by clump mass, density at the surface, and impinging FUV
field. From the simple geometry the average density is 1.92 times the
density at the surface and we get a fixed relation between clump mass,
$M_{\rm clump}$, density at the surface, $n$, and average column
density
\begin{equation}
    \langle N \rangle = 1.25\times 10^{20} \mathrm{cm}^{-2} \frac{M_{\rm clump}/M_\odot}{\pi (r_{\rm clump}/\mathrm{pc})^2}
    \label{eq:avcolumn_deriv1}
\end{equation}
with 
\begin{equation}
    r_{\rm clump}=40.5 \mathrm{pc} \left(\frac{M_{\rm clump}/M_\odot}{4\pi/3 \times 1.92\times n/\mathrm{cm}^{-3}}\right)^{1/3}
    \label{eq:avcolumn_deriv2}
\end{equation}
combining to a fixed relation between column density and density at the surface 
\begin{equation}
    \frac{n}{\mathrm{cm}^{-3}} = \left(\frac{\langle N \rangle}{1.348\times 10^{19} \mathrm{cm}^{-2}}\right)^{3/2} \Bigg/ \sqrt{\frac{M_{\rm clump}}{M_\odot}}.
    \label{eq:avcolumn}
\end{equation}

From Eq.~(\ref{eq:avcolumn}) we obtained for a column density of
$10^{21}$~cm$^{-2}$ (average value for the 4 positions in Draco) a
density at the surface of 2020~cm$^{-3}$ when using a clump mass of
0.1~$M_\odot$ or 640~cm$^{-3}$ for a clump mass of 1~$M_\odot$. This
is the density range we expect to be seen in the PDR modeling of the
observed lines.

\begin{table}
\caption{Density and FUV field from the PDR model.} \label{tab:pdr-results}
\begin{center}
\begin{tabular}{l|c|c|c}
\hline
Source  & $n(\mathrm{H}$) [cm$^{-3}$] & FUV [G$_\circ$] & $\chi^2$ \\
\hline
Draco F1 & 311$\pm$1  &  0.23$\pm$0.0008 &  4.4 10$^{-4}$\\
Draco F2 & 370$\pm$85 &  0.17$\pm$0.05   &  2.91  \\
Draco N1 & 636$\pm$18 &  0.23$\pm$0.005  &  0.15  \\
Draco N2 & 272$\pm$78 &  0.17$\pm$0.04   &  6.65 \\
Spider   & 293$\pm$297 & 0.16$\pm$0.02	 &  116\\
Polaris  & 3838$\pm$52 & 0.35$\pm$0.004	 & 7.3 10$^{-3}$  \\
\hline
\end{tabular}
\end{center}
\end{table}

We show here the line intensities and ratios of \CII, CO and
\CI\ emission and fit the line ratios to obtain the most likely value
for density and FUV field. We only use the velocity components that
correspond to the velocity of the \CII\ line (for example $-$25 km
s$^{-1}$ for Draco Front 1a and $-$22 km s$^{-1}$ for Draco Front 2a,
respectively). However, each source has a different set of lines
available so that a comparison is difficult. We emphasize that the CO
lines are very sensitive to the assumed total depth of the cloud. The
\CII\ intensity and the surface temperature trace surface properties
while the CO emission only arises from the layers deeper in the cloud
where CO can form.  It is a simplified approach to consider a single
density only and not a density distribution. A surface tracer like
\CII\ traces somewhat thinner material than the molecular
lines. However, Fig.~\ref{fig:pdr-draco} shows that this has no
significant impact on our results. The curves for the \CII\ intensity
have a very shallow density dependence. Because of the low critical
density of the \CII\ transition, a shift in density has little effect
on the \CII\ lumninosity and consequently our fit results do not
depend on the density of the \CII\ emitting material. \\
KOSMA-$\tau$ provides models for different clump masses integrating
over that mass. Using the {\sl Herschel} total hydrogen column
densities with a beamsize of 70$''$ they are all well below 1
M$_\odot$, except for Polaris with a value of $\sim$1 M$_\odot$.
Specifically, we determine a mass of 0.21 M$_\odot$, 0.14 M$_\odot$,
0.31 M$_\odot$, and 0.32 M$_\odot$ for the Draco F1, F2, N1, and N2
positions, 0.09 M$_\odot$ for Spider, 0.08 M$_\odot$ for Musca, and
0.8 M$_\odot$ for Polaris. For simplicity, and because these models
are directly available in the open access PDR toolbox and all results
can thus easily be verified, we used pre-calculated models with M = 0.1
M$_\odot$ for Draco, Spider, and Musca and with M = 1 M$_\odot$ for
Polaris. Note that using models with M = 0.3 M$_\odot$ (R\"ollig,
Ossenkopf priv. comm.) do not significantly change the results. With
the higher mass the solution for the absolute intensities of the CO
and \CI\ lines shift by about a factor of two to lower densities while
the \CII\ intensity and the ratios remain almost
unchanged. \\
Figures~\ref{fig:pdr-draco} and \ref{fig:pdr-sources}
display the results of the PDR model in a parameter space of FUV field
and hydrogen density for all sources.  The observed lines and line
ratios are presented as isocontours, along with their associated
errors. It should be noted that for \CII\ in Spider, Polaris, and
Musca, only the noise level is available, effectively serving as an
upper limit. We show in Fig.~\ref{app-pdr} in
Appendix~\ref{appendix:pdr} the full model results for \CII\ emission
for masses M = 0.1 M$_\odot$ and M = 1 M$_\odot$ and in
Fig.~\ref{app-co} the model result (as an example) for CO 2$\to$1.  In
terms of representing the FUV field, a dashed red horizontal line
corresponds to the field determined through the translation of the 160
$\mu$m fluxes, while a blue line signifies the field established via
the stellar census. Notably, data from \citet{Xia2022}, which tend to
be higher than other values, are omitted. This omission does not alter
the interpretation of the PDR modeling. \\
A solution for the density and FUV field from the PDR plots is defined
by a common crossing point of all lines. Because the absolute line
intensities depend on geometrical details like beam dilution effects,
line ratios are more reliable. We thus used the line ratios in the
LineRatioFit method in the PDR toolbox to determine the most likely
values of total hydrogen density density and FUV field including
errors and $\chi^2$. These values are listed in
Table~\ref{tab:pdr-results}.

For the Front1 position, the observations align with a very low FUV
field of around 0.4 G$_\circ$ and densities of approximately 500
cm$^{-3}$ by eye-inspection of Fig.~\ref{fig:pdr-draco}. The FUV field
and the density from the fit are lower, that is, 0.23 G$_\circ$ and
311 cm$^{-3}$. \\
For the Front 2 position, the \CII\ line crosses the CO intensities
for a density of $\sim$2$\times$10$^3$ cm$^{-3}$ at an UV field of 1
G$_\circ$. The CO 2$\to$1/1$\to$0 ratio points toward a lower UV field
of around 0.4 G$_\circ$. The fitted values are much lower with a FUV
field of 0.17 G$_\circ$ and a density of 370 cm$^{-3}$.\\
For the Nose 1 position, we also consider the CO 3$\to$2 line
intensity (Heithausen priv. comm.) and the \CII/CO 3$\to$2 and CO
3$\to$2/2$\to$1 line ratios and the \CI\ line intensity. We derived
the latter by averaging over the 4 positions observed in Draco close
to the Nose 1 position \citep{Heithausen2001} and arrive to a value of
4.3$\times$10$^{-7}$ erg s$^{-1}$ sr$^{-1}$ cm$^{-2}$. The CO
3$\to$2/2$\to$1 ratio falls essentially outside the parameter space
(very low FUV field and densities) and is not visible in the plot. On
the other hand, the \CII/CO 3$\to$2 ratio, the \CI\ line, and the CO
2$\to$1 and 1$\to$0 intensities could align for densities of
approximately 3$\times$10$^3$ cm$^{-3}$ for a low FUV field of around
1-2 G$_\circ$. Fitting only the line ratios, however, leads to a lower
FUV field of 0.23 G$_\circ$ and a density of 636 cm$^{-3}$. \\
For the Nose 2 position, the individual CO line intensities intersect
with the \CII\ line at densities around 10$^3$ cm$^{-3}$ at an UV
field of around 0.6 G$_\circ$. In contrast, the line ratio fitting
leads to a much smaller density of 272 cm $^{-3}$ and a FUV field of
0.17 G$_\circ$.  It is mostly the CO(2$\to$1)/(1$\to$0) ratio that
leads to these low values.

\begin{figure*} 
\centering
\includegraphics[width=7cm,angle=0]{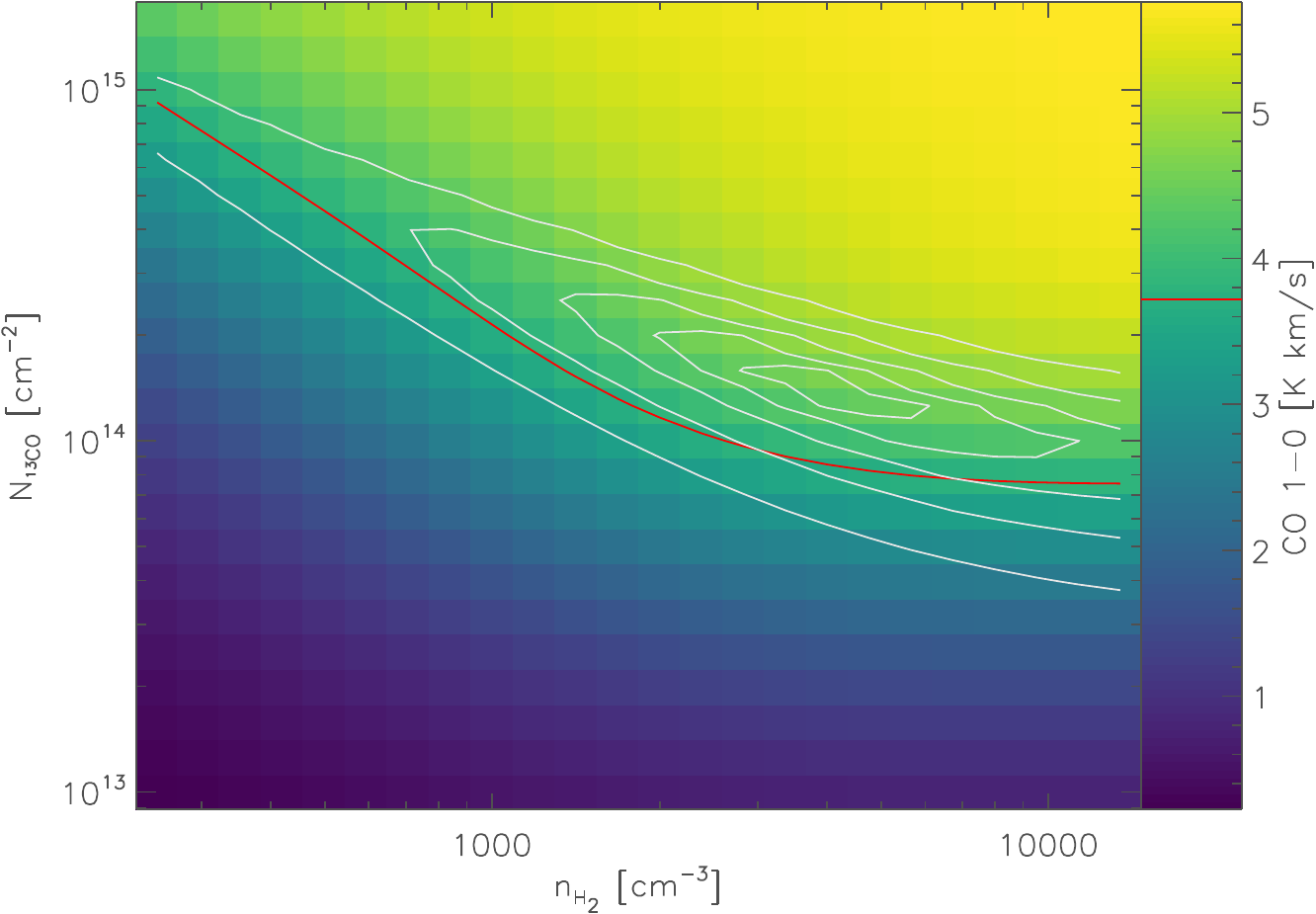}\hspace{1cm}
\includegraphics[width=7cm,angle=0]{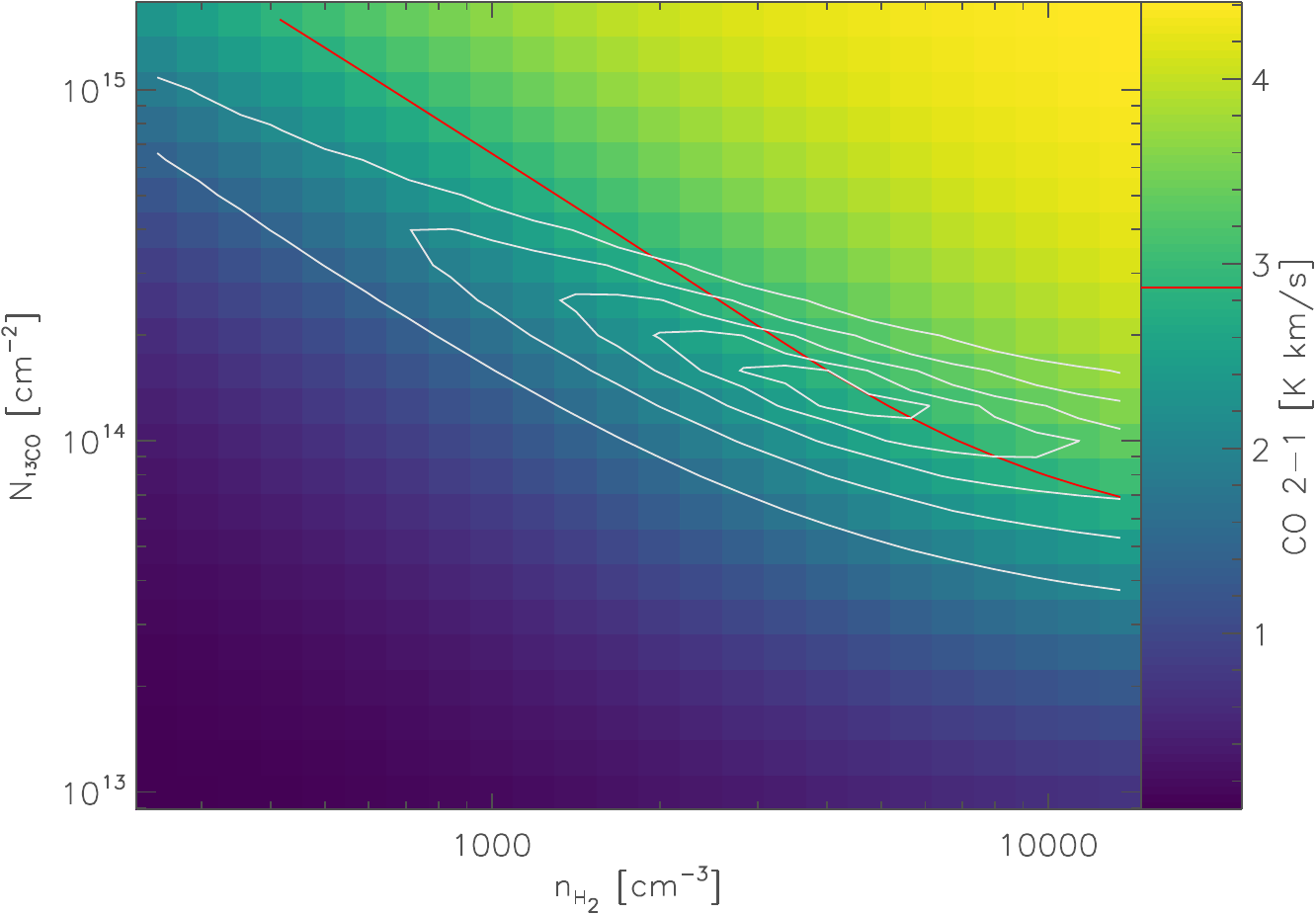}
\includegraphics[width=7cm,angle=0]{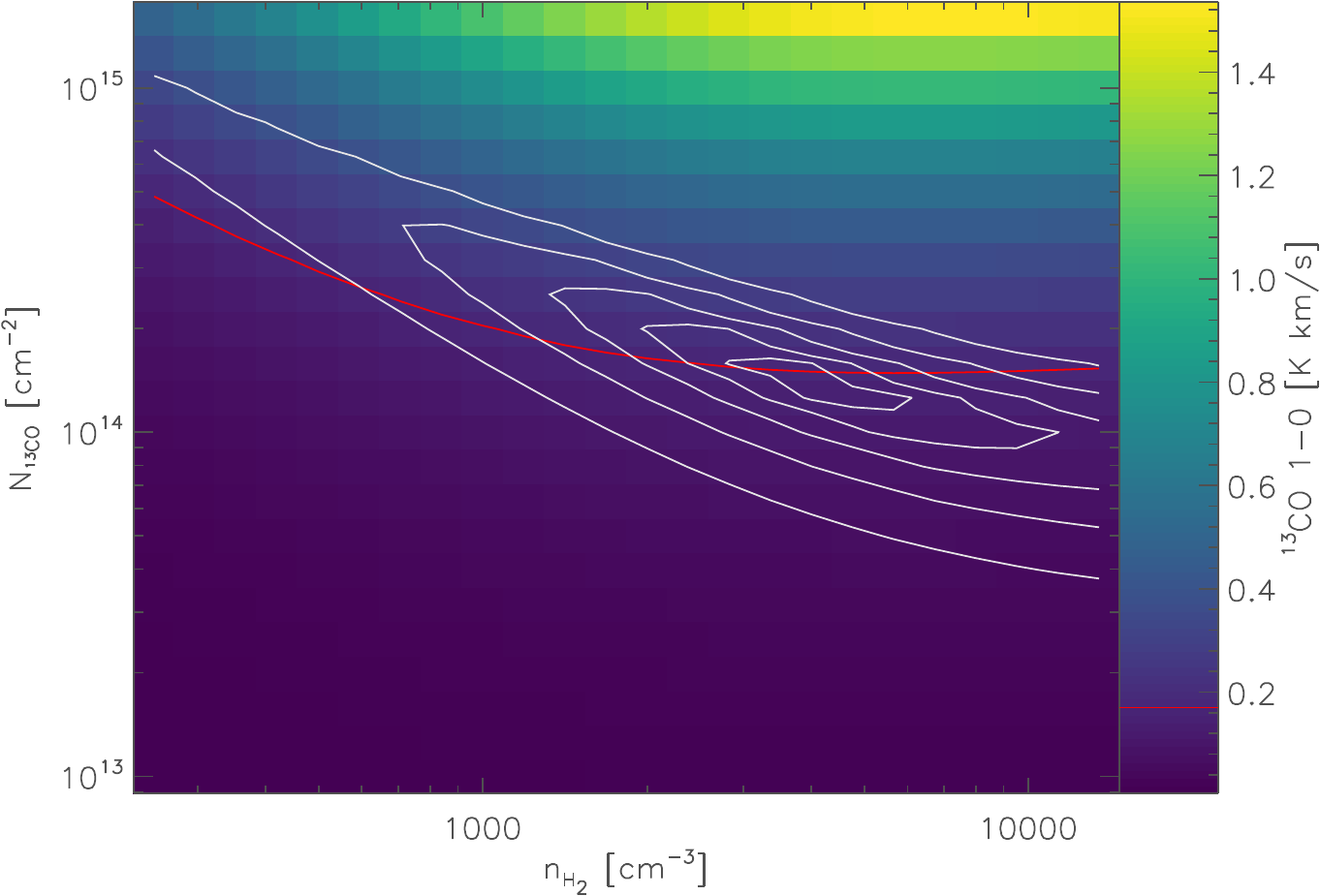}\hspace{1cm}
\includegraphics[width=7cm,angle=0]{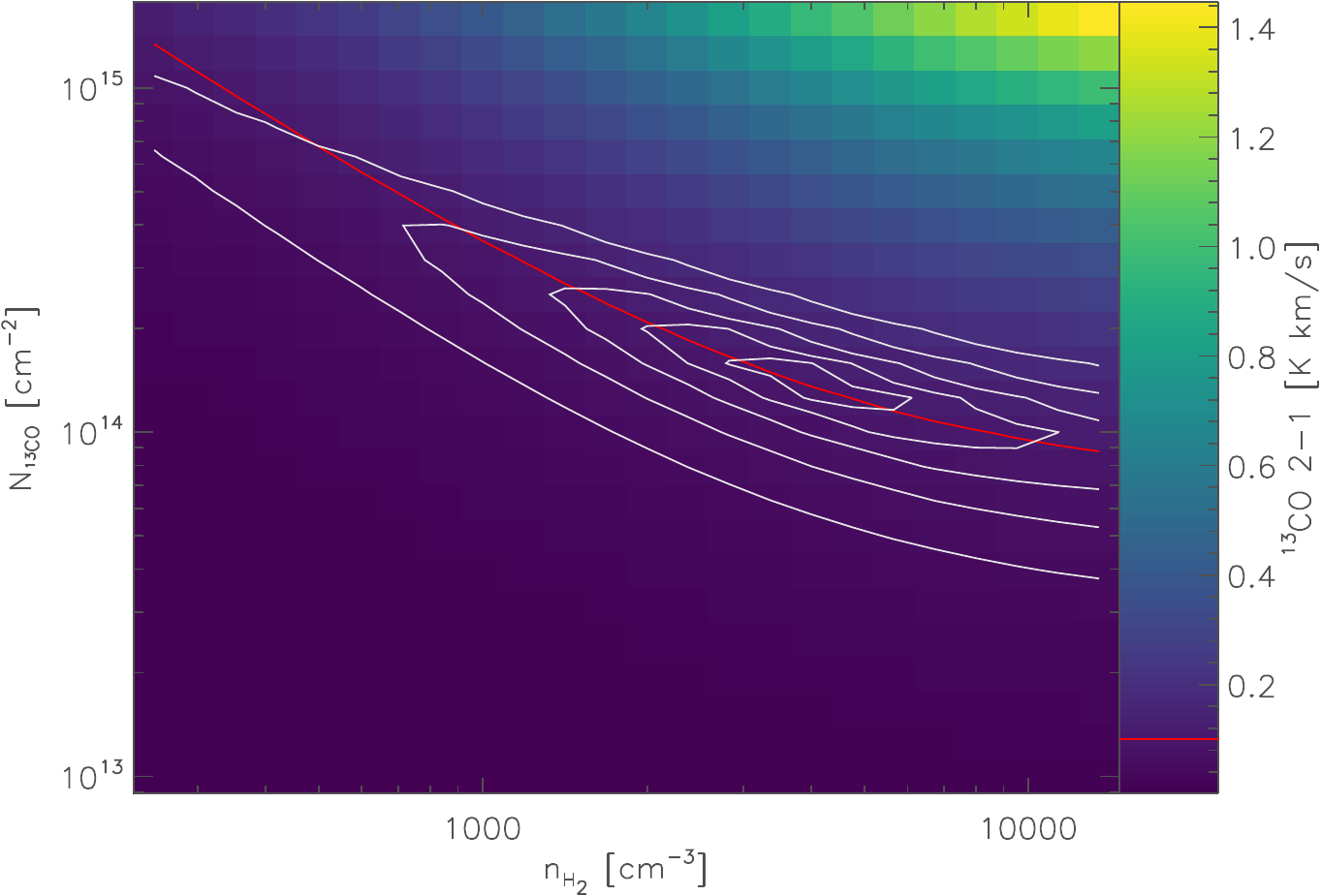}
\caption{Visualization of the RADEX fit to the four CO and $^{13}$ CO
  lines at the F1 position in Draco. Each plot shows a constant
  temperature cut through the three dimensional distribution of line
  intensities and $\chi^2$ values. The kinetic temperature of the
  $\chi^2$ minimum, that is, 9~K (see Table~\ref{tab:radex}) was
  used. The colors in the plot show the integrated line intensities
  for the four lines, the observed value is marked by a red line. The
  gray contours give $\chi^2$ values of 8, 16, 32, 64, and 128. As the
  $\chi^2$ distribution is a global property, those contours are the
  same in all four subplots.}
\label{fig:radexfit1}
\end{figure*}

\begin{figure*} 
\centering
\includegraphics[width=7cm,angle=0]{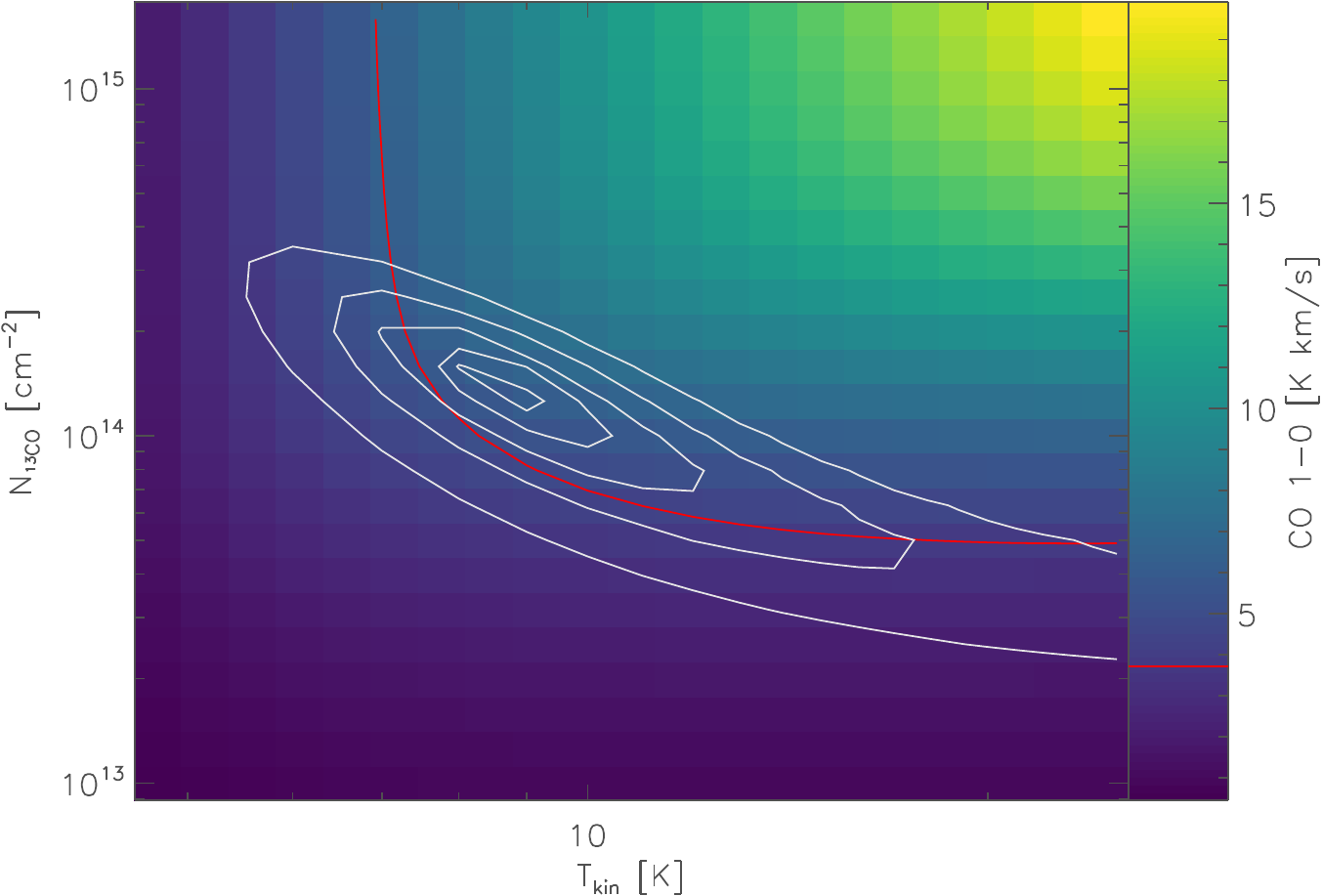}\hspace{1cm}
\includegraphics[width=7cm,angle=0]{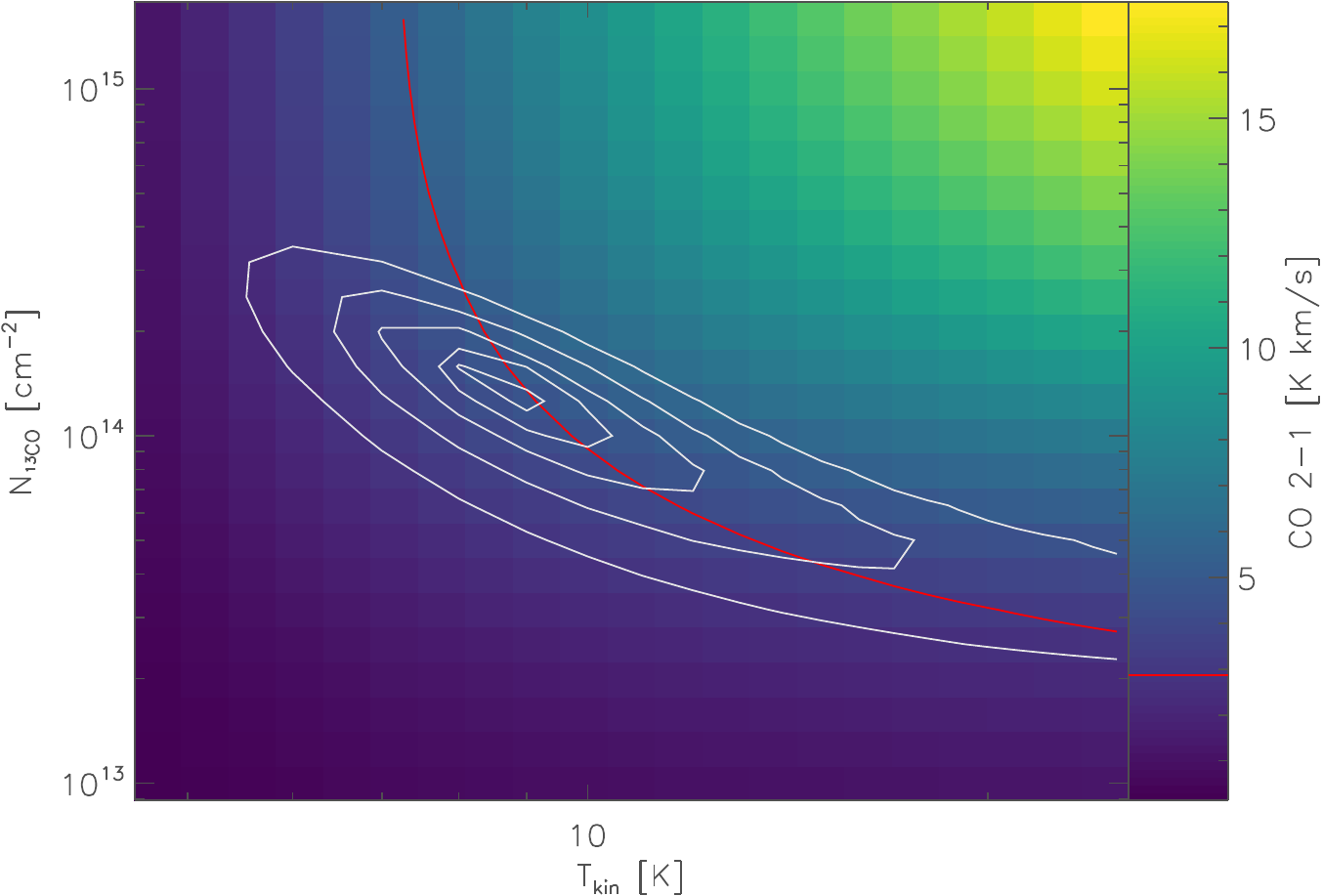}
\includegraphics[width=7cm,angle=0]{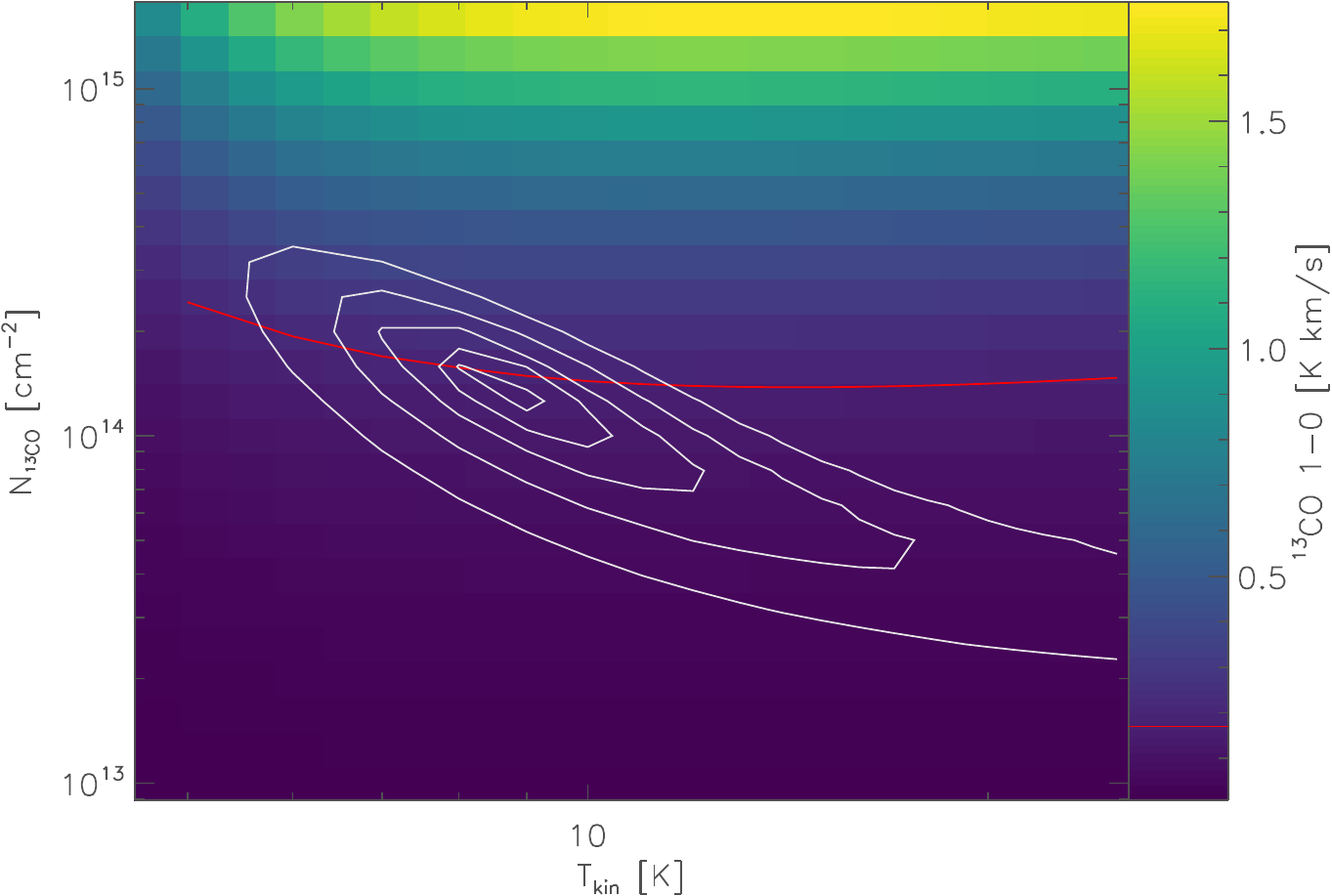}\hspace{1cm}
\includegraphics[width=7cm,angle=0]{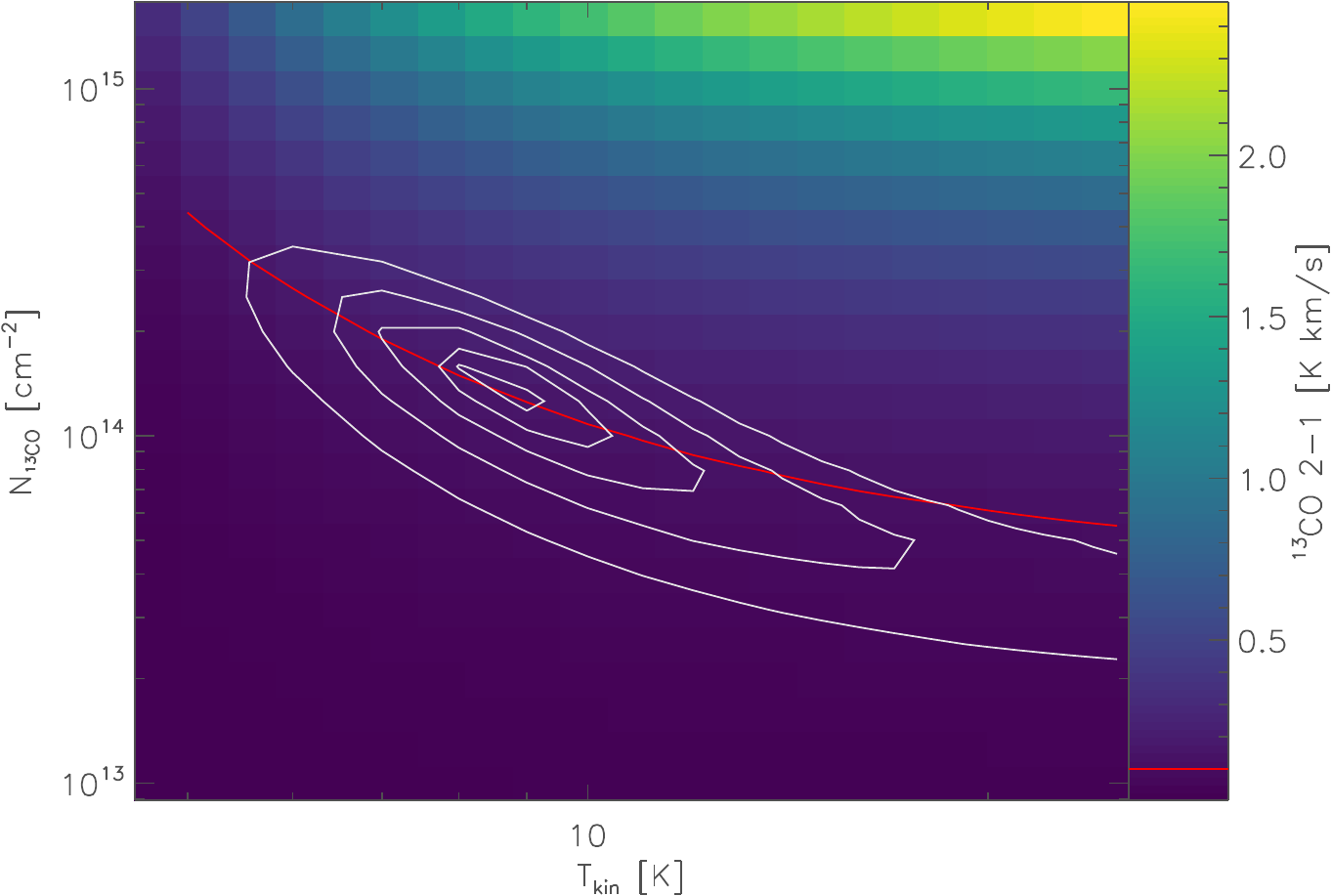}
\caption{Same as Fig.~\ref{fig:radexfit1} but for a cut at the
  constant density of the $\chi^2$ minimum of $4.9\times
  10^3$~cm$^{-3}$ (see Table~\ref{tab:radex}).}
\label{fig:radexfit2}
\end{figure*}

Figure~\ref{fig:pdr-sources} displays the PDR modeling results for
Spider, Polaris, and Musca. Note that the \CII\ value is only the
noise level and that there are fewer complementary lines. The FUV
field needs to be around 0.1 G$_\circ$ for Polaris to explain the
observed lines and ratios and then yields a density of 2$\times$10$^3$
cm$^{-3}$. The line ratio fitting gives a higher value of 3838 cm
$^{-3}$ for the density but a lower value of 0.35 G$_\circ$ for the
FUV field. The same mismatch was already noticed by \citet{Bensch2003}
when analysing [CI] observations of Polaris. They proposed some kind
of preshielding of the gas to explain the low FUV field.  For Musca,
it is very difficult to make any definitive statements since we only
have two line intensities and no ratios. However, the tendency is that
the FUV field is very low ($<$ 0.1 G$_\circ$) and the density is
around 10$^3$ cm$^{-3}$.

Summarizing, it becomes obvious that the line ratio fitting in the PDR
model always arrives to a very low FUV field ($<$ 0.35 G$_\circ$) for
all sources, which is neither supported by the FUV field determined
from the 160 $\mu$m flux by us and by \citet{Xia2022} nor by the
theoretical prediction of 1.2 - 1.6 G$_\circ$ we obtained using the
procedure outlined in \citet{Parravano2003} or the census of the
stars. The densities are also lower than in the case where we also
consider line intensities.

\begin{figure*} 
\centering
\includegraphics[width=9cm,angle=0]{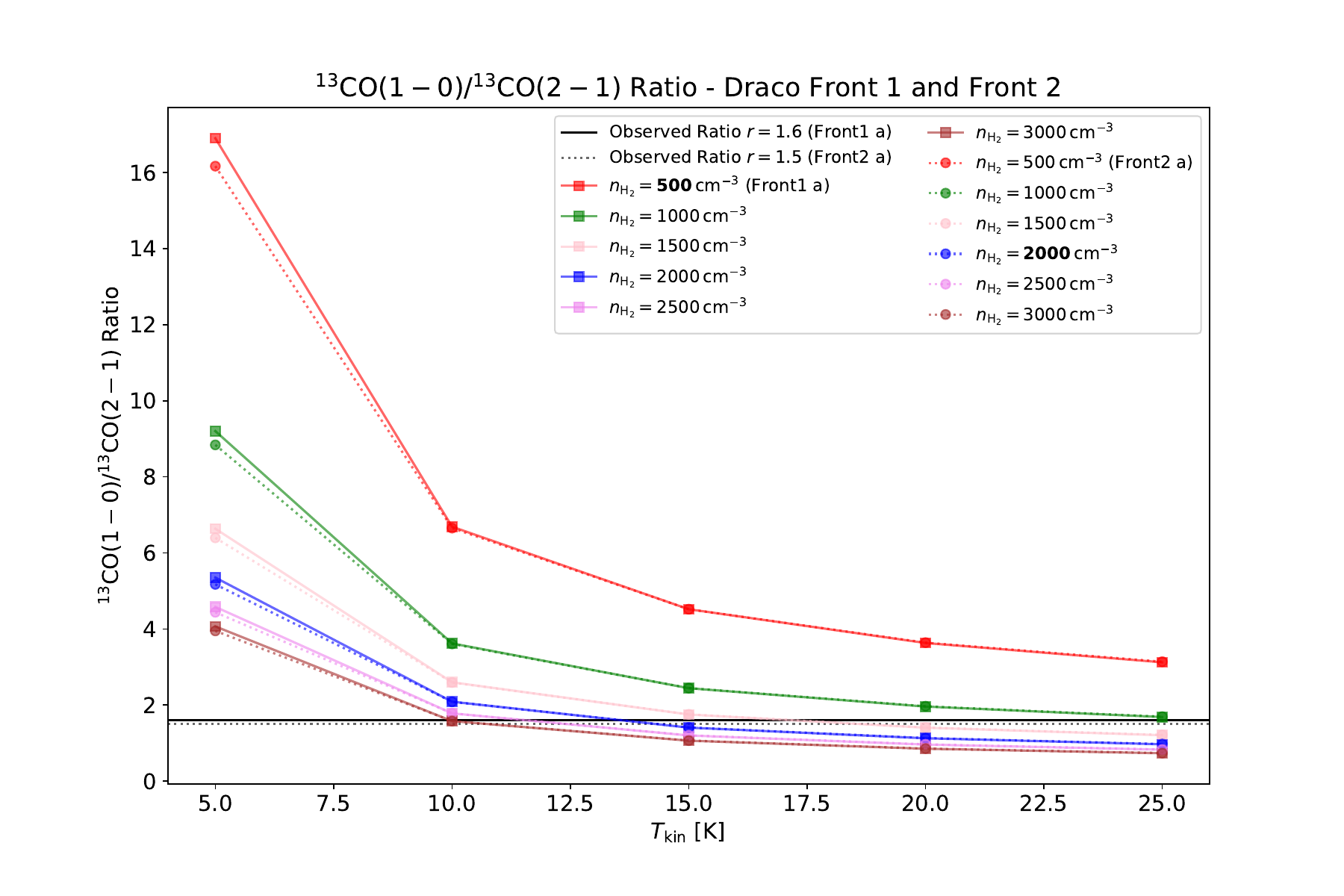}
\includegraphics[width=9cm,angle=0]{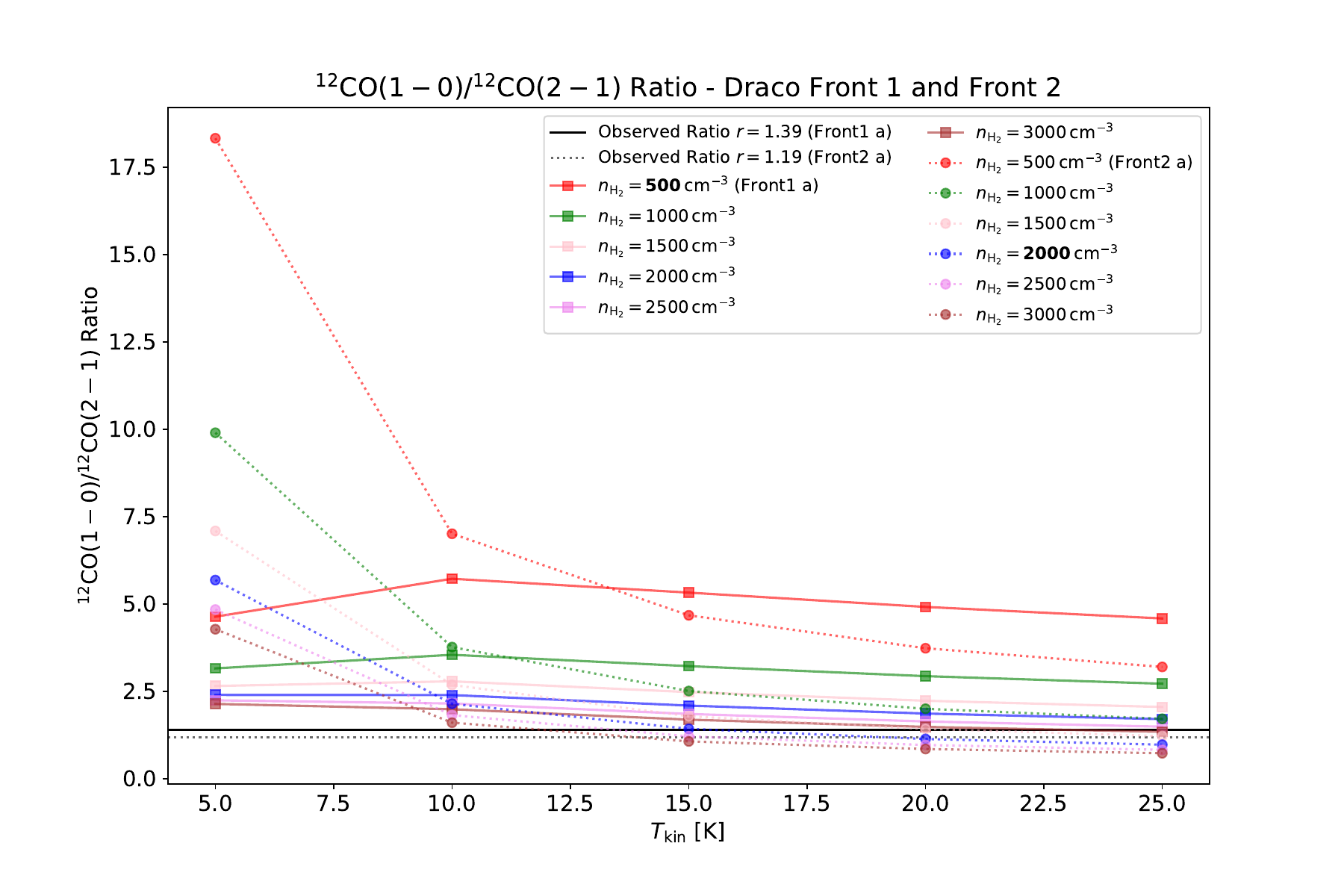}
\includegraphics[width=9cm,angle=0]{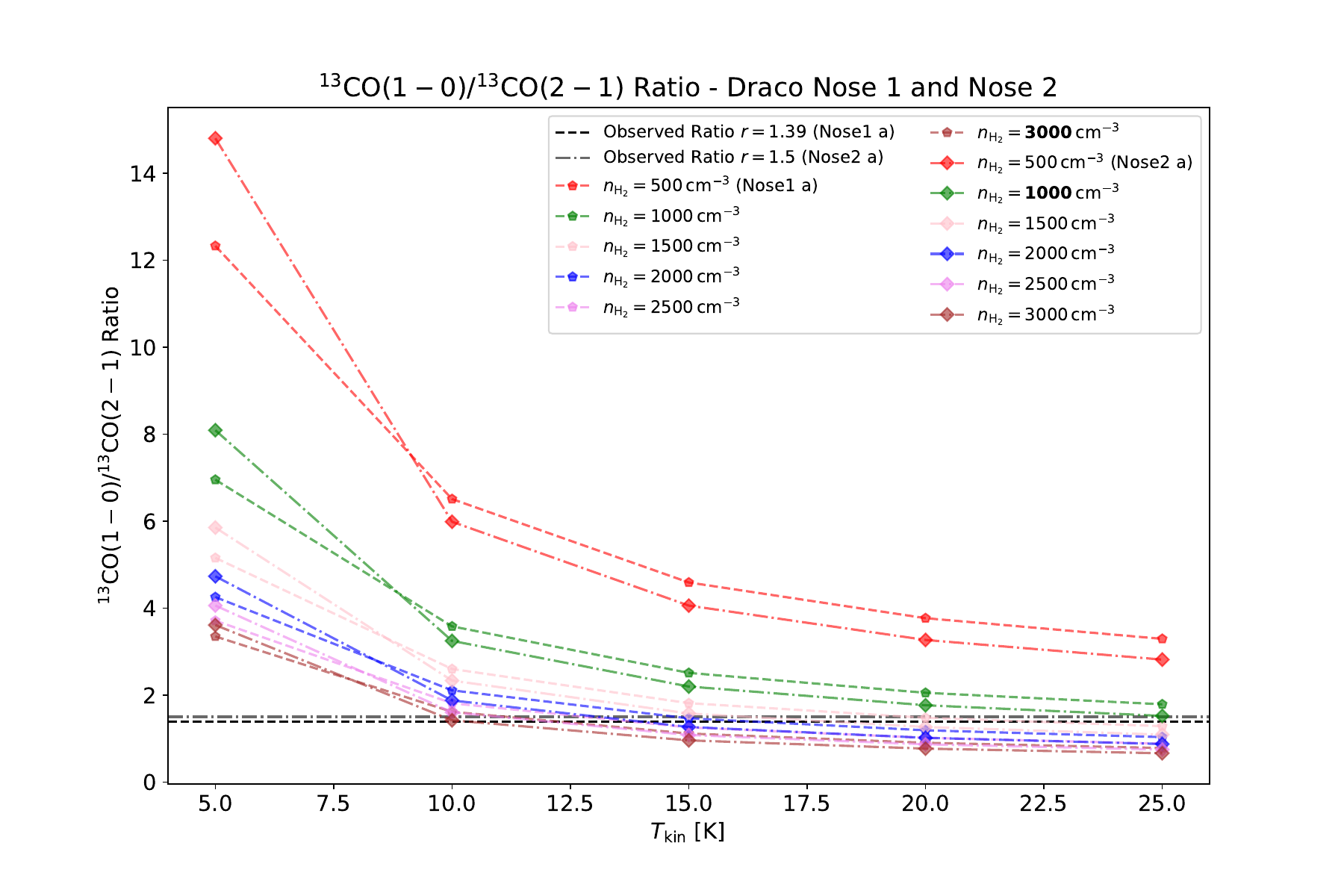}
\includegraphics[width=9cm,angle=0]{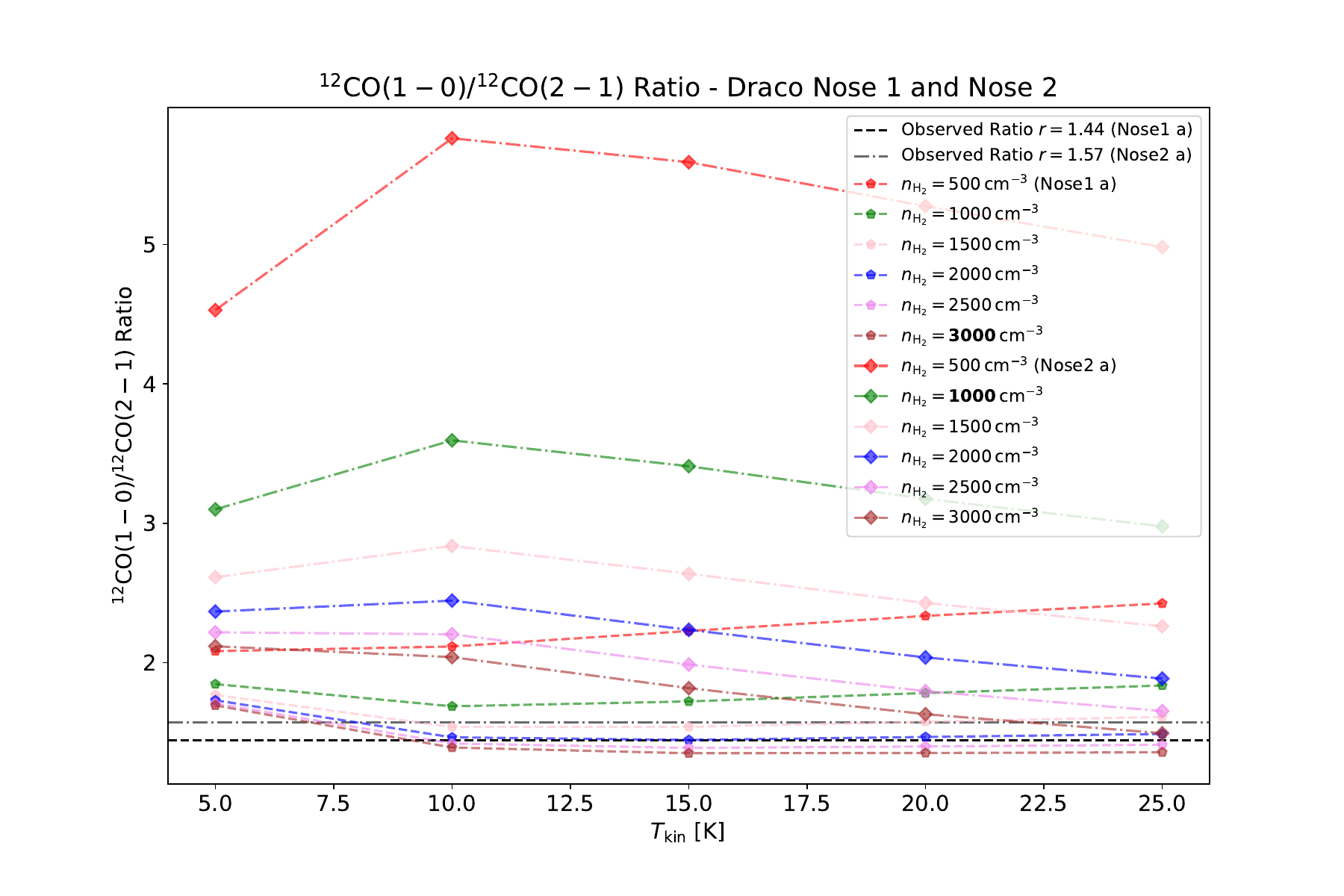}
\caption{RADEX results for the $^{13}$CO 1$\to$0/2$\to$1 and $^{12}$CO
  1$\to$0/2$\to$1 brightness ratios as a function of density and
  temperature for the Draco front and nose positions. The observed
  ratios are shown as a solid and dashed black line. The density
  obtained with PDR modeling is marked in bold.}
\label{fig:radex}
\end{figure*}

\begin{figure*} 
\centering
\includegraphics[width=9cm,angle=0]{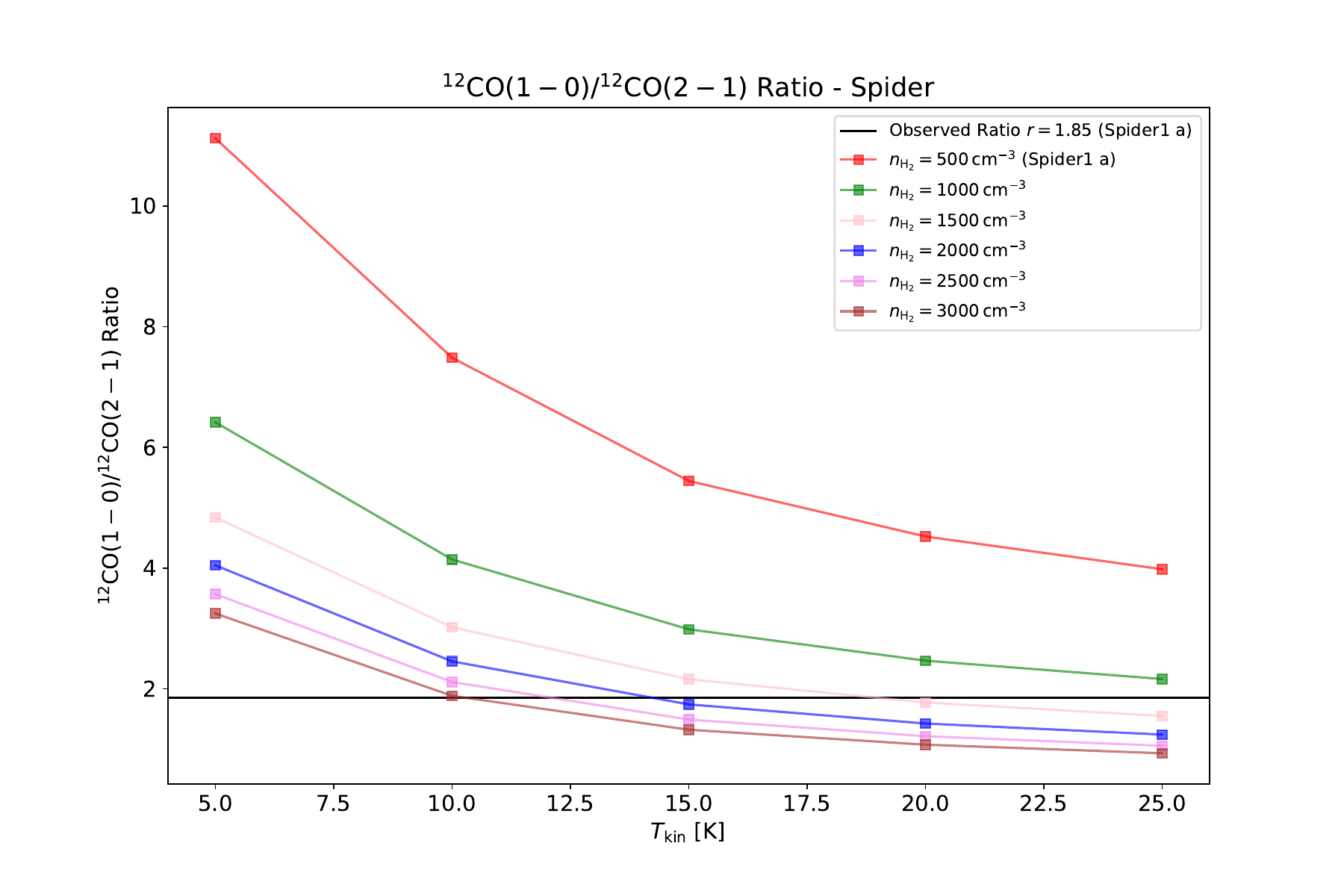}
\includegraphics[width=9cm,angle=0]{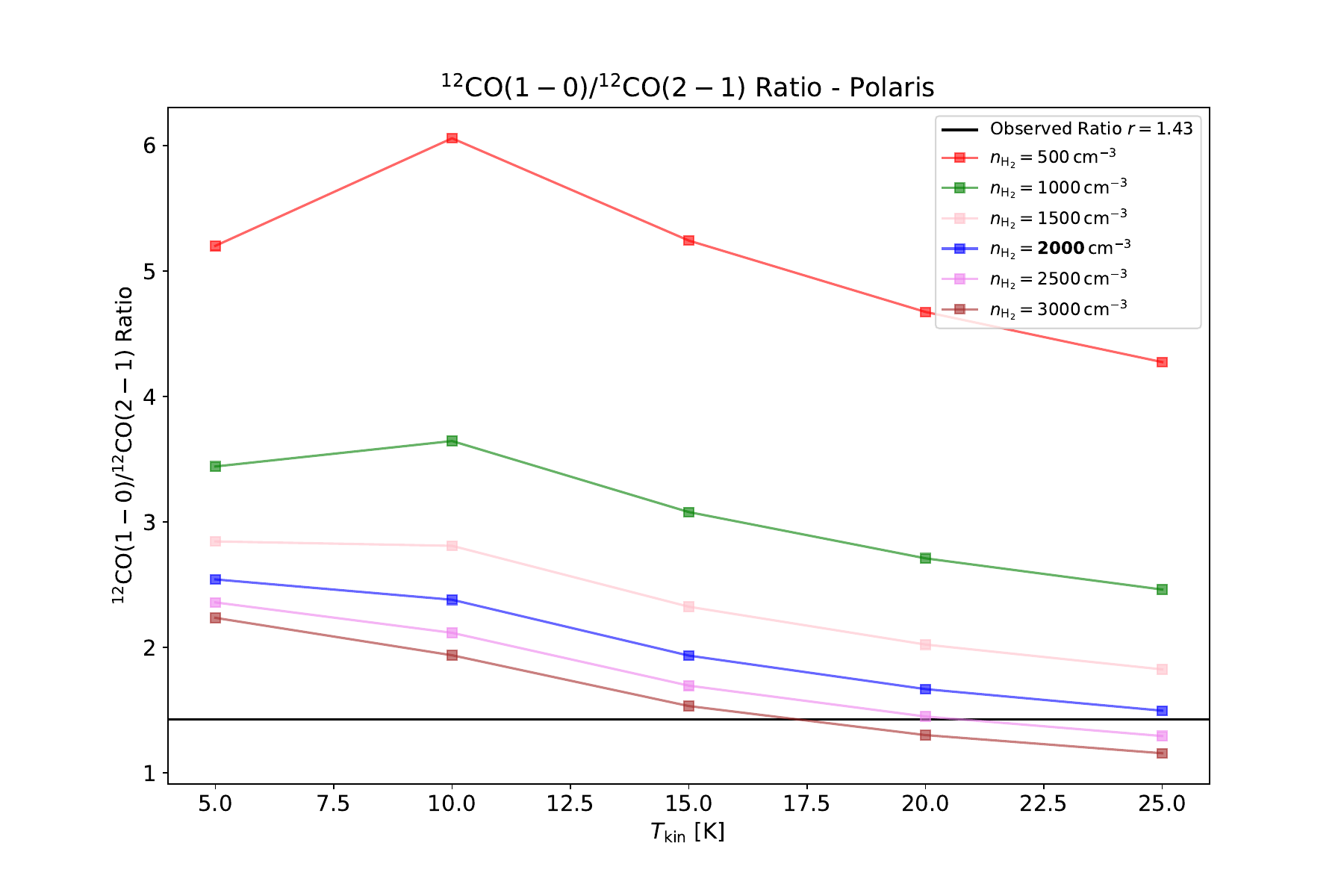}
\caption{RADEX results for the $^{12}$CO 1$\to$0/2$\to$1 brightness
  ratios as a function of density and temperature for the Spider 1a
  position (left panel) and the Polaris positions (right panel). The
  observed ratios are shown as a solid and dashed black line. The
  density obtained with PDR modeling is marked in bold.}
\label{fig:radex2}
\end{figure*}

\subsection{Non-LTE line analysis with RADEX} \label{subsec:radex}

The PDR modeling provided results for observations of warm surface gas
tracers such as \CII\ and \CI\ as well as CO lines for the cooler
interior of the gas clump. We used RADEX \citep{vanderTak2007} to
determine the physical properties of the cooler, molecular gas. We
investigated whether the observed $^{13}$CO 1$\to$0, 2$\to$1 and
$^{12}$CO 1$\to$0 and 2$\to$1 lines and their ratios can be reproduced
with this non-LTE molecular radiative transfer code. We excluded the
\CII\ line since it is not possible to determine independently the
density, temperature and column density only from one line
intensity. In other words, the determination of the \CII\ column
density is possible, but not independent of the line brightness and
vice versa.

RADEX computes the line intensities as a function of temperature,
H$_2$ density (assuming that H$_2$ is the main collision partner) and
the column density of the species for the configuration of an
isothermal homogeneous medium with a given velocity dispersion in the
simplified shape of a uniform sphere. Using an $^{12}$CO/$^{13}$CO
abundance ratio of 70 \citep{Langer1990} and a total carbon abundance
of $X$(C)/$X$(H)$=2.34\times 10^{-4}$ \citep{SimonDiaz2011} we
translated the column densities of $^{13}$CO and CO into minimum
hydrogen column densities by assuming all carbon is in CO.

To estimate the velocity dispersion of the molecular gas we used the
mean of the line width of the two $^{13}$CO lines observed. For all
calculations, we only used the CO velocity component
(Table~\ref{draco-obs2}) that corresponds to the \CII\ velocity
component. For example, the Front 1 position only has a velocity
component for \CII\ at $-$25.6 km s$^{-1}$ (Tab.~\ref{draco-obs1}),
while the CO lines have two components (Tab.~\ref{draco-obs2}). We
thus used only the Front 1a velocity component at $-$25.1 km with a mean
velocity dispersion of 0.96~km s$^{-1}$ in the RADEX fit.

We display the results for RADEX in two different ways, either using
the line intensities or the ratios.  The line fit for the intensities
computed a three dimensional $\chi^2$ distribution in the space of the
parameters. The minimum gives the numerically best solution. This is
visualized in Figs. \ref{fig:radexfit1} and \ref{fig:radexfit2} for
the data from the F1 position in Draco. The figures show perpendicular
cuts through the 3-D cube of column density, density and temperature
at the location of the $\chi^2$ minimum. Each plot shows the
integrated intensity of one of the four CO and $^{13}$CO lines in
colors and the global $\chi^2$ distribution. The observed values are
marked by a red line. From the crossing of the intensity iso-contours
it is obvious that the density is best constrained by the combination
of the CO 2$\to$1 and the $^{13}$CO 1$\to$0 lines but that the fit
overpredicts the intensity of the CO 1$\to$0 line. The kinetic gas
temperature is well constrained by the combination of the CO lines and
the $^{13}$CO 1$\to$0 line. The column density is also well
constrained by the total intensities.

The best fit parameters for all four positions in Draco are given in
Table~\ref{tab:radex}. The corresponding plots for the other three
positions are very similar to Figs. \ref{fig:radexfit1} and
\ref{fig:radexfit2}, this is why we omit showing all positions. The
intensities just have to be shifted according to the values given in
Table~\ref{draco-obs2}. The minimum $\chi^2$ values fall between 4 and
13, which indicates a reasonable but imperfect fit by this model
with one degree of freedom, based on four observed line intensities
and three model parameters. The assumption of a homogeneous medium is
obviously an oversimplification but the parameters are quite well
constrained. It is interesting that the column density of gas traced
by the CO isotopologues falls significantly below the column density
measured through the dust emission (Table~\ref{draco-obs1}). This is
partially due to the fact that the RADEX fit considers only one
velocity component but in particular for the F1 and the N2 positions
the difference is so large that it indicates a large fraction of
atomic or CO-dark molecular material that is not traceable through
CO. This is in agreement with the estimate from \citet{Schneider2022}
that identified only 11\,\% of the gas mass is molecular.

\begin{table*}
\caption{Parameters of the minima of the RADEX fits to all four measured intensities in Draco.} \label{tab:radex}
\begin{center}
  \begin{tabular}{l|c|c|c|c|c}
    \hline
    Position & $T_\mathrm{kin}$ [K] & $n_\mathrm{H_2}$ [10$^3$ cm$^{-3}$] & $N_\mathrm{^{13}CO}$ [10$^{14}$ cm$^{-2}$] & $N_\mathrm{H,mol}$ [10$^{20}$ cm$^{-2}$] & $\chi^2$ \\
\hline
F1 & 9.03 &  4.92 &  1.30 &  0.39 & 3.94 \\
F2 & 13.7 &  1.65 &  4.92 &  1.47 & 4.41 \\
N1 & 15.9 &  1.97 &  9.94 &  2.97 & 12.1 \\
N2 & 11.1 &  3.56 &  2.04 &  0.61 & 12.9 \\
\hline
\end{tabular}
\end{center}
\end{table*}

Unfortunately, we cannot perform the same RADEX fit for the other
clouds where no $^{13}$CO data are available as a three-dimensional
problem cannot be constrained from two measured values only.

For a more intuitive visualization we also plot the RADEX output
brightness ratios of $^{13}$CO 1$\to$0/2$\to$1 and $^{12}$CO
1$\to$0/2$\to$1 as a function of temperature and H$_2$ density in
Fig.~\ref{fig:radex} for the Draco positions. Here, the column
densities were determined in the classical way from the $^{13}$CO
1$\to$0 line integrated intensity using an excitation temperature
T$_\mathrm{ex}$ that was determined from the $^{12}$CO 1$\to$0 line
peak brightness, assuming optically thick emission \citep{Mangum2015}.
The upper panels show the results for the Front 1 and Front 2
positions, the lower ones for the two nose positions. The model ratios
for Draco are compared to the observed ones (indicated as a solid
black and dashed line) and have the advantage that the ratios are
independent of beam filling to first order. We can start from the PDR
modeling densities to compare our ratios with the RADEX results. For
that, we use the upper limits from the PDR modeling obtained by eye
inspection of the observed line intensities and ratios. With the PDR
  model density of 500 cm$^{-3}$ for Front1 we obtain no solution at
  all. The density must be at least 3$\times$10$^3$ cm$^{-3}$ at a
  temperature of around 10 K. Note that from Table~\ref{tab:radex}, we
  derive a density of 4.9$\times$10$^3$ cm$^{-3}$ at a temperature of
  9 K. The Front 2 position can be reproduced by the upper limit PDR
  model with a density of 2$\times$10$^3$ cm$^{-3}$ (note that the
  density from the line ratio fitting is only 370 cm$^{-3}$) and leads
  to a consistent temperature of 15 K for both, $^{13}$CO and
  $^{12}$CO line ratios. This also fits with our values from
  Table~\ref{tab:radex} with the most likely density of
  1.65$\times$10$^3$ cm$^{-3}$ at a temperature of 14 K. The Nose 1
  position has the highest PDR model density of 3$\times$10$^3$
  cm$^{-3}$ (but only 636 cm$^{-3}$ from the line ratio fitting) that
  would give a low temperature of 10 K, again for both, $^{13}$CO and
  $^{12}$CO line ratios. From RADEX, we obtain slightly smaller
  densities of around 2$\times$10$^3$ cm$^{-3}$ at a temperature of 16
  K.  For the Nose 2 position, it is more difficult to obtain a fully
  coherent output from the analysis: a PDR model density of 10$^3$
  cm$^{-3}$ requires a temperature of 25 K for $^{13}$CO. But the
  RADEX $^{12}$CO line ratio is much higher (around 3) at this
  temperature (and also for lower temperatures) than the observed one
  of 1.57. Only high densities of around at least 3$\times$10$^3$
  cm$^{-3}$ make the observed $^{12}$CO line ratio fitting with low
  temperatures of around 10 K.

Note that PDR model densities are by no means the 'exact' values. They
only give an estimate of the density in case of a very low FUV field
and we observe a discrepancy between the densities and FUV field
determined from line intensities and ratios and from line ratios
alone. For the optical depths, we derived values below 1 for temperatures
above 5 K for the $^{13}$CO lines.  For the $^{12}$CO lines the values
range between the optical thin case with $\tau\lesssim0.3$ for Front 1
and the optical thick case with mostly $\tau>1$ for the Front 2
position over all temperatures.  In both nose positions the opacity
lies above 1 throughout the whole parameter range of temperatures.
 
Summarising, we have a parameter space of possible solutions for all
positions from the RADEX model alone.  The densities are between 1650
and 4900 cm$^{-3}$ at a temperature of 9 - 16 K
(Table~\ref{tab:radex}). Overall, these values align mostly with the
ones of \citet{Deschenes2017} who found molecular clumps in Draco at a
temperature of 10 K and an average density of $\sim$10$^3$ cm$^{-3}$
(note that the density of individual molecular clumps is sometimes
higher).  The dust temperatures in Draco are low (around 13 K, see
Table~\ref{draco-obs1}), so that in case the dust and gas is well
mixed, the low gas temperatures are consistent.

Figure~\ref{fig:radex2} displays the RADEX results for the $^{12}$CO
1$\to$0/2$\to$1 ratio in Spider and Polaris since only these lines are
available. We omit Musca because we only have the $^{12}$CO 2$\to$1
line brightness. It is noteworthy that for Spider, the $^{12}$CO
1$\to$0 line is optically thin with values between $\sim$0.2 and
$\sim$0.5 for a temperature range of 10 - 20 K and the $^{12}$CO
2$\to$1 line is optically thick with values above $1$ for the whole
parameter space, whereas for Polaris, the optical depth is mostly
above 1 for both $^{12}$CO lines. This discrepancy renders the
interpretation of the results somewhat less reliable.  The PDR
modeling for Spider did not yield reasonable density estimates, even
for very low FUV fields (outside the model's range). We anticipate
that the densities and temperatures are low in this relatively diffuse
region with a substantial atomic contribution. Assuming the curve with
the lowest density for the CO clump in Spider that would fit with our
observations ($n$=1.5$\times$10$^3$ cm$^{-3}$) yields a temperature of
20 K. Interestingly, the dust temperature in Spider is also
approximately this value ($\sim$18 K), slightly higher than in all
other sources.  For Polaris, the PDR modeling provides a density of
$\sim$2$\times$10$^3$ cm$^{-3}$ and even 3838 cm$^{-3}$ from the line
ratio fit for a very low FUV field. The observed $^{12}$CO ratio then
fits well with a temperature of around 25 K. However, it is important
to note that the $^{12}$CO lines in Polaris are optically
thick. Hence, we propose that the resulting possible temperature and
density range is approximately 10 - 25 K at densities of around
2-3$\times$10$^3$ cm$^{-3}$.

\subsection{Shock modeling of line emission in Draco} \label{subsec:shock}

For modeling a potential shock-origin of the \CII\ emission, we
utilized the Paris-Durham shock
code.\footnote{\href{https://ism.obspm.fr/shock.html}{https://ism.obspm.fr/shock.html}}
Our objective was not to conduct an exhaustive modeling but rather to
assess whether the observed quantities can be preliminary explained by
shocks. The pre-calculated model
grids\footnote{\href{https://app.ism.obspm.fr/ismdb}{https://app.ism.obspm.fr/ismdb}}
were employed for this purpose. The shock models are executed using
shock code version 1.1.0, revision 115, and encompass C-type shocks at
moderate velocities ($<$100 km s$^{-1}$), or J-type shocks at low
velocities ($<$ 30 km s$^{-1}$), both propagating in weakly magnetized
environments.  The shocks are implemented as propagating through a
layer of gas and dust, manifested as stationary, plane-parallel,
multifluid shock waves, in environments with or without external UV
radiation. The parameters that are parameterized include the shock
velocity (2 - 90 km s$^{-1}$), initial density of the preshock medium
(100 to 10$^8$ cm$^{-3}$), intensity of the UV radiation field (0 to
1700 G$_\circ$), cosmic ray ionization rate (10$^{-17}$ to 10$^{-15}$
s$^{-1}/{\rm H}_2$), and the magnetic field strength, expressed in
terms of $\beta$ and ranging from 0.1 to 10. $\beta$ defines the
initial transverse magnetic field strength in $\mu$G and is equal to
$\beta \times \sqrt{n(H) [\rm {cm}^{-3}]}$. The fractional abundance
of PAHs is fixed at 10$^{-8}$, and the chemistry incorporates 140
species through a chemical network involving about 3000
reactions. Surface reactions, adsorption, and desorption processes are
not considered except for H$_2$ formation. For further elaboration,
refer to \citet{Gusdorf2008}, \citet{Lesaffre2013},
\citet{Goddard2019}, \citet{Lehmann2020,Lehmann2022} and the
references therein.

The code generates output containing the thermodynamic and chemical
structure of the shock, including temperature, fluid velocities,
abundances, column density profiles, and line intensities of H$_2$ as
well as several ionized or neutral atoms. We utilized this output to
analyze \CII\ line intensities and the total hydrogen column density,
N(H), in comparison with Draco observations. Specifically, we applied 
the observed \CII\ line intensities and {\sl Herschel} hydrogen column
density values from Table~\ref{draco-obs1} to positions Draco Front 1
and 2, and Nose 1 and 2.  The \CII\ intensities at the Front 1, Front
2, Nose 1, and Nose 2 positions are 1.55, 2.60, 2.67, and 2.53,
respectively, measured in units of 10$^{-6}$ erg s$^{-1}$ sr$^{-1}$
cm$^{2}$. For our model, we fixed the radiation field at 1.7 G$_\circ$
(note that available values are 0.17, 1.7, 17, .. G$_\circ$) and the
preshock density at $n$ = 100 cm$^{-3}$ which is the lowest possible
density to select.
We tried also higher densities but could not find a match between the
shock model results and the observations. We thus used a preshock gas
of 100 cm$^{-3}$, though the preshock density can indeed be
lower. However, densities lower than 1000 cm $^{-3}$ are a reasonable
assumption since the density of the densest molecular clumps currently
found in Draco is between 1000-4900 cm$^{-3}$ (Sects. \ref{subsec:pdr}
and \ref{subsec:radex} and \citet{Deschenes2017}).  We then explored 
variations in other parameters, such as the cosmic ray rate, $\beta$,
and shock speed, within reasonable ranges. Different cosmic ray rates
(10$^{-15}$, 10$^{-16}$, and 10$^{-17}$ s$^{-1}$/H$_2$) have limited
impact on the \CII\ intensity and H column density. We choosed 
10$^{-16}$ s$^{-1}$/H$_2$ as it is suggested as the average value for
the diffuse ISM in the Milky Way \citep{Dalgarno2006,Indriolo2012}.
The magnetic field strength, characterized by low values of $\beta$
(0.1, 0.3), results in diminished \CII\ intensities and H column
densities. Higher values of $\beta$ (3, 10) produce larger hydrogen
column densities, but these values do not align well with the
\CII\ emission.  In summary, our analysis provides insights into the
impact of various parameters on the \CII\ emission and hydrogen column
density, revealing the influence of factors such as the radiation
field, density, cosmic ray rate, and magnetic field strength in the
Draco region.

The best match between model values and observations is found for the
Front 1 position, with a shock velocity of 20 km s$^{-1}$, preshock
density $n = 100$ cm$^{-3}$, and $\beta = 1$. The model predicts a
\CII\ intensity of $1.63 \times 10^{-6}$ erg s$^{-1}$ sr$^{-1}$
cm$^{2}$ at a total hydrogen column density of $3.73 \times 10^{20}$
cm$^{-2}$. It is important to note that the column density is
calculated for a single layer, representing a lower limit.  When
comparing all Draco positions, the \CII\ intensity is roughly
consistent across all of them. Running models with different preshock
densities would probably better match the observed intensities, but
this is for the moment out of the scope of this paper.  The hydrogen
column density is approximately four times higher than the model's
predicted values for the nose positions. However, this discrepancy
could be attributed to the limitations of the single layer model.  It
is evident that higher shock speeds are not feasible. Beyond a
threshold of v = 25 km s$^{-1}$, the \CII\ intensity experiences a
significant drop. J-type shocks prove ineffective, while only C-type
shocks with velocities $\leq 20$ km s$^{-1}$ yield reasonable
results. Nonetheless, reducing the velocities further leads to a
decline in the \CII\ intensity.


\begin{table}
\caption{Summary of physical parameters.} \label{tab:all}
\begin{center}
  \begin{tabular}{l|c|c|c|c|c}
    \hline\hline
{\tiny Source}    & {\tiny I$_{CII}$} & {\tiny F$_{160}$} & {\tiny FUV }  & {\tiny N(H)}  & {\tiny  n}\\
          & {\tiny [K km s$^{-1}$]} & {\tiny [MJy/sr]} & {\tiny [G$_\circ$]} & 
          {\tiny [10$^{20}$ cm$^{-2}$]} &  {\tiny [10$^3$ cm$^{-3}$]} \\
          \hline
{\tiny Draco$^a$} & {\tiny 0.3325}   & {\tiny 44}  & {\tiny 3.6}  &  {\tiny 10.0}  & {\tiny 3.0}\\ 
{\tiny Spider}    & {\tiny $<$0.06}  & {\tiny 38}  & {\tiny 2.9}  &  {\tiny  6.3}  & {\tiny 2.0$^b$}\\ 
{\tiny Polaris}   & {\tiny $<$0.03}  & {\tiny 18}  & {\tiny 1.5}  &  {\tiny 32.2}  & {\tiny 20-50$^c$}\\ 
{\tiny Musca}     & {\tiny $<$0.03}  & {\tiny 61}  & {\tiny 5.8}  &  {\tiny 34.3}  & {\tiny 7.0$^d$}\\ 
\end{tabular}
\end{center}
\vskip0.1cm \tablefoot{Comparison between the \CII\ intensity (column 2), the 160 $\mu$m flux (column 3),
  the FUV field from the 160 $\mu$m flux (column 4), the total hydrogen column density (column 5),
  and the volume density (column 6) for the observed sources.}\\ 
$^a$Average value from the 4 positions observed in Draco.\\
$^b$From \citet{Barriault2010b}. \\
$^c$From \citet{Grossmann1992,Heithausen1995,Derek2010}. \\
$^d$From \citet{Bonne2020a}. \\

\end{table}

\section{Discussion} \label{discuss}

The main result of this study is that the \CII\ 158 $\mu$m line was
detected at several positions in the Draco cloud, but not in any of
the other quiescent clouds (Spider, Polaris, Musca), even though these
regions have similar column and volume densities, 160 $\mu$m fluxes,
and FUV fields (see Table~\ref{tab:all}). We note that the volume
densities are similar for all sources (a few times 10$^3$ cm$^{-3}$),
only Polaris has a higher density of 2-5$\times$10$^4$ cm$^{-3}$ and
should thus emit even stronger in \CII\ than Draco because subthermal,
optically thin \CII\ intensities scale with $n^2$
\citep{Goldsmith2012}.  As we explained in Sect.~\ref{subsec:pdr},
though the density of the gas is higher in the interior of a molecular
clump than at the surface, we can consider similar densities for the
CO and \CII\ emitting gas because the \CII\ line shows a very weak
dependency on the density. \\
It is important to emphasize that the accurate derivation of the FUV
field is a critical point.  As discussed in Sect.~\ref{subsec:fuv},
different methods yield variations in the estimated FUV field. A
census of Galactic OB-stars and a continuous approximation
\citep{Parravano2003} using the distribution and birthrates of OB
stars in the Milky Way from \citet{McKee1997} and considering the
extinction toward our sources, yields values between 1.2 and 1.6
G$_{\circ}$.  On the other hand, the calculation based on the 160
$\mu$m flux and a dust model \citep{Xia2022} results in larger values
(at least a few G$_\circ$) when assuming that the dust is only heated
through the FUV radiation.

We checked if the FUV field determined from the 160 $\mu$m emission
may be underestimated because the wavelength range below 160 $\mu$m is
not taken into account.  This is not the case. There is nearly no or
only weak 70 $\mu$m emission visible in the {\sl Herschel} maps of all
sources and only very weak 12 $\mu$m emission
\citep{Meisner2014}. There is also no overestimation of the 160 $\mu$m
flux due to the contribution of cold thermal dust emission from the
molecular cloud. It is thus difficult to understand why the 160 $\mu$m
fluxes are so high in all sources if it is not an external FUV field
that heats the dust.

When looking for alternative mechanisms for the heating of the dust we
can quickly exclude cosmic rays. At a typical cosmic ray ionization
rate of $2\times 10^{-16}$~s$^{-1}$ and heating efficiencies below
10~eV per ionization event \citep{Glassgold2012} the cosmic ray
heating rate falls more than three orders of magnitude below the UV
heating rate at one Habing field so that even an enhanced cosmic ray
rate is probably insufficient to explain the observed dust
heating. Instead shock heating from cloud collisions can in principle
inject enough energies.  For example, with a rather low hydrogen gas
density of 500~cm$^{-3}$ (Front 1 position in Draco), a kinetic
temperature of 100~K, and relative velocities of 20~km~s$^{-1}$, like
the difference between the IVC and LVC in Draco, the
thermalization time, $\tau_{\rm therm}$ is at maximum 30~years when
assuming perfectly elastic collisions \citep{Sauder1967}. Inelastic
contributions may shorten this. The compressive heating from the
shock, $\Gamma=n\times m_{\rm H}/2 \times v_{\rm shock}^2 / \tau_{\rm
  therm}$ is then more than four orders of magnitude above the UV
heating rate at one Habing field. Higher densities would even increase
$\Gamma$. Although the details of the shock physics may modify this
rate relative to the idealized value, the order of magnitude estimate
shows that shock heating at our conditions could easily feed enough
energy into the dust to explain the enhanced 160 $\mu$m emission.
Note that referring to \citet{Lehmann2020,Lehmann2022} up to 10-30 \%
of shock energy can be irradiated away in LyAlpha (which then can be
absorbed and reemitted by the dust, potentially resulting in an
increase of the 160 $\mu$m emission (see also
\citet{Bonne2022}). This, however, requires shock velocities around 30
km s$^{-1}$. Nevertheless, lower velocity shocks may still have a
measurable impact.

The 'cloud collision' in Draco that can give an explanation to the
increased 160 $\mu$m emission and the \CII\ emission can be seen as an
interaction of the atomic/molecular IVC with mostly atomic gas
(the LVC) while the Draco cloud moves through the
ISM. Notably, local gas within the velocity range of $-$10 to 30 km
s$^{-1}$ contributes to these interactions, while the high-velocity
gas ($-$200 to $-$100 km s$^{-1}$) remains excluded from this process
(refer to Fig.~\ref{app-draco-hi} and \ref{app-pv}).
Figure~\ref{fig:map-draco} presents a clear representation of the
complex velocity structure of \HI\ in Draco, with a minimum of two
components observed between approximately $-$16 and $-$30 km
s$^{-1}$. At the forefront, the two positions in Draco distinctly
exhibit two \HI\ components, out of which only one corresponds to the
\CII\ line, situated around $-$26 and $-$24 km s$^{-1}$ for the Front
1 and Front 2 positions, respectively. The Nose 1 and Nose 2
\HI\ spectra are more complex and do not show clearly two
components. This is not surprising because the gas is here mostly
molecular and denser.  Conversely to Draco, the other sources
primarily exhibit inconspicuous dynamics. Musca stands out as having
more pronounced dynamics due to filament formation \citep{Bonne2020a,
  Bonne2020b}. The dissipation of turbulence through the emergence of
dense structures results in low-velocity shocks that elevate both gas
and dust temperatures and may thus also explain the rather high 160
$\mu$m flux. However, it is noteworthy that no detection of
\CII\ occurred in Musca.  Spider and Polaris have lower 160 $\mu$m
fluxes compared to Draco and Musca, but the emission is mostly at
local velocities and does not seem to involve cloud collisions.

The PDR modeling indicates that the \CII\ and CO lines can only be
accounted for by a significantly small UV field, much smaller than 1
G$_\circ$. Among the positions studied, only Draco Nose 1, Nose 2, and
Front 2 positions exhibit some compatibility with a model where CO
emission originates from the inside of high-density (1-3$\times$10$^3$
cm$^{-3}$), cold (T$\sim$10-20 K) molecular clumps.  The CO emission
is then a result of thermal excitation within the dense gas clumps,
which likely possess a limited filling factor. \CII\ may then arise
from the PDR surfaces at those densities but it can also stem from a
mostly atomic interclump component.  As for the Draco Front 1
position, as well as Polaris and Musca, their \CII\ and CO line
emissions (and the corresponding upper limits) are reconcilable with a
PDR model featuring very low FUV fields, less than 0.2 G$_\circ$,
coupled with densities around 10$^3$ cm$^{-3}$. Only the Front 1
positions falls out of this with its density of only 500 cm$^{-3}$.
The Spider position cannot be modeled with very low FUV intensities.
RADEX modeling of the observed CO line ratios in all sources
determines mostly higher densities than estimated from the PDR
modeling (except of the low densities for the Draco Front 1 position),
but we cannot make a statement about \CII.  The puzzling result is
still why the \CII\ line was observed in Draco and not in all other
quiescent sources. The only differences between Draco on one hand, and
Spider, Polaris, and Musca on the other hand, is the higher dynamics
of Draco because of a possible cloud interaction.  An approach of
shock modeling applied to the \CII\ emission in Draco reveals that the
observed intensities are compatible with conditions akin to a preshock
density of 100 cm$^{3}$ and a shock velocity of 20 km s$^{-1}$.
The shock velocity is not very high which may explain why the
\CII\ line widths (between 1.3 km s$^{-1}$ for Front 1 and 4 km
s$^{-1}$ for Front 2) are not very broad and not significantly larger
than the ones of CO. \\
In consideration of these findings, it is plausible that Draco
contains dense, cold, molecular clumps that are enveloped by a diffuse
atomic phase ($\simeq$10-100 cm$^{-3}$, T$>$50 K), serving as the
source of the \CII\ emission. This emission, in turn, acts as a
cooling line to dissipate kinetic energy stemming from interactions
between \HI\ clouds or as the IVC descends onto the galactic disk. In
such a scenario, the primary factor driving this process is the ram
pressure exerted on the cloud by the escalating halo density, as
stated by \citet{Desert1990}.

\section{Summary} \label{sec:summary}

The \CII\ 158 $\mu$m line was observed with SOFIA at selected
positions in the quiescent clouds Draco, Spider, Polaris, and
Musca. Emission on a level of $\sim$0.2 to 0.4 K km s$^{-1}$
(S/N$\sim$4) of the \CII\ line was only detected in four positions in
the Draco cloud, and CO lines ($^{12}$CO and $^{13}$CO 1$\to$0,
2$\to$1) were observed with the IRAM 30m and APEX and mostly detected
in all sources. The flux at 160 $\mu$m is rather high in all sources
(18 MJy/sr for Polaris; 44 MJy/sr for Draco; up to 61 MJy/sr for
Musca), while the FUV field is very low. Converting the 160 $\mu$m
flux gave values of 1.5 to 6 G$_\circ$, while the FUV field from a
stellar census is around 1.3 - 1.6 G$_\circ$. A theoretical
determination considering the distribution of OB stars in the galaxy
and extinction yielded a value of 1.2 - 1.6 G${_\circ}$. The PDR
modeling (\CII, \CI, and CO) and RADEX modeling (CO) can partly
explain the observed emission arising from clumps with a density of a
few 10$^3$ cm$^{-3}$ at a temperature of 10-20 K in Draco. The PDR
model, however, requires a very low UV field, much lower than 1
G$_\circ$, which is not provided by the different methods of our FUV
field determinations.  For Draco, heating by collisions of \HI\ clouds
could explain the high level of the 160 $\mu$m flux and the
\CII\ emission, which is reproduced by a shock model with a preshock
density of 100 cm$^{-3}$ and a C-shock with a velocity of 20 km
s$^{-1}$. We propose that the shock arises from the interaction of the
\HI\ clouds associated with Draco.
 
\begin{acknowledgements}

This study was based on observations made with the NASA/DLR
Stratospheric Observatory for Infrared Astronomy (SOFIA). SOFIA is
jointly operated by the Universities Space Research Association
Inc. (USRA), under NASA contract NNA17BF53C, and the Deutsches SOFIA
Institut (DSI), under DLR contract 50 OK 0901 to the University of
Stuttgart. upGREAT is a development by the MPIfR and the
KOSMA/University Cologne, in cooperation with the DLR Institut f\"ur
Optische Sensorsysteme.\\
N.S. acknowledges support from the FEEDBACK-plus project that is
supported by the BMWI via DLR, Projekt Number 50OR2217.  \\
S.K. acknowledges support by the BMWI via DLR, project number 50OR2311. \\
This work is supported by the Collaborative Research Center 1601 (SFB
1601 sub-project A6 and B2) funded by the Deutsche
Forschungsgemeinschaft (DFG, German Research Foundation) – 500700252.

\end{acknowledgements}

\bibliography{draco_CII.bib}

\begin{appendix} 

\section{Velocity integrated HI maps of Draco and Spider} \label{appendix:maps}

\begin{figure*}
\centering
\includegraphics[width=5cm,angle=0]{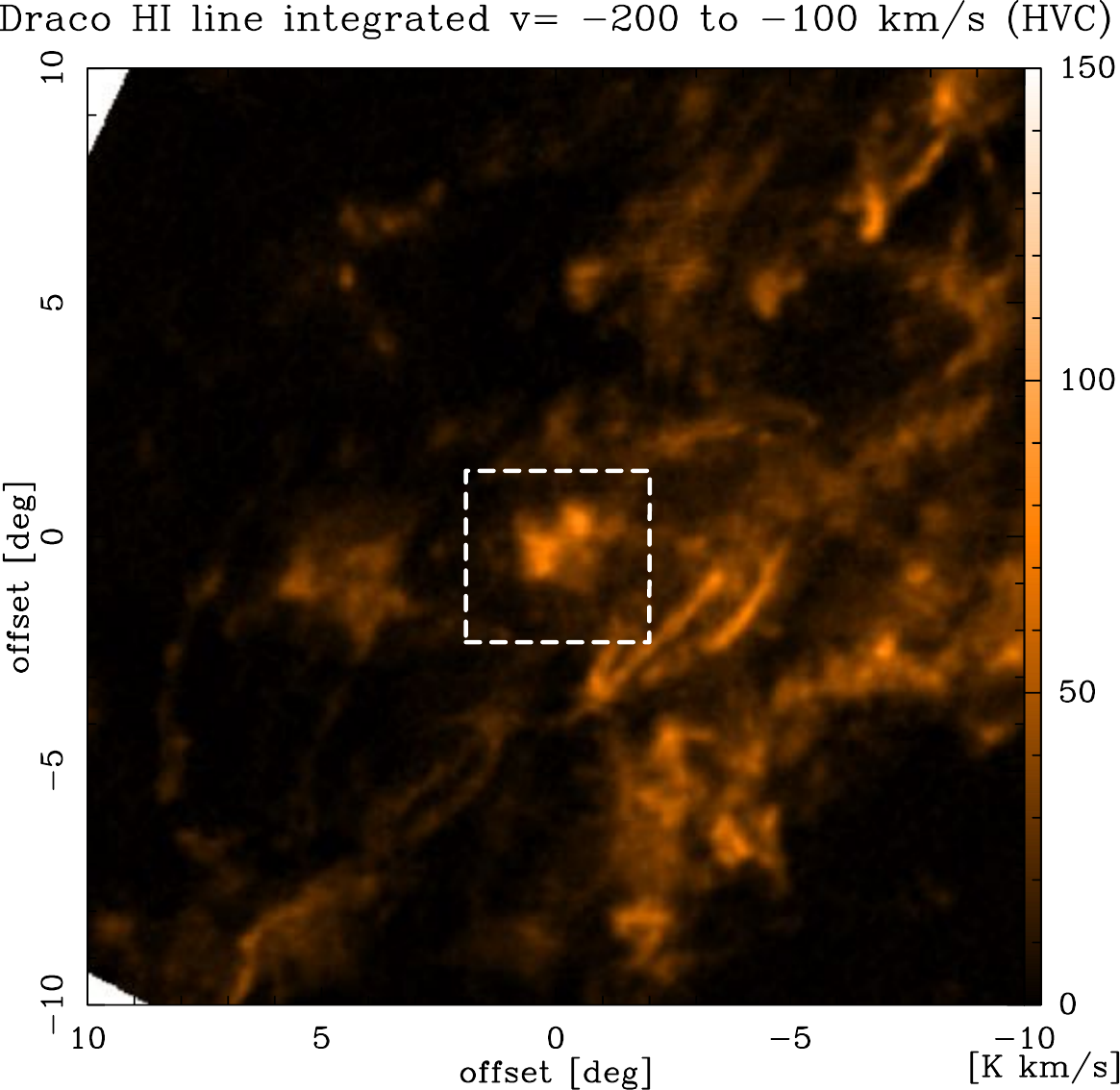}
\hspace{0.5cm}
\includegraphics[width=5cm,angle=0]{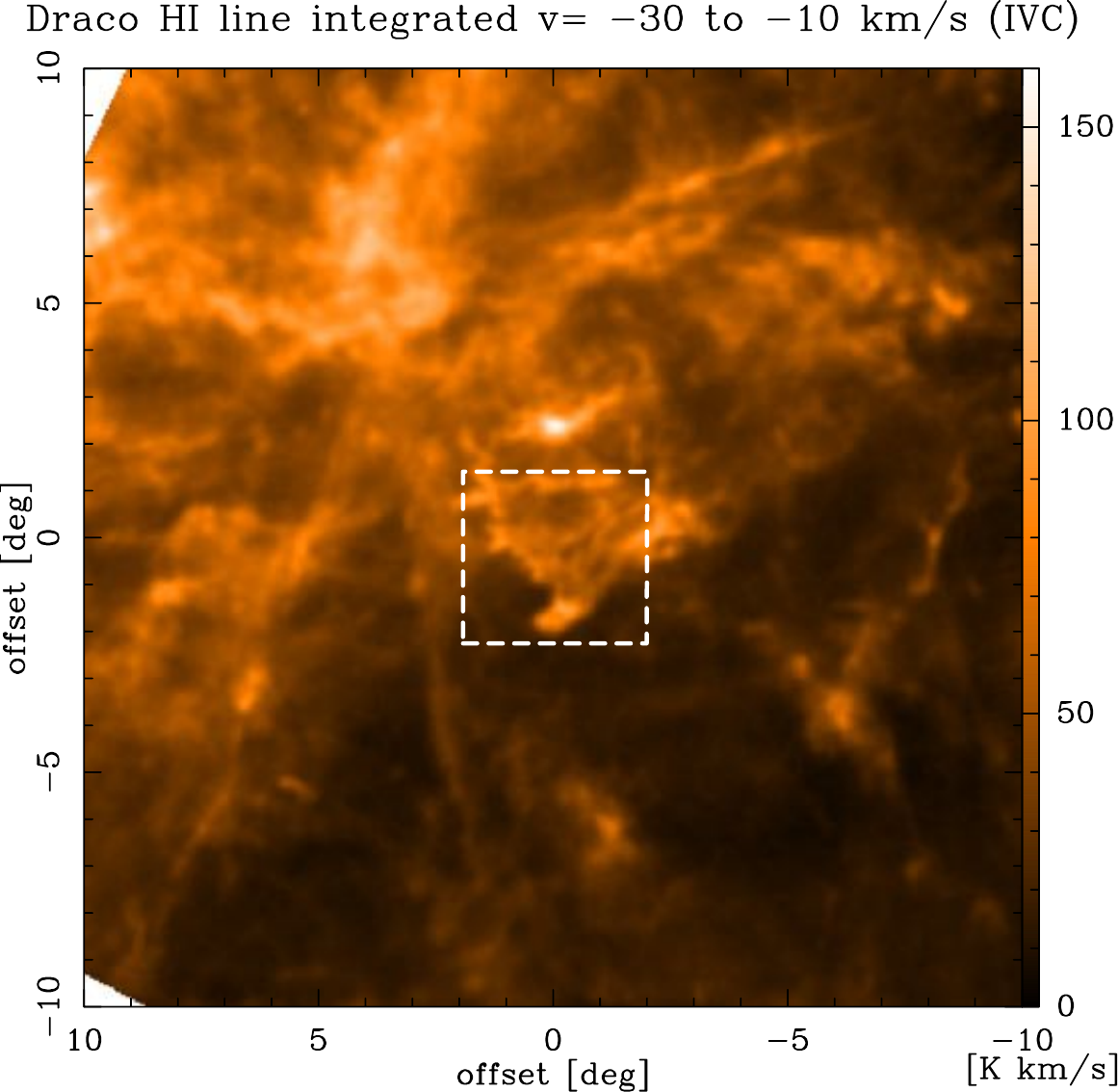}
\hspace{0.5cm}
\includegraphics[width=5cm,angle=0]{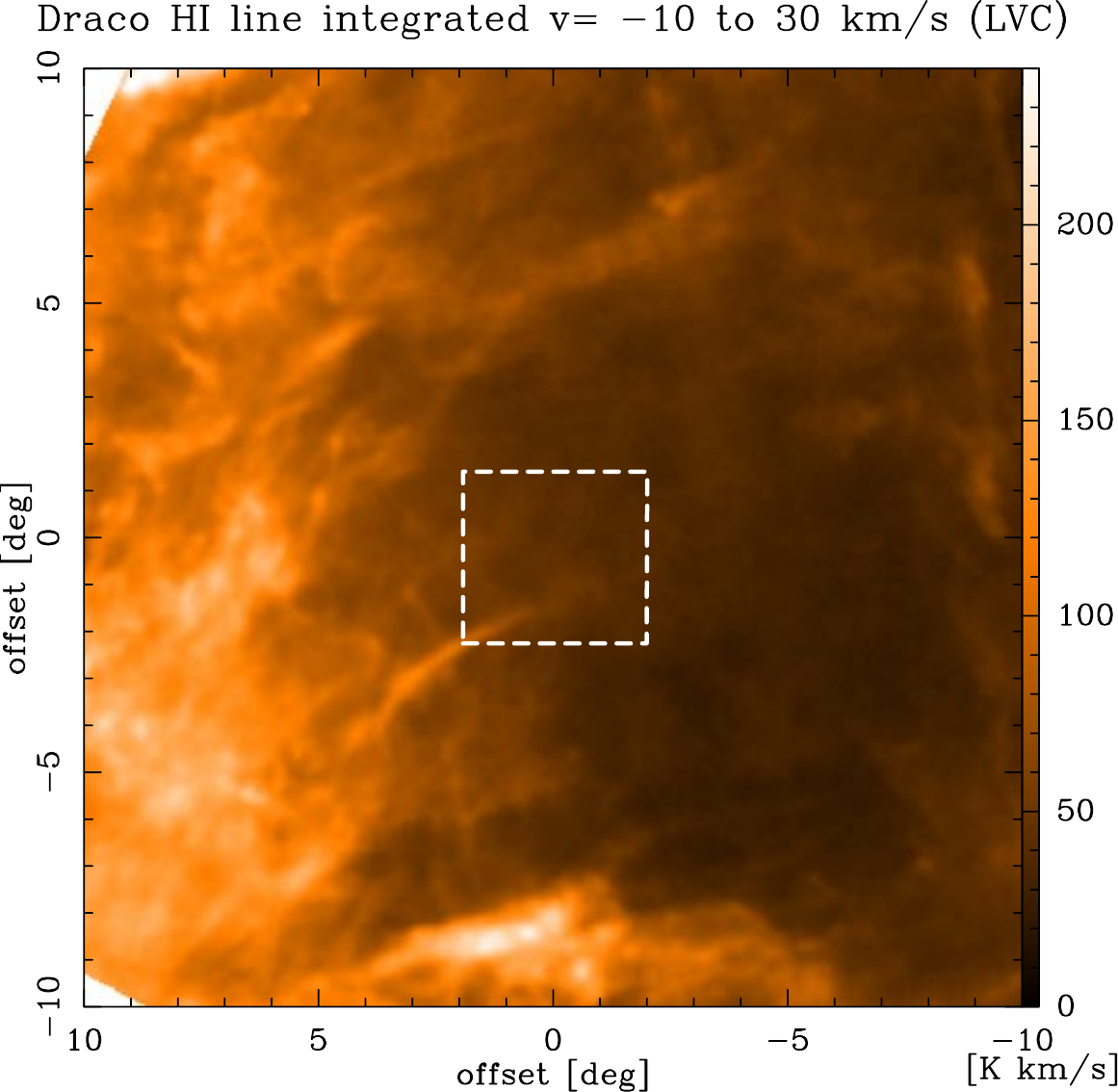}
\caption{Velocity integrated \HI\ maps of the Draco region from the
  EBHIS survey.  The panels show the \HI\ line integrated emission
  over all velocity ranges, covering the HVC (left, v=-200 to -100 km
  s$^{-1}$), the IVC (middle, v=-30 to -10 km s$^{-1}$), and the LVC
  (right, -10 to 30 km s$^{-1}$. For better comparison, the scale in K
  kms$^{-1}$ was kept constant.}
\label{app-draco-hi}
\end{figure*}

\begin{figure*}
\centering
\includegraphics[width=7cm,angle=0]{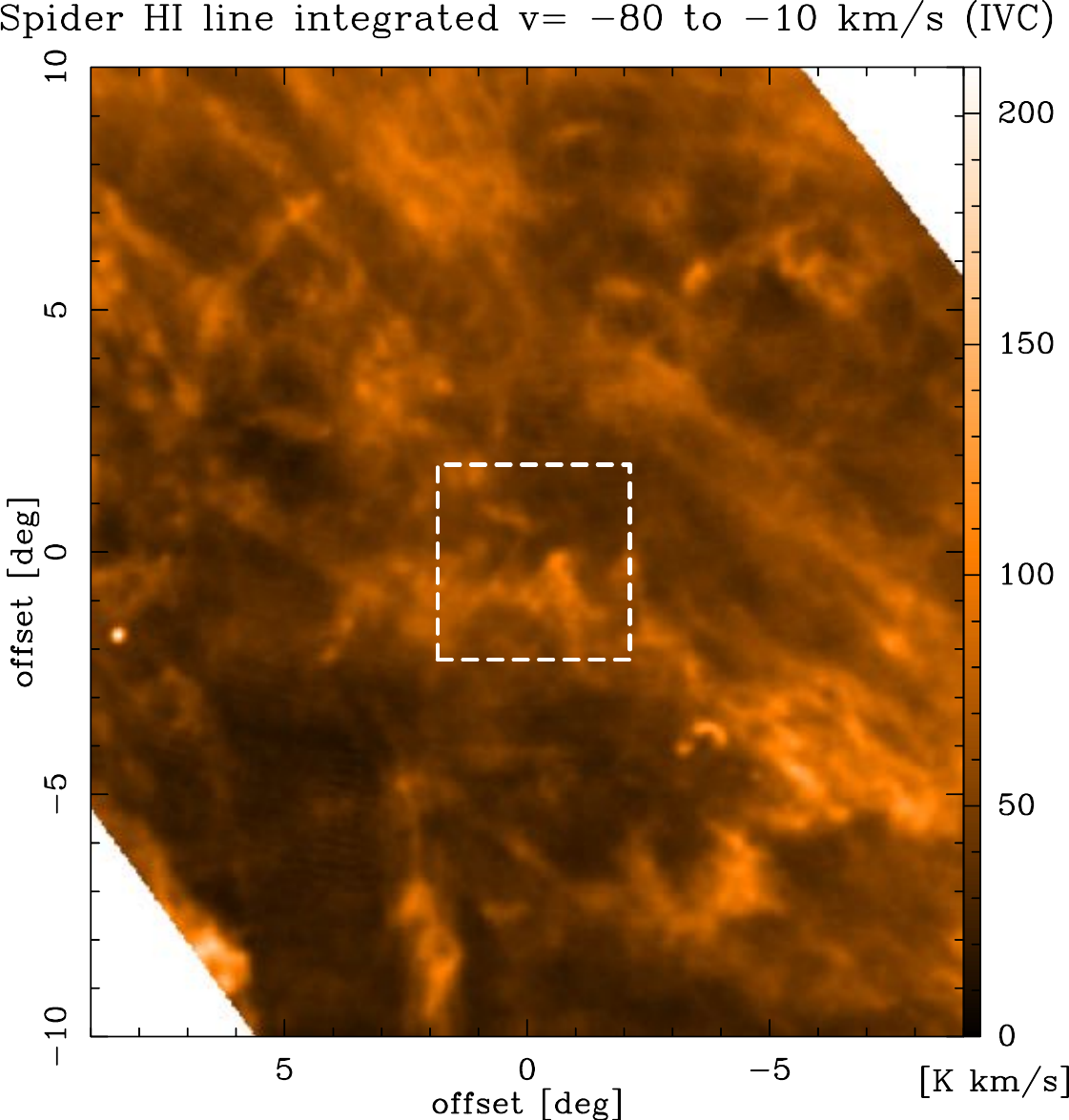}
\hspace{0.5cm}
\includegraphics[width=7cm,angle=0]{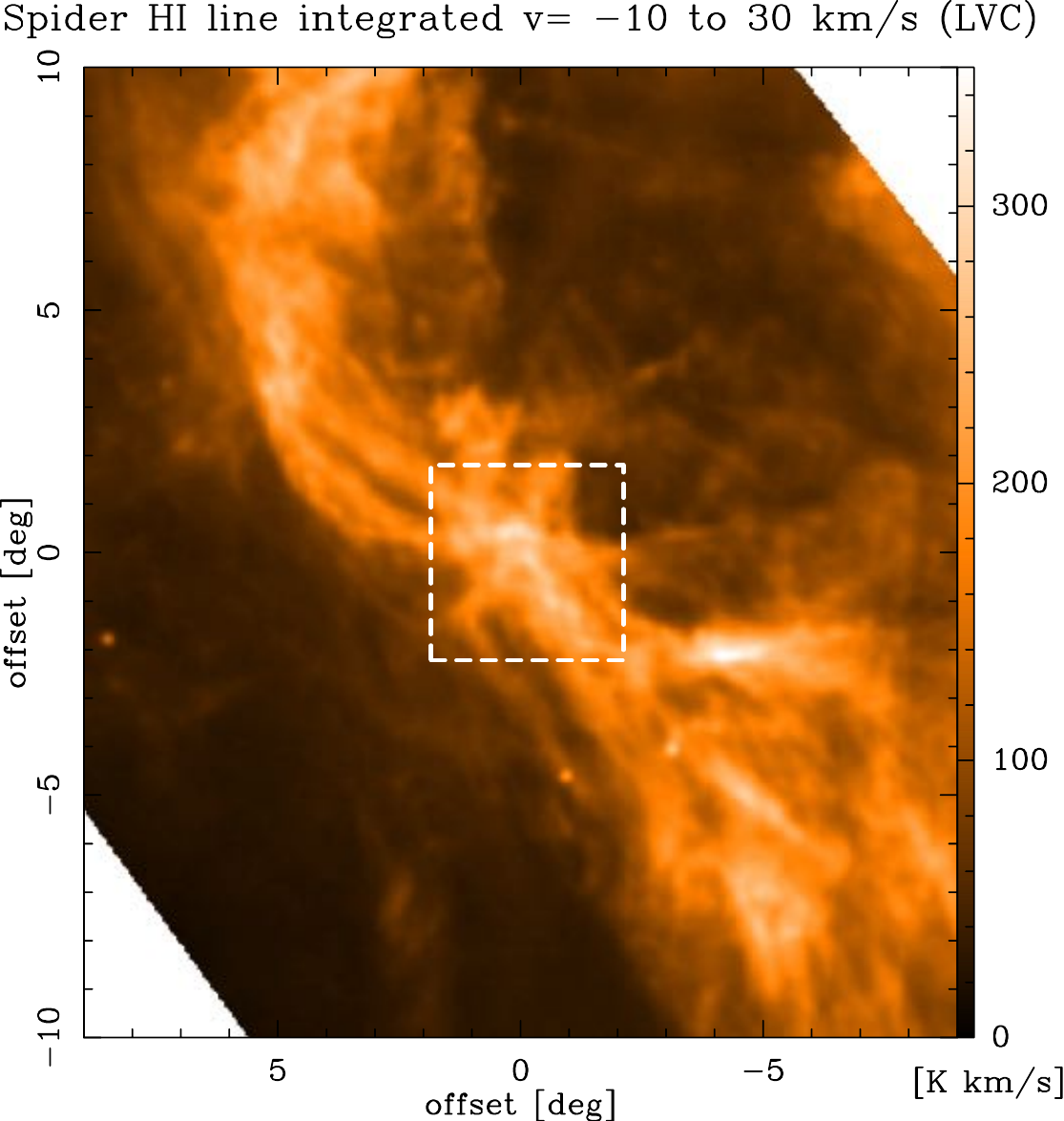}
\caption{Velocity integrated \HI\ maps of the Spider region from the
  EBHIS survey.  The panels show the \HI\ line integrated emission
  over all velocity ranges, covering the IVC (left, v=$-$80 to $-$10
  km s$^{-1}$), and the LVC (right, $-$10 to 30 km s$^{-1}$).}
\label{app-spider-hi}
\end{figure*}

Figures \ref{app-draco-hi} and \ref{app-spider-hi} show the column
density contributions of the different velocity components to the
total column density in Draco and Spider derived from the
\HI\ observations.  They should be compared to the dust column
densities in Figs. 4 and 5. Comparing the morphology we can assign the
individual velocity components to the analyzed dust maps. For Draco we
find a clear correspondence of the dust column with the IVC ($-$30 to
$-$10 km s$^{-1}$, for Spider to the local velocity component ($-$10
to 30 km s$^{-1}$).

\section{Position-velocity cut for Draco} \label{appendix:pv}

\begin{figure} 
\centering
\includegraphics[width=8.5cm,angle=0]{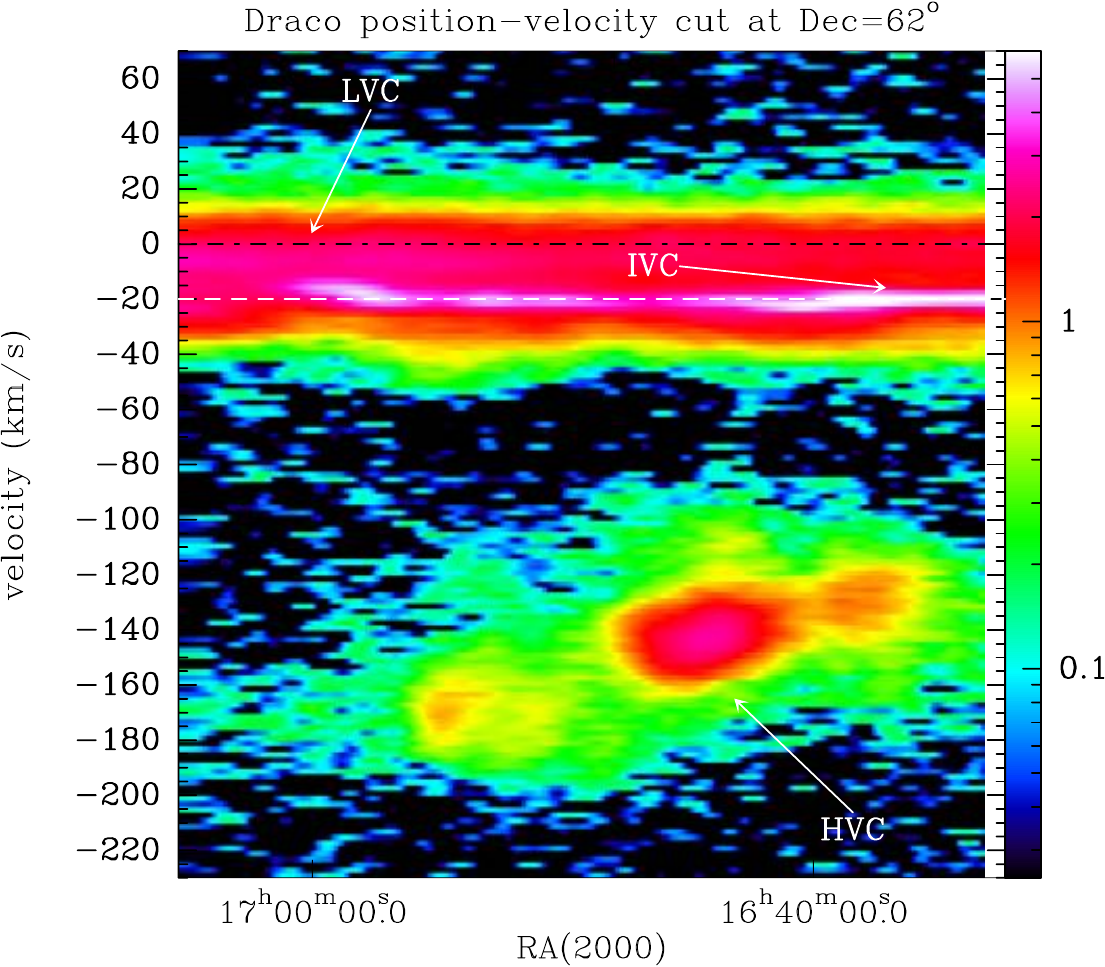}
\caption{Postion-velocity cut in Draco at a constant declination of 62
  deg in the \HI\ line emission.  The full velocity range comprising
  the LVC, IVC, and HVC is shown and indicated in the plot.  There is
  no clear velocity bridge between the HVC and LVC.}
  \label{app-pv}
\end{figure}

Figure~\ref{app-pv} displays a position-velocity cut in Draco at
constant declination of 62 deg in the \HI\ line emission on a
logarithmic scale. The LVC, IVC, and HVC, respectively, are
indicated in the plot.  While LVC and IVC are closely connected in
velocity space, there is no velocity bridge between the HVC and
LVC/IVC.
  
\section{PDR toolbox results for \CII\ emission} \label{appendix:pdr}

\begin{figure} 
\centering
\includegraphics[width=8.5cm,angle=0]{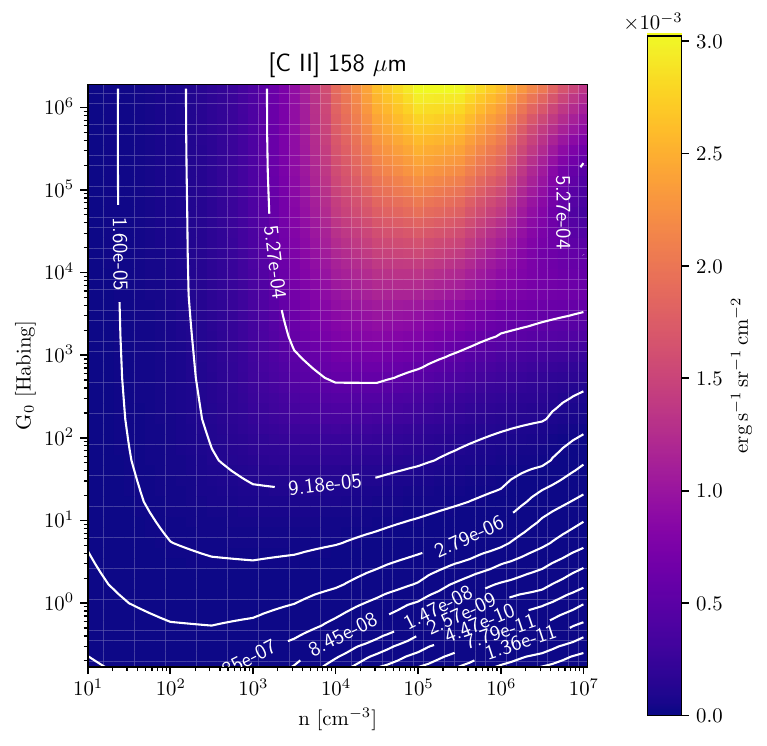}
\includegraphics[width=8.5cm,angle=0]{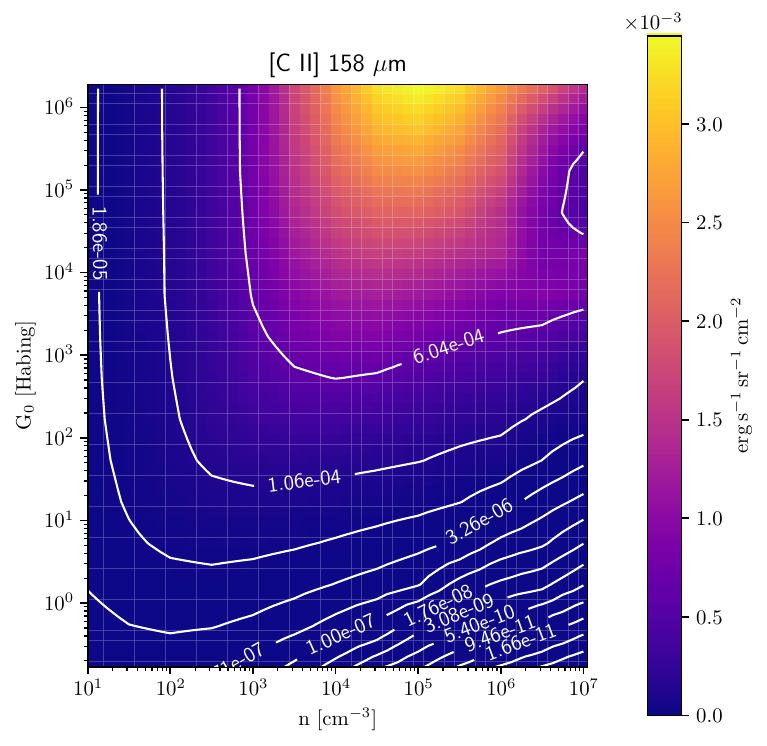}
\caption{PDR toolbox predictions for the density and FUV field from
  \CII\ 158 $\mu$m emission in erg cm$^{-2}$ s$^{-1}$ sr$^{-1}$ using
  the KOSMA-$\tau$ model.  The model using a mass of M = 0.1 M$_\odot$
  (Draco, Spider, Musca) is shown in the top panel, the one with M = 1
  M$_\odot$ (Polaris) in the lower panel.}
  \label{app-pdr}
\end{figure}

\begin{figure} 
\centering
\includegraphics[width=8.5cm,angle=0]{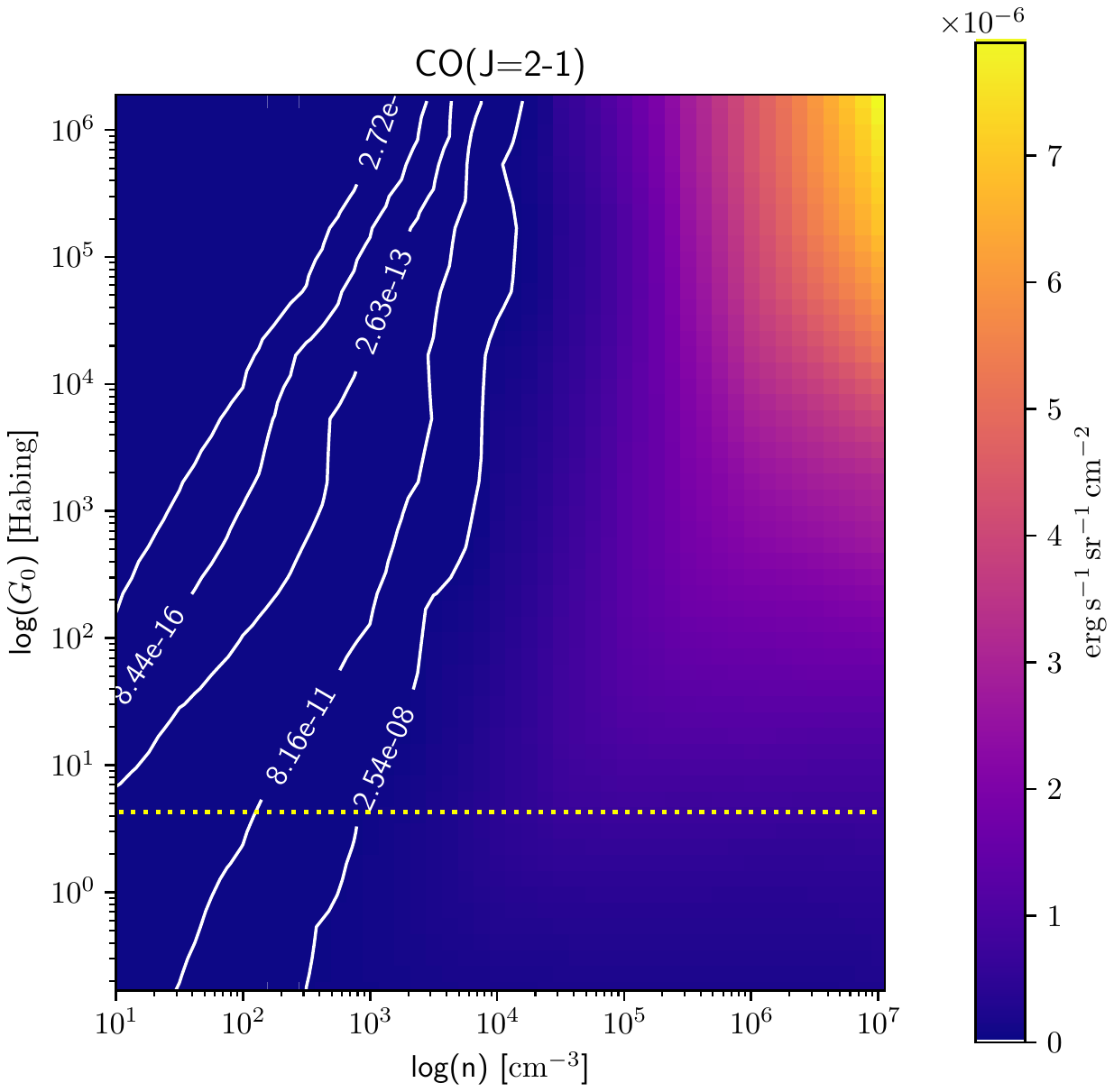}
\caption{PDR toolbox predictions for density and FUV field from
  CO(2$\to$1) emission in erg cm$^{-2}$ s$^{-1}$ sr$^{-1}$ using the
  KOSMA-$\tau$ model with the model using a mass of M = 0.1
  M$_\odot$. }
  \label{app-co}
\end{figure}

Figure~\ref{app-pdr} displays the calculated parameter range of
density and UV field from the KOSMA-$\tau$ model for \CII\ 158 $\mu$m
emission for masses of M = 0.1 M$_\odot$ and M = 1 M$_\odot$ in units
of erg cm$^{-2}$ s$^{-1}$ sr$^{-1}$. Figure~\ref{app-co} displays as
one example (M = 0.1 M$_\odot$) such a plot for $^{12}$CO 2$\to$1.
Since the \CII\ observations in Polaris, Musca, and Spider only
represent the noise level, these figures illustrate in which direction
the density and UV field goes in case of higher/lower \CII\ limits.

\end{appendix}


\end{document}